# The mathematical law of evolutionary informational dynamics and an observer's evolution regularities

Vladimir S. Lerner
13603 Marina Pointe Drive, C-608, Marina Del Rey, CA 90292, USA vslerner@yahoo.com

*Abstract*

An interactive stochastics, evaluated by an entropy functional (EF) of a random field and informational process' path functional (IPF), allows us modeling the evolutionary information processes and revealing regularities of evolution dynamics. Conventional Shannon's information measure evaluates a sequence of the process' *static* events for each information state and do not reveal hidden *dynamic* connections between these events.

The paper formulates the mathematical forms of the information regularities, based on a minimax variation principle (VP) for IPF, applied to the evolution's both random microprocesses and dynamic macroprocesses.

The paper shows that the VP single form of the mathematical law leads to the following evolutionary regularities:

-creation of the order from stochastics through the evolutionary macrodynamics, described by a gradient of dynamic potential, evolutionary speed and the evolutionary conditions of a fitness and diversity;

-the evolutionary hierarchy with growing information values and potential adaptation;

-the adaptive self-controls and a self-organization with a mechanism of copying to a genetic code.

This law and the regularities determine unified functional informational mechanisms of evolution dynamics.

By introducing both objective and subjective information *observers*, we consider the observers' information acquisition, interactive cognitive evolution dynamics, and neurodynamics, based on the EF-IPF approach.

An evolution improvement consists of the subjective observer's ability to attract and encode information whose value progressively increases. The specific properties of a common information structure of evolution processes are identifiable for each particular object-organism by collecting a behavioral data from these organisms.

The applications of this approach are illustrated by the experimental facts and computer simulations with the software.

Universal nature of the *information process'* evolution dynamics and regularities of the information observers have many applications not only in cognitive and intelligent systems, but also in different biological, social and economic systems.

*1. Introduction*

In spite of numerous publications on evolution theory Dawkins 1976, Dyson 1999, Hedrick 2005. Joyce 1992, Kastler 1967 , Kimura 1983, , Michod 1999, Nicolis and Prigogine 1977, Wright 1968-69, 1977-78, others, the central questions concerning the existence of a general evolutionary law for Life, remains unanswered.

E. Schrödinger 1944, analyzing a physical aspect of Life, concludes that an organism supports itself through orders that it receives from its environment via the *maximization of a negative entrop*y, as a general principle, connecting its dynamics and stochastics.

Darwinian law's modern formula E.Mayr 2004, stating that "evolution is a result of genetic variances through ordering by the elimination and selection", focuses on a competitive "*struggle* for Life" amongst organisms.

G. Dover 2000 emphasizes on the genes' *cooperative* phenomena: "the central feature of evolution is one of tolerance and cooperation between interacting genes and between organisms and their environment… Genes are born to cooperate".
Are there any connections between all these concepts and formulas, covering them under a general law?
Most publications in mathematical biology Murray 2002, Turchin 2003, do not allow a quantitative prognosis of the law phenomena, and only support the law by experimental and/or simulation data. For example, Schwammle and Brigatti 2003 describe the result of *simulation*, depending on priory *chosen* parameters, without a detailed insight of the *dynamic* mechanism of micro-macroevolution. In such approaches, many essential evolutionary phenomena are missing.

A principal aspect of evolutionary law is an ability to *predict* the process development and phenomena based on the law mathematical forms. The existing mathematical formalism does not satisfy these requirements; the evolution equations are not bound by a general principle related to a unique law. The modern evolution theory is dominated by diverse assumptions, concepts, methods, and hypotheses. The unique formalism of evolutionary law, as well as general systemic regularities of evolution, expressed in information form, are still unknown.

We generalize different forms of evolution process considering a *flow of information* as an *information process independently of its (physical ,biological, economical, other) origin;* the flow is *freely distributing in a bounded space–time environment, which is randomly affecting the flow*.

Such approach focuses on the *process' information dynamics and its creation from stochastic dynamics*.

The questions are: What are the regularities of this process, expressed in an information form? Can these regularities be revealed and extracted through the process' observations? What are the most informative space-time observations of this flow (providing a maximum of extracted information to reveal the seeking regularities)? Which principle could be used to solve these problems?

Because of the universal nature of information, such an information approach is *applicable* to some physical processes, represented by their information models (Lerner 1999,2010,2011).

This paper goal is to find *general systemic regularities* of evolutionary process, based on *information dynamic* approach with a single variation principle as a mathematical law, and *apply* them to an information observer.

Approaching this problem with a broad and rather formal point of view, we introduce the *informational* evolutionary dynamics with the information functional mechanisms of evolution not dependable on specific material substances.

An observed *evolutionary* process is affected by *stochastic* perturbations, which activate the process dynamics, involving a wide diversity of superimposing process' of distinct nature.

Such a complex random observer can be modeled by *information* interactions, making the information description a *universal* language for disclosing the observer's information regularities.

Searching a law, which governs *complex dynamics of interacting proce*ss, led us to the *process' statistical dynamics* and then to finding a *variation principle* that, according to R. Feynman 1963, might describe regularities of such dynamics, possibly, at a macroscopic level. To get a math law, we look at the fundamental minimum principle, trying to formulate it in an information form: "Among all possible paths of this flows in this time-space, choose such one (an optimal) for which total entropy (along the path) reaches a minimum." While maximum entropy is associated with flow disordering, minimum entropy means its ordering, which implies the regularity.

The information bi-level model with a stochastic process at the microlevel and dynamic process at the macrolevel, following from the solution of variation problem for information path functional (IPF) (Lerner 2004, 2007), embraces the essential regularities of the evolutionary dynamics, such as creation of an order from stochastics, evolutionary hierarchy, others, considered here. It shows that a minimax variation principle (VP) for the IPF, as a single mathematical form of information law, defines the above regularities, and it is capable of prognosis of the evolutionary dynamics and its specific components: the evolution potentials, diversity, speed, and genetic code.

An informational aspect of evolution in Lerner 1999, 2001, 2003, 2010 symbolizes "the cooperation and ordering with a creation of information code that transfers the novelties during an evolutionary cycle."

We show that instead of a "punctuated equilibrium" Gould and Eldredge 1993, evolution depends on a punctuated *nonequilibrium* where the microlevel's stochastic fluctuations contribute to the macrolevel's dynamics, generating a dynamic potential of evolution. The applied variation principle leads to a dynamic model of *open system* with an *irreversible* macroprocess, originated by a random microprocess.

The known *variation* problems produce the equation of a *close* system with a *reversible* processes.

The observed process' path functional information measure allows revealing a *systemic* connection of the *process' events* and building an *information network* for a complex observer *multi-dimensional process*, which models their systemic *regularities*. The specifics of our approach consists in both presenting new concepts of evolutionary dynamics and in establishing them on the base of proven mathematical foundations.

Introducing an information observer, as a formal model of emergent information interactions, we disclose the observer's inner dynamics at the acquisition of external information. These dynamics follow from the interaction between the external information process with the observer's inner information process of obtaining this information. It is shown that the interactive processes of the information observer have a discrete form of information quanta, leading to both a *necessity* to create an observer's inner code and the actual *encoding* of the observed information process in this code.

The observer's informational cognitive dynamics, based on the information dynamic principles, reveal (1)*information analogs* of cognition's physical features; (2)specific neurodynamics mechanisms; (3)encoding an *observed information process in the evolution dynamic network.*

These are essential in many applications, including information decoding and recognition in learning, in biology, economics, and so on. The paper is organized as follows:

Part I. describes information regularities of a macrodynamic evolutionary process, which include

 -A study of the information process and the approach's basics (sec.1.1) ;

- The main informational dynamic regularities of evolution process (sec.1.2);

- Thermodynamics, physics, and stability of the evolution process (sec.1.3);

- Mathematical forms of the evolutionary regularities (sec1.4) and the evolution cyclic process (sec.1.5), finalized with the schema of the evolutionary information functional mechanisms (sec.1.6).

Part II introduces an information observer with its evolutionary dynamics, which include

-Notion of an information observer and its main features (sec. 2.1)*;*

-Acquision of information (sec 2.2);

-An information observer's cognitive and neurodynamics (sec.2.3) with a summary (sec.2.4)*.*

Focusing the paper on the evolutionary information regularities and the observer's cognitive evolution dynamics, we describe their essence in the main text and place the basic mathematical results in the Appendix (A0-A5).

**I. Information regularities of a macrodynamic evolutionary process**

*1.1. The studying information process and the approach's basics.*

Conventional information science generally considers an information *process*, but traditionally uses the probability measure for the random *states* and Shannon's entropy measure as the uncertainty *function* of the states (Shannon 1948, Jaynes 1963, Kolmogorov 1968, others).

A *process* is not an arbitrary sequence of states. Rather, it performs mathematical and logical operations to achieve the required goal, implemented by a mutually connected sequence of symbols. These logical connections form a composition of a finite binary tree (Pawlak 1963), or three-argument relations (Blikle 1965), characterizing the process logical structures. A logical structure of a mutually *connected information* symbols *defines* an *information process*, which takes into account its inter-symbol's statistical dependencies.

We study an observed information process using a controlled Markov diffusion process (A0) as a formal *model* of a random nonstationary interactive information process, which can *approximate* some other classes of microprocesses (Stratonovich 1966). Matthäus et al 2011, Petrovskii et al 2011 show that such shochastic interactions are a *natural* phenomenon generated by variety of biological organisms.

Following R. P. Feynman's path functional approach (Feynman and Hibbs 1965) we introduce integral *functional measure* on trajectories of the diffusion process (an entropy functional EF), which accumulates the *inner connections* between the information states of the process. It is shown (A1) that cutting off this integral information measure on the separated states' measures decreases the quantity of process information by the amount which was concealed in the connections between the separate states. The integral functional's measure accumulates *more process information than the sum of the information measures in its separated states*.

Such a measure could evaluate the process' meaningful information related to the acceptance, cognition, and perception of the information. Modeling of human multiple expressions of personal thoughts allows us a formal evaluation of these thoughts and a human personality, as well as to minimize a difference between such expressions and their meaning.

This opens of possibility of a formal measuring of both a human understanding and semantics of the information. Applying this information measure allows us modeling the *cooperative* information structures, which *enclose the concealed information and the hidden connections in evolutionary dynamics*.

The Shannon's entropy measures generally the current process' *states*, selected from a *stationary* process.

This measure evaluates a sequence of the process' *static* events for each information state and do not reveal *hidden dynamic* connections between these events.

To extract information regularities from the observed diffusion process and reveal the process' information dynamics, we apply to the EF a variation principle (VP). The VP solutions determine an information path functional (IPF) (defined on its *dynamic* trajectories) as a dynamic equivalent of the EF (defined on the random process).

The IPF set of extremal trajectories approximate the considered diffusion process (as a microprocess) with a maximal probability, forming the macrotrajectories of the microprocess. By analogy with R.P. Feynman's *path functional* approach, which includes a variation principle to obtain the equations of quantum mechanics, we obtain the equation of informational macrodynamics (IMD), following from the considered VP and IPF.

The applied VP *minimizes* the uncertainty-entropy of the process, expressing its minimum through an information path functional (IPF) on the extremal trajectories. While maximum entropy is associated with disordering and complexity, minimum entropy means ordering and simplicity. The VP extremal trajectory *minimizes a maximum* of an observable information with reaching a *minimum* of the entropy functional on this trajectory, providing maximum of delivered information with its minimum lost. These essentials carry out the VP minimax entropy principle. *Below we analyze the properties of the information process, whose evolution dynamics are described by the IMD optimal solutions.*

### 1.2. The main informational dynamic regularities of evolution process.

According to the VP solution, the extremal trajectory is divided on the extremal *segments* at the "punched" *discretely* selected points (DP), where VP imposes a dynamic constraint on the extremal solutions (A2). Between the extremal segments (at DP) exists a "*window*", where the random information affects the dynamic process, creating its piece-wise *dependency* upon both the *observed* data and the *changes*, with the dynamic model's possibility of forming an optimal *piece-wise control* (satisfying the VP), which are applied to the diffusion process.

The VP solutions allow selecting the set of states $x(\tau) = \{x(\tau_k)\}, k = 1,...,m$ at the DP-localities, forming the *boundary* points of a diffusion process: $\lim_{t \to \tau} \tilde{x}(t) = x(\tau)$, where the macrodynamics arise from stochastics and determine some *boundary* conditions, limiting the above set (Dynkin 1960, Prochorov and Rozanov 1973).

This brings a *quantum character of generation for both the macrostates and the macrodynamics information at the fulfillment of the variation condition.* The DP boundary points (with quantum macrostates) accompany an influx of maximum entropy from the stochastic process, which gives the start to a *nonequilibrium* process along each extremal segment, with reaching a local entropy's minimum by the segment end.

The above specifics follow also from the information form of Schrödinger's equation (Lerner 2010), with the solution for information waves of the IPF variation problem, restricted by a *dynamic constraint* on EF.

The synthesized optimal controls (A.3) start at the beginning of each segment, act along the segment, and *connect* the segments in the macrodynamic optimal process, where the discrete interval of the applied control is associated with the segment's length between the DP. While the EF collects the *integral information* of the random *process*, the IPF (via the VP) reveals the *dynamic regularities* of the EF collected information, allowing a dynamic *prediction* of random process.

The VP invariant conditions (A2) allows selecting and evaluating an extremal *segment*, which approximates a related segment of the random (microlevel) process (between the DP punched localities), as a segment of a *macroprocess*.

Fig.1a illustrates the selection of the process' *portions* of information carried by the VP extremal segments, separated by the "*windows*" between them.

The macromodel's dynamic operator (A3.3) is identified through the process' multi-covariations, connecting the process' states. Such an operator reveals a mutual dynamic information between the states.

At DP-localities, the collected quanta $[x(\tau)]$ determine the VP information invariants $\mathbf{a}_{io}[x(\tau)]<0$ and $\mathbf{a}_{i}[x(\tau)]>0$, evaluated a quantum information, generated by both the segment's internal macrodynamics and a step-wise control's external action (A3). (Both invariants depend on the spectral frequency parameter $\gamma$ of the segment).

Fig. 1b shows an impulse control's (IC) (A1), consisting of two-step controls (SP1, SP2) (A3), whose joint action initiates both influx of the maximum entropy and the assembling of segments, but only at the windows. Action of the SP2 control continues by the end of the extremal movement, when new pair of these controls is generated.

Possessing the invariant's information leads to a *prediction* of both each DP and the subsequent movement of the process, based on information which is collected at each previous DP.

As a result, the extremals of the macroprocess provide a dynamic prognosis of the evolution of random process.

Following from the VP, a sequence of the quantum information of the process' invariants determines an algorithm, which enables it to encode the random process.

Therefore, the VP applied to EF establishes a *mathematical form of an information law*, which is able to generate the information process and the IPF dynamic regularities, selected from a random process.

Imposing the VP implies a *discrete* sequence of the IPF extremal's segments, allowing a piece-wise prediction of the macrodynamics with the quantum measurement of the random process between the segments (at DP) and the following encoding of the process, according to the EE collected information. The process' discreteness is performed by an external control, applying to a random process and acting on each segment's border.

That is why the initial random process requires a control, which is necessary to implement the VP as the law.

Such an external control models an emerging interaction in a discrete form, which, as the law's fundamental component, exists as long as the law is fulfilled. In Lerner 2009 we detailed the macrodynamic modelling of a random process.

The time-space informational dynamics are described by a sequence of the IPF extremal segments, which form the spiral trajectories, located on a conic surface, while each segments represents a three dimensional extremal.

And the following mathematical results determine the *essentials* of the process' evolution.

*Proposition 1.1.* Current time course of the controllable information process, satisfying the VP, is *accompanied* with a sequential *ordering of both the macromodel's information spectrum* and the *time intervals* of the extremal's segments.

Specifically, for the ranged spectrum of the model operator's eigenvalues, selected at a beginning of each process' $n$ segments: $\alpha_{1o}, \alpha_{2o}, \alpha_{3o}, ..., \alpha_{io}, ..., \alpha_{no}$, where $\alpha_{1o}$ holds a maximal information frequency and $\alpha_{no}$ holds a minimal information frequency, it is *required to prove*:

(1) That $\alpha_{1o}$ *is selected from the process' shortest segment's time interval, and* $\alpha_{no}$ *is selected from the process' longest segment's time interval;* and

(2) The process' current time course consists of a sequence of the segment's *ordered time intervals* $t_k^1, t_k^2, t_k^3, ..., t_k^i, ..., t_k^n$, *with* $t_k^1$ *as a shortest segment's time interval and* $t_k^n$ *as a longest segment's time interval, while* $t_k^1$ *is the first time interval at the process beginning.*

The proof of (1) follows from invariant $\mathbf{a}_{io} = \alpha_{io} t_k^i$ (A3), which implies that each maximal $\alpha_{1o}$ corresponds to a minimal $t_k^1$, or vice versa, each minimal $\alpha_{no}$ is selected from a maximal $t_k^n$.

The proof of (2) is a result of reaching of each local minimum of the VP functional's (A1.1) derivation at the end of each segment:

$$E \mid \frac{\partial S_{ik}}{\partial t} \mid = \mid \alpha_{ik} \mid \to \min, i = 1,...,n \qquad (1.1)$$

(along the time course of the optimal process), and reaching of global minimum for this functional's derivation at the process end: $E \mid \frac{\partial S_k}{\partial t} \mid = \mid \sum_{i=1}^{n} \alpha_{ik} \mid \to \min \qquad (1.2)$

*where* $\alpha_{1k}, \alpha_{2k}, \alpha_{3k},..,\alpha_{ik},...,\alpha_{nk}$ are the segment's ending eigenvalues, connected with the eigenvalues at the segment's beginning, by the invariant relation $\alpha_{ik} = \alpha_{io} \exp \mathbf{a}_{io}$. From that, it follows that minimum of $\alpha_{1k}$ leads to minimum for $\alpha_{io}$. Thus, the eigenvalues' minimums decline consecutively, following the decrease of the functional's derivations along the process' time course. This leads to both sequential ordering of the eigenvalues $\{\alpha_{io}\}$ along the process time course, which starts with its maximal $\alpha_{1o}$ (corresponding to a relative maximum for the functional's derivation, at the process' beginning, among its declining local minimums during the time course); *and* to the ordering of the corresponding segment's time intervals $\{t_k^i\}$, starting with its minimal $t_k^1$, at the process beginning. The proposition is proved. •

<u>Collorary 1.1.</u> (1) Each process' eigenvalue $\{\alpha_{io}\}$, identified during the time course, includes the *current* process' data, coming from each punched locality DP(i) of the random process, where the identification takes place;

(2) The influx of the data starts with the identified maximal eigenvalue $\alpha_{1o}$ (at the segment with minimal time interval $t_k^1$), continues consequently with the time course, and ends with the identified minimal $\alpha_{nk}$ (at maximal time interval $t_k^n$);

(3) An hierarchy of the IN nodes, originated from maximal $\alpha_{1o}$, collects a current process information, starting with the process' minimal time interval and ending with minimal $\alpha_{nk}$ and maximal time interval $t_k^n$ - at the IN *final node*, which collects a *total* amount of data coming from all previous nodes;

(4) The IN information, acquired by each node, satisfies the VP invariant relations (A2.7, A3.5-3.7);

(5) The controls, implemented the above relations, progressively increase the number of cooperating IN's nodes (enclosed into its final node), which leads to growing the model's cooperative complexity (Lerner 2006). •

A sum of the spectrum of the model operator's $A_t$ eigenfunctions $\sum_{i=1}^{n} \alpha_{it} = TrA_t$ describes a summary relative information flow, whose integral $\Delta S = E[\int_t TrA_s ds]$ measures a *collective information* of the flow.

An ensemble of the process extremals (Fig.1b) is joining in an elementary binary unit (doublet) and then in a triplet, producing a spectrum of coherent frequencies. However such a collectivization is activated *only* at the locations of the "windows", which restricts formation of the cooperative structures. The triplet's cooperative dynamics join the extremal segments by three, or coupling sequentially seven segments in a joint triplet.

A manifold of the segments, cooperating in the triplet's optimal structures, forms a sequence on the information network (IN) nodes (Fig.1), where the IN accumulated information is conserved in the invariant form.

Local entropy minima are enclosed through sequential cooperation of the IN nodes creating an information structure which condenses the total minimal entropy produced (evaluated by the VP functional's derivation (A2)) at the end of each segment. The VP coordinates a balance between the structural information of the IN and the amount of external information. An influx of the VP entropy maxima coincides with the maximum of optimal acquisition of external information used for an external observation and prediction of the process' observed phenomena.

Thus, increasing structural information reduces the amount of external information needed.

An influx of the VP entropy maximums coincide with the maximum of acquisition of an observer's external information, used for an external observation and prediction of the process' observed phenomena (part II).

New information arises at the transferrance from the current segment, where existing information is accumulated, to a newly formed segment. The transformation is implemented by a control action, which, being initiated by the accumulated information, interacts with the microlevel information and transfers the emerged (or renovated) information to a new segment. Such new information is a result of an interaction between an existing (accumulated) information and the information currently delivered through the microlevel's windows.

The final IN node collects a total amount of data coming from all previous nodes.

The dynamic process within each segment's extremal, where the VP applies, is reversible. Irreversibility arises at the random window between segments, before the segments are joined by the controls at the DP.

Assembling of each of the three segments into a triplet is accompanied by a local instability, arising a chaotic oscillation, initiating a chaotic resonance (details in sec.1.3). The connected triplet's chain generates a *collective* resonance, where the contributions of all triplets sound in unison. This procedure synthesizes a *harmony* of the assembled triplets.

The information transformed from each IN previous to the following triplet has an increasing value, because each following triplet encapsulates and encloses the total information from all previous triplets.

The node location within the IN hierarchy determines the value of information encapsulated into this node.

A sequence of the successively enclosed triplet-nodes, represented by discrete control logic, creates the IN code (A3), as a virtual communication language and an algorithm of minimal program to design the IN.

It has been shown (A3.17-3.19) that the *quantity of information*, accumulated by the inner connections of a process, is *encoded* by the code word's length of an *optimal algorithm program*, determined by the IPF invariants which evaluate these connections. This allows us encoding the EF-IPF *mathematical and logical operations* of the *information process* (following from the VP) in a form of the IMD software (Lerner 2010).

The optimal IN's code has a *double spiral* (helix) triplet structure (DSS) (Fig.2), which is sequentially enclosed in the IN final node, allowing the reconstruction of both the IN dynamics and topology.

The IN information geometry holds the node's binding functions and an asymmetry of triplet's structures.

In the DSS *information geometry,* these binding functions are encoded, in addition to the encoded nodes' dynamic information. The DSS specifics depends on the structure of the EF functions drift and diffusion in (A0.3).

The IMD embraces both the individual and collective regularities of the information model and its elements, forming multiple connections and a variety of the IN information cooperative networks, with growing both concentration of information and its volume; while the interactions of the IN nodes are able to produce a new information.

The IN, identified for a particular process, characterizes its dynamic hierarchical *information stricture* of *collective* states, representing a *collective dynamic motion*. The IMD interactive cooperative dynamics posseses a macrosystemic *cooperative complexity* (MC) measured by an increment of quantity of information, divided on the related increment of information volume, which is formed a each cooperation (Lerner 2006, 2008). The information invariants of both information dynamics and information geometry *and* the number of joint elements evaluate the MC. In the IN, MC evaluates a total *hierarchical complexity* of the triplet's connection by both the invariants or the related DSS code.

Encoding (A3.19) provides a *shortest optimal program* (in bits) for each *IPF segment* of *the information process* that connects the considered approach to *algorithmic information theory* (AIT) (Kolmogorov 1968)*, which* measures the complexity of an object by the size in bits of the smallest program for computing it. The shortest program, which encodes the MC, connects the MC with AIT, that could be used for the evaluation of cooperative binging.

Therefore, the VP allows establishing not only information connections between the current changes (sec.A1), identified by specific sources and related information measures, but also *finding the regularities* of these connections in the form of a complex *causal-consequence relationship* and an information network, carried by EF-IPF analytics and the following logic operations, implemented by the observer (secs.2.1-2.3).

Specifically, there are three forms of causalities within the IN (Fig.1): (1) Local –within each VP extremal segment when an information force (and a control at the segment's beginning) initiates (causes) a related information flow along the segment. This is a symmetric deterministic causality, accompanied by a potential reversibility in the time of this local causal action; (2) Interim –within a "window", initiated by some external interactive actions (including a control), which cause the effect at a next extremal segment ad joint to the window. This is an asymmetric casualty, accompanied by irreversible time course and generally nondeterministic causal relationships; (3) Global–arises along the IN at the information transformation from each previous to the following node. This causality includes both the local causality (within each path between the nodes) and the interim causality (at each transformation from the node to the above path). However, the main specific of this causality consists of arising a total *sequential* causal relationships along the IN hierarchy: from the IN starting events to a first triplet's node, which joins the first three causes, and then to the following nodes, which sequentially bind all previous node information, and, finally, encloses it in the IN ending node, which accumulates all IN causalities. Thus, a mutual connection of symbols in the information process leads to a self-collectivization them into the segments and then to a *self-organization* in the IN hierarchy, with its nodes, as information objects (observers), having causal-consequence relationships.

*1.3. Thermodynamics, physics, and stability of the evolution process*

Most publications on this subject employ the models of the linear phenomenological irreversible thermodynamics, using fluctuations from a stationary state, or a quasi equilibrium process (De Groot 1962, Dyarmati 1974).

Foundation of nonlinear irreversible thermodynamics in Stratonovich 1985 is based on the *n*-dimensional correlators of Markov random process and their connections to the measured physical macrovariables.

Goldenfeld and Woese, 2011 have reviewed the evolution collective dynamics in non-equilibrium statistical mechanics, providing the genetic elements, growth of complexity, and the essentially "self-referential nature of evolutionary dynamics in biology". Some publications use Shannon's informational approach to a process of self-organization, applying a control's *parameter* for the evaluation of irreversibility in the state's transition (Zaripov 2002).

Applying the Shannon entropy measure to *n*-dimensional random *process* with the statistical dependent events leads to the unsolved problem for measuring the long terms *n*-dimensional correlations, while these events are a *naturally connected* by the entropy path functional.

We consider *information* approach with the macroprocess' irreversibility arises from a random Markov process at the entropy functional's punched localities (DP), whereas the relations for preservation energy might not be fulfilled.

The main problem consists of math difficulties of applying a *macro* evolution approach to a random process and random entropy. The invariant variation condition (A2.7, A.3.5-3.7) allows selecting and evaluating an extremal *segment*, which approximates a related segment of the random (microlevel) process (between these punched localities), as a *macroprocess*. Based on information, collected at each previous DP-locality, the macroprocess' extremals provide a *dynamic prognosis* of an evolution of the Markov process.

The considered principle of a *minimax entropy's path integral (IPF Minimax)* provides a *minimum* of entropy functional (EF) along an observed random process through an extreme of information path functional (IPF), whose extremal trajectories approximate the EF minimum with a *maximal* probability, defined along the trajectories' path, with a summary *entropy minimum* by the path end.

Maximum entropy principle (MaxEntr) (Guiasu and Shenitzer 1985, others) tends toward the entropy maximization for *evolving* states and the *maximum entropy production*. Principle MaxEntr and the related thermodynamic law mean *disorder;* the *evolution*, directed by this law, will not increase order and organization in the species, but leads to their destruction. We (as a human being) would not exist at all.

The controls, implementing the IPF Minimax, divide the extremal trajectory on segments with windows between them, where an observed information affects each segment's dynamics, and *provide a maximum entropy influx* at starting moment of each segment and a *minimum entropy production* at the segment end.

The window allocates the boundary points of an observed diffusion process, where each control's impulse (consisting of two step controls, pushing the dynamic constrain down and up accordingly) intervenes to the process, cutting it and transferring the captured information to the beginning of each following segment's dynamics. The boundary point $x(\tau-o)$ prior the new impulse control starts, is predicted by VP, and it is located at the end of the preceding extremal segment . The starting step-down control (that terminates the segment's dynamics by turning the dynamic constrain off) is applied to the diffusion process at the moment $\tau_o$, which cuts the process under this control action initiating a path from $\tilde{x}(\tau_o)$ through the control's jump-wise selection. The following step-up control action, applied to the diffusion process at the moment $\tau_d$, puts an end $\tilde{x}(\tau_d)$ to the process' controlled path between $\tilde{x}(\tau_o)$ and $\tilde{x}(\tau_d)$ and starts the dynamic process at the next extremal segment (by turning the constrain on) at a boundary point $x(\tau +o)$. A width of the control impulse between these moments on interval $\tau =(\tau_o,\tau_d)$ (defined by a function of diffusion for the controlled process on this interval) is identified through the correlations $r_{od}=E[\tilde{x}(\tau_o),\tilde{x}(\tau_d)]$ between these moments.

A selected copy of random states $-2\tilde{x}(\tau_d)$ is transferred to classical dynamic states $-2x(\tau +o)$ by a deterministic step-up control action $v=-2x(\tau +o)$, which starts the next segment's dynamics.

In the IPF Minimax, only specific path between states $[\tilde{x}(\tau_o),\tilde{x}(\tau_d)]$, predicted by this principle's VP, are selected from an observed process (at its punched localities-windows), to satisfy a *maximum* entropy (probability). Information, collected at this local path, gives a start to a *minimal entropy path*, chosen by the observer to find the path's *ending states* (producing a minimum entropy) as a stable and the most *preferable* states.

Such *optimal chosen path* reflects the *current observations with maximum probability* and instructs the observer's further evolution to be encoded in the observer's structural genetics.

The IPF minimax predicts the localization of *measurement,* when an observer's initiated measurement *interacts* with the environment, which entangles a measured Markovian path (at its quantum level) (see also Meyer *et all*, 2001), creating the mixed states of the quantum path with a corresponding quantum probability $\hat{P}=|\hat{\psi}|^2$, ($\hat{\psi}$ is a complex amplitude of a wave function), and a quantum information measure $\hat{H}=-Tr\hat{P}\ln\hat{P}$.

The information of this quantum path determines the IPF *quantum information*, which is selected at each punched locality of a random process. At the predicted moment $(\tau -o)$ of the entanglement, $\hat{P}[\hat{x}(\tau -o)]$ reaches a minimum, compared to the probabilities of related pure states $\hat{P}[\hat{x}(t)], t<(\tau -o)$.

According to the VP, the dynamic trajectory approximates the observer's measured random process with a maximal probability (on the process random trajectories). Hence, the selected information states satisfy the predicted maximum probability (in probability theory) but might not satisfy a quantum probability of chosen quantum states.

W.H.Zurek 2007 demonstrates that a predicted wave's information value encloses specific properties of the observer's measured process with a following jump-cut action on the process.

The VP predicted control jump, acting at the moment $\tau_o$, discontinues (decohers) an observed process (at these punched locations) and selects $\tilde{x}(\tau_o)$ with the related the quantum state $\hat{x}(\tau_o)$ holding maximal probability for both $P[\tilde{x}(\tau_o)]$ and $\hat{P}[\hat{x}(\tau_o)]$, while breaking a symmetry of a mixed quantum wave at $\hat{x}(\tau -o)$. Since, the control jump dissolves the correlations of the nearest states, these states become independent (orthogonal), and thus could be measured by other observers. Moreover, randomness of an observable process assumes the existence of many random paths, which can be measured at the same moment. However, de-correlation of the random process between the path states' locations $\tilde{x}(\tau_o)$ and $\tilde{x}(\tau_d)$ reduces the quantity of information bound by these states and thus, diminishes the future information outcomes from this process. The measurement along the path, brought by the impulse control action, satisfies the definition of information (A0) for each of the path's distinguished event.

The orthogonality of a current measurement $\hat{x}(\tau_o)$ also leads to selection of the orthogonal (eigen) states; the orthogonality diagonalizes the operator, which provides both the step control and this measurement with the selection of the operator's real eigenvalues. This also leads to diagonalizing of the identified correlation matrix. The controls' action at the state $\tilde{x}(\tau_d)$ replicates this state, reapplying it to the dynamic trajectory at the moment $(\tau - o)$, and also breaks a symmetry of the dynamics at these points. The selected states keep maximal probability of the random process, and their copies, produced by the step-up controls, which transfer them to *classical dynamic macrostates* on the IPF extremal trajectories. Both breaking a quantum symmetry, following the entanglement, and the de-correlation of nearest random states of the initial correlated random process, depict the interactive effect of on the observed environment.

According to Props. 2-4 (sec.2.2), the control, implementing above operations, binds the observer with its environment during the interactive dynamics and spends the quantity of information $\mathbf{a}_o^2(\gamma)$ on the process measurement, transferring this information to the dynamics, and starting each dynamic movement.

The above operations, working insight of an interactive process, link up randomness, quantum information with classical dynamics.

Increasing the order in evolutionary processes requires an external flow of entropy (negentropy) in *open thermodynamic systems*, associated with the *irreversibility and dynamics of the evolution*.

The IPF requires a maximum influx of the external entropy, implemented by the impulse control, acting on the path's beginning (during a very short time), holding the minimal entropy production along the optimal path and the process' balance and stability. This provides a gradual entropy's minimization and reaching the entropy's minimum by the path end and corresponds to Schrödinger's 1944 concept: "the only way to contradict that destructing Second Law is to compensate it by a minimization of entropy and/or entropy production, what all living and evolving species do".

According to the principle of minimum entropy production, proposed by Prigogine 1977, any non-equilibrium process evolves to a stationary state (process) with a *minimal entropy production* (if the system is stable by Lyapunov Criteria, which is satisfied in the evolving and *living* species).

The entropy functional has an analogy with the Onsager-Machlup functional in Nonequilibrium Irreversible Thermodynamics (NIT) (Durr and Bach 1978), and IPF-IMD equations represents the *informational form* of the NIT equations, considered under the action of control functions:

$$\dot{x} = \overline{L}X, \ \dot{x}_t = I, X = \frac{\partial S}{\partial x}, \ \overline{L} = 2b(x,t), \tag{3.1}$$

where $I$ is a vector of a information flow, $X$ is a vector of a information force, $\overline{L}$ is a kinetic operator (which generally could be asymmetrical and a nonlinear). In *irreversible kinetic macroequation* (3.1), the irreversibility is a consequence of the jump-wise changing of the equation's operator at the DPs with transferring process' diffusion into kinetics and vice versa along with the renovating operator on a new extremal. In the IMD model, the DPs are the crucial points of changes in a dynamical evolution with the fixed path functional's production rate:

$$E[-\frac{\partial S^i}{\partial t}(\tau)] = E[H(\tau)] = Tr[A(\tau)] = \sum_{i=1, j=1}^{i=n, j=m} \lambda_i(\tau_j) > 0, \tag{3.2}$$

given by the sum of positive Lyapunov characteristic exponents (LCE) $\lambda_i(\tau_i)$ (the operator positive eigenvalues) which coincides with the Kolmogorov entropy (Kolmogorov 1968) at these crucial points.

The IMD model holds the qualities of an *open system* such as the nonlinearity and irreversibility at the DPs and the stationary and reversibility within each extremal segment, corresponding to a conservative system. These "punched" DP-localities provide a *conjugation between the kinetics and diffusion(* $\sigma\sigma^T$ *)*: at $\overline{L}(\tau-o) \geq \sigma\sigma^T$ the kinetic flow transfers to diffusion; at $\sigma\sigma^T \geq \overline{L}(\tau+o)$ the diffusion flow transfers to kinetics (Lerner 2006a).

It implies a spontaneous *evolution transition* from the most probable microstate ensemble to a macrostate and vice versa. Function of dispersion $b = b(\tilde{x}(\tau)) = 1/2\sigma\sigma^T$ identifies the dynamic operator at macrolevel (A3.3a):

$$|A(\tilde{x}(\tau))| = b(\tilde{x}(\tau))(2\int_{\tau-o}^{\tau} b(\tilde{x}(\mu))d\mu)^{-1}, \quad (3.3)$$

which is responsible for the macrolevel's dynamics, including the *deviations* of macrolevel trajectory, being reversible within each extremal segment and irreversible between the segments. These results are confirmed experimentally in the examples with hydrodynamic fluid (Hurtago et al 2011). Relations (3.4), following for VP, provide the equations of the *detailed balance* for the optimal path along a random process at each punched locality $\tau_j$:

$$b_i(\tilde{x}_i(\tau_j))(\int_{\tau_j-o}^{\tau_j} b_i(\tilde{x}_i(\upsilon_j))d\upsilon_j)^{-1} = b_k(\tilde{x}_k(\tau_j))(\int_{\tau_j}^{\tau_j+o} b_k(\tilde{x}_k(\mu_j))d\mu_j)^{-1}, i,k=1,...n; j=1,...,m. \quad (3.4)$$

According Prop.1.1, the operator's eigenvalues decrease along the evolution of optimal trajectory, which leads to decreasing also the corresponding relative *deviations* of macrotrajectory, defined at each DP (3.3).

A sequential cooperation of the extremal segments converges them into a joint movement with a common dynamic trajectory having a single-dimensional dispersion's *deviations*. These peculiarities of informational macrodynamics have important applications in irreversible nonlinear thermodynamics (Toyabe *et al* 2010), others.

For a physical model, each its dimension represents a particular physical process, such as chemical, electrical, diffusion, mechanical, etc.; and their interaction creates new cross phenomena like thermoelectrical, electrodiffusion, electrochemical, and electromechanical phenomena.

The *multi-dimensional* superimposition embraces a potential *inner* "needle" control action (Fig.1b), associated with modeling of the process' jump-wise interactions. In a chain of the superimposing processes $x_i(t,\tau), i=1,2,.....n$, each current process ($i$) controls one or more of the following chain's processes ($i+1=j$) with a possibility of changing the operator in the NIT macroequation

$$\frac{dx_i}{dt} = \sum_{i,j=1}^{n} L_{ij} X_j, i \neq j = 1,...,n. \quad (3.5)$$

The mutual cross phenomena, modeled by the applied control functions, connect the chain by relations

$$\frac{dx_i}{dt} = L_i X_i + u_i, \ u_i = \sum_{j=1}^{n} L_{ij} X_j, i \neq j, \quad (3.5a)$$

where the controls at the current *i*-segment:

$$u_i = u_i(\delta v(t',l')) \quad (3.5b)$$

include the corresponding needle control, applied at the DP-space –time *localities* $(t',l')$:

$$\delta v(t',l) = -2x(t',l) + 2x(t'+o,l), \quad \delta v(l',t) = -2x(l',t) + 2x(l'+o,t), \qquad (3.6)$$

which arrange the cross connections and change the operator components with a potential operator's renovation.

This means that, in modeling complex dynamics, the considered controls performs *function of a cross-interacting superimposing processes,* connecting the segments. The vice versa is also true: in complex dynamics, the superimposing processes are modeled by the above control's functions (see also Lerner 2009).

These controllable self-cooperations generate the IN dissipative structures.

In fact, the waves with the natural *high pulses*, analogous to the impulse control, have been created according to the Peregrine's solution to the non-linear Schrödinger equation (Chabchoub et al 20011).

The kinetic operator for time-space movement:

$$\bar{L}(x(t',l')) = 2b(x(t',l')) = E_{x(t,l')}[x(t,l)\frac{\partial x}{\partial t}(t,l)^*] = E_{x(t',l)}[\frac{\partial x}{\partial t}(t,l)x(t,l)^*] \qquad (3.7)$$

is changing by a jump at each time-space point ($t'$, $l'$) of applying the control where the relations (3.4) in the form

$$\frac{I_i}{X_i}(t_j',l'_j) = g_i(t_j',l'_j) = g_k(t_j',l'_j) = \frac{I_k}{X_k}(t_j',l'_j), \qquad (3.7a)$$

take pace; here $g_i$, $g_k$ are the subsequently equalized components of the generalized transient conductivity (admittance). The sequence of the chain dependable $n$-controllable components of these conductivities can be reduced to a single currently controllable conductivity, for example, to electrical conductivity (which is proportional to the diffusion conductivity) and whose measuring is the simplest. Following that and the connection between the diffusion and electroconductivity, an average function (3.2) of the controlled process in the IMD equations:

$$\frac{\partial \Delta \hat{S}}{\partial t} = 1/2\dot{\sigma}_e \sigma_e^{-1}, \quad \int_t^{t+\tau} b(t)dt = \sigma_e(\tau), \qquad (3.8)$$

can be expressed via the nonlinear electroconductivity $\sigma_e = \sigma_e(\tau)$, measured at the discrete points $\tau$; where (3.8) serves as an indirect *indicator* of the IPF's averaged Hamiltonian (3.2).

The nonlinearity of matrix $\bar{L}$ (3.7) is a result of interactions, new effects and phenomena at the superimposition, which are modeled by the consolidation process of forming the cooperative macrostructure.

In the space-consolidated model, these processes involve diagonalizing of the dynamic operator under the periodical rotations of a symmetrical transformation (Lerner 2008). Such a procedure also decreases the number of components of the path functional's Lagrangian, minimizing the entropy production. This means that the space movement, directed toward diagonalizing of the dynamic operator (as an attribute of the space consolidation process), is a source of the generation of an *additional* negentropy for the cooperation.

The cooperation brings a physical analogy of the states' superposition into a *compound* state, accompanied by a nonsymmetry of the formed *ordered* macrostructures, the *irreversibility* at each DP, and the emerged nonlinear phenomena, accompanied by the creation of new properties.

Applying the IPF leads to revealing complex regularities and uncertainties of a random process by building a system of the INs with the information invariants and encoding a chain of the events, covered by the random process.

The IN's evolutionary function (3.2) forms a ranged sum, satisfying the VP. Its maximal and minimal values characterize the maximal and minimal speeds of evolutionary process, which are sequentially connected by the IN structure.

This allows evaluate the cooperative complexity for all process, as well as its specific measure (3.2) at each stage of evolution.

The IN final node's eigenvalue characterizes both the system's terminal evolutionary speed and the system cooperative complexity (Lerner 2006).

The amount of information, being spent on transformation from EF to IPF, measures the conversion from disorder to order and from nonsymmetry of the random processes to symmetry of the time –forward and time-backward dynamics at each extremal segment. Each segment's local consolidation breaks this symmetry at DPs and restores it after the control starts the following extremal's motion on this segment. The amount of information, evaluated this control action, measures the above break of symmetry and of the order.

The IPF measures the process uncertainty by the entropy functional and allows minimizing uncertainty by the optimal control actions. The IPF Hamiltonian $H = -\partial \Delta S / \partial t$, which determines both the instant entropy production and the macromodel operator, also defines the Lyapunov function and the LCE of the process stability, connecting the stability to the process uncertainty. The process optimization by the controls actions changes the LCE sign at the DP enabling the stability of cooperative process concurrently with the minimization of its uncertainty.

The above results connect the model's Uncertainty, Regularity, and Stability.

These phenomena allow applying the IMD model for a wide class of real systems, which can exhibit the above behaviors at different stages of dynamic evolution (Prigogine 1977,1980).

***1.4. Mathematical forms of the evolutionary regularities and a unified law of evolution.***

We finalize the above results by the following mathematical regularities and a unified law.

1. *The mathematical law of creation the dynamic regularities from stochastics* is defined by the dependency of both the kinetic operator and a local *gradient of dynamic potential $gradX(\tau)$* (A.2.8) on a dispersion matrix in the form

$$L_k(x,\tau+o)=1/2b(x,\tau+o), \quad (\int_{\tau-o}^{\tau} b(t)dt) = 2M[\tilde{x}(\tau)\tilde{x}(\tau)^T] \qquad (4.1)$$

$$gradX(\tau+o) = (\int_{\tau-0}^{\tau+o} \sigma\sigma^T dt)^{-1}, \sigma\sigma^T = 1/2b, \qquad (4.2)$$

which define also the related drift vector in the entropy functional (A1.1):

$$a^u(\tau+o) = \sigma\sigma^T(\tau+o)X(\tau+o) \qquad (4.3)$$

and the dynamic operator in (3.3), where the potential connects the dynamics and stochastics.

At a growing of the dynamic gradient (as a feedback from dynamics), the diffusion part in (4.2) tends to decrease, which diminishes an effect of randomness on the dynamics, and vice versa. The dynamic potential, which exists only along the extremal segments, depends on the diffusion, identified at the DP punched localities between the segments. These Eqs connect the dynamics and stochastics, explaining how do the dynamics inherit the randomness, selected at the DPs, and retain them in the information invariants, which determine the dynamic movement renovated at each extremal segments .

2. *The law for the gradient of dynamic potential, local evolutionary speeds, the evolutionary conditions of a fitness, and a life-time*.

At the beginning of each extremal, the gradient of potential reaches a local maximum (as the inpulse control dissolves correlations in (4.1)), and then evolves on the extremal, according to (4.1,4.2) in the inverse proportion to the correlation *on* the extremal:

$$gradX(t) = (E[x(t)x(t)^T])^{-1}. \quad (4.4)$$

Considering each extremal movement $x(t)$ within the segment time interval $t_k = (t_k^i)$ (between the moment $t_o \leq t < t_k$) and using the equation of extremal (A.3.1), we get

$$E[x(t)x(t)^T] = E[x(t_o)x(t_o)^T]exp(2At), A = A^T, A < 0, t_o = \tau + o, \tau = \{\tau_k\}, t_o = \{t_o^i\}, t_k = \{t_k^i\}, i = 1,...,n, n \geq m,$$

which at the moment $t_k$ brings the gradient to the form

$$gradX(t_k) = gradX(t_o)exp(2At_k), gradX(t_o) = gradX(\tau + o). \quad (4.4a)$$

This means that, at a stable extremal movement (with $A < 0$), the local maximum of the gradient (at the DPs, corresponding $A(\tau) > 0$) declines, reaching by the segment's end the local minimums according to (4.4a).

Along the evolution dynamics, represented by a chain of the extremals, the local minimums of the potential's gradients are diminishing, as well as the related local eigenvalues are lessening (according to Prop.2.1).

During the extremal movement, function (1.1) has the form:

$$E[\frac{\partial S_i}{\partial t}(t)] = c_i \lambda_i(t), |\lambda_i(t)| = \alpha_i(t) \pm \beta_i(t), c_i \leq 4; \lambda_i(t) = diagA(t), t \leq t_k, (4.5)$$

which decreases, reducing $|\lambda_i(t)|$ that approaches (1.2) with real eigenvalues (at $t = t_k$, $\beta_i(t_k) = 0$).

The real eigenvalues $\alpha_i(t)$, determined by function (A3.4) and invariant $|\alpha_{io}t| < |\alpha_{io}t_k| = \mathbf{a}_o$, at any $t < t_k, \alpha_{io} < 0$ decline, approaching a minimum at $t = t_k$:

$$\min E | \frac{\partial S_{ik}}{\partial t}| \to \min |\alpha_i(t_k^i)|, i = 1,...,n \ . \quad (4.6)$$

The invariant condition of the minimum derivation (4.6) at the end of each extremal segment leads to relation

$$\min \alpha_i(t_k^i) = \min \alpha_j(t_k^j), (\alpha_i(t_k^i), \alpha_j(t_k^j)) = diagA(t_k), \quad (4.7)$$

which allows joining the extremals in the IN (Fig.1) with the local information speeds, satisfying (4.6,4.7).

Such a cooperation is constrained by the eigenvalues (eigenvectors) of the *same* matrix operator *A*.

The gradient of potential is connected with the matrix *A* by relation

$$\frac{d}{dt}\ln[gradX(t)] = \frac{d}{dt}gradX(t_o) + 2A = \frac{d}{dt}\ln[gradX(t)] = 2A, t \leq t_k, (4.8)$$

following from (4.4, 4.4a). At a negative *A*, this relative speed of the gradient is negative on the extremal.

From (4.8) we get the invariant condition analogous to (4.7) and (3.4) at these points in the form:

$$\min | \frac{d}{dt}[gradX_i(t_k^{i-1})]\{[gradX(t_k^{i-1})]\}^{-1}| = \min | \frac{d}{dt}[gradX_i(t_k^i)]\{[gradX(t_k^i)]\}^{-1}, i = 1,...n. \quad (4.9)$$

These conditions (connecting the *local* process' *speeds*) express a requirement for the VP extremals to *fit* each other in the *evolutionary cooperative* dynamics.

While each local maximum of the dynamic gradient intensifies an impact of the following dynamics on the evolutionary development, its decrease by each segment's end tends to decrease the influence of the diffusion part in (4.2) (as a feedback from dynamics); this diminishes an effect of randomness on the dynamics, and vice versa.

The law for an *average information speed* along the trajectory of *evolutionary dynamics* (1.1):

$$E\left|\frac{\partial S(t_k)}{\partial t}\right| = \left|\sum_{i=1,k=1}^{n,m} \alpha_i(t_k^i)\right| \to \min, \quad (4.10)$$

depends on the entire spectrum's eigenvalues, or a total dynamic potential, representing the whole system (as a population), which accumulates an emergence of new random features for a whole population.

The speed grows with the addition of each following eigenvalue, even though this eigenvalue decreases.

A local speed, measured by a *decreased eigenvalue* in the above trace, *declines,* which weakens the evolutionary dynamics' information speed. A minimal *acceptable* eigenvalue, corresponding to a local maximum of the dynamic gradient, limits the evolutionary speed. The evolution speed grows with enlarging the population's dimension.

The evolutionary model possesses two scales of time: a reversible time equals to a sum of the time intervals on the *extremals* segments: $\sum_{i=1}^{n} t_k^i = T_e^r$, and the irreversible time, counted by a sum of the irreversible time elementary intervals *between* the segments: $\sum_{i=1}^{n} \delta t_i^k = T_\tau^{ir}$ at $\tau_k \cong \delta t_i^k = (\Delta S_i^\delta / \mathbf{a}_o^2 - 1) t_i^k$. (4.11)

A sum $T_e^r + T_\tau^{ir} = T_m^l$ (4.11a) defines the model lifetime.

Here $\Delta S_i^\delta$ is an elementary information contribution between the segments, generated by a random process. If each $\Delta S_i^\delta$ between the segments is compensated by the needle control action with $\Delta S_i^\delta = \mathbf{a}_o^2$, then the irreversible time does not exist, means that the macroprocess' dynamics cover the random contributions. At any $\Delta S_i^\delta > \mathbf{a}_o^2$, the stochastics affect dynamics bringing the macroprocess' irreversibility. Both reversible and irreversible time intervals depend on the quantity and quality of information accumulated within these discretes, which, measured by the considered information invariants, reveal the *time's quantum informational nature*.

Both of these times evolve with the evolution of the invariants and the system information dynamics.

4. *The law for diversity, variations, stability, a maximal potential of evolution, and complexity*.

The relative increment of the averaged information speed (4.10):

$$E\left|\frac{\partial S_i^*}{\partial t}\right| = 4 \sum_{i=1}^{i=n_i \pm \Delta n_i} \left|\frac{\Delta \alpha_{io}}{\alpha_{io}}\right|, \alpha_{io} = \operatorname{Re}\lambda_{io}; D = \sum_{i=1}^{i=n_i \pm \Delta n_i} a_{io}^*(n,\gamma), a_{io}^* = \left|\frac{\Delta \alpha_{io}}{\alpha_{io}}\right|. \quad (4.12)$$

is determined by diversity $D$, measured by the sum of the *admissible variations* for the system's eigenvalues spectrum, which *preserves* the spectrum *stability* at current derivations. Whereas each variation is produced by the increment of diffusion according to (4.1), or the related dynamic potential (4.2).

A max $D = D_o$ defines the spectrum's maximal variation, limited by a system's ability to sustain the spectrum extended *dimension* $n_{im} = n_i \pm \Delta n_i$. Relation $D_o = P_e$ measures a maximal potential of evolution $P_e$ and brings a maximal increment of the average information evolutionary speed

$$H_\Delta^o = \max E \mid \frac{\partial S_i^*}{\partial t} \mid = P_e = D_o \qquad (4.13)$$

for a given macrosystem's dimension, related to the size of a population. Extending the population's size by increasing its dimension $n$ (satisfying the stability) raises the potential of evolution. The model limitations on the potential of evolution $P_e = \max \sum_{i=1}^{i=n} a_{io}^*(\gamma)$, which counterbalances to the admissible eigenvalue's deviations, are determined by the fulfillment of the following equations, preserving each triplet's dimension:

$$\alpha_{io}^* = 1 - \frac{\alpha_{i+1,o}^t \alpha_{i-1,o}^t}{(\alpha_{io}^t)^2} = 1 - \frac{\gamma_{i-1}^\alpha}{\gamma_i^\alpha} = \frac{\Delta \gamma_i^\alpha}{\gamma_i^\alpha} = \varepsilon, \; \mathbf{a}(\gamma) \gamma_i^\alpha(\gamma) = inv,$$

$$\gamma_i^\alpha(\gamma) = \frac{\alpha_{io}^t}{\alpha_{i+1,o}^t} = \frac{t_{i+1}}{t_i} = inv, \; \frac{t_{i-1}}{t_i} = \frac{t_i}{t_{i+1}} = inv, \qquad (4.14)$$

where $\gamma_{i-1}^\alpha(\gamma), \gamma_i^\alpha(\gamma)$ are the triplet eigenvalue's ratios and $t_i, i = 1,...n$ are the segment's time intervals.

A triplet's ability to sustain the admissible eigenvalue's deviations will be compromised, if either $\alpha_{io}^t$ approaches $\alpha_{i-1}^t$, or $\alpha_{i+1}^t$ approaches $\alpha_{i+1,o}^t$, *because* if both $|\alpha_{io}^t - \alpha_{i-1,o}^t| \to 0, |\alpha_{io}^t - \alpha_{i+1,o}^t| \to 0$, the triplet disappears, and the macromodel dimension decreases according to $m = \frac{n-1}{2}$. The above minimal eigenvalue's distance is limited by the admissible minimal relative $|\frac{\Delta \alpha_{io}^t}{\alpha_{io}^t}|_o \cong 0.0072$ at perturbations $\alpha_{it}^*$. The maximal distance between both $\alpha_{i-1,o}^t$ and $\alpha_{io}^t$, $\alpha_{io}^t$ and $\alpha_{i+1,o}^t$ at the fixed $\alpha_{i-1,o}^t$, $\alpha_{i+1,o}^t$ satisfies to the known condition of dividing a segment with $(\alpha_{i-1,o}^t, \alpha_{io}^t, \alpha_{i+1,o}^t)$ in the mean and extreme ratio: $\frac{\alpha_{i-1,o}^t}{\alpha_{io}^t} = \frac{\alpha_{io}^t}{\alpha_{i+1,o}^t}$, which coincides with the above formulas for the triplet's invariant (4.14).

The triplet's *information capacity* to counterbalance the maximal admissible deviations defines the triplet's potential $P_e^m$:

$$P_e^m = \max \frac{\Delta \gamma_i^\alpha}{\gamma_i^\alpha} = \max |\varepsilon(\gamma)|. \qquad (4.15)$$

For an entire macromodel with $m$ such triplets, the total potential is $P_e^n = m \, P_e^m$. The macromodel potential $P_e^n$ is limited by the maximum *acceptable* increment of dimension that sustains the macrostates' cooperation:

$$P_e^n \cong 1/3 \, m, \; m = \frac{n-1}{2}. \qquad (4.16)$$

A *maximal* $\varepsilon(\gamma)$ at the permissible deviations of $\gamma_i^\alpha(\gamma)$, which preserves the triple macrostates' *cooperation* and the triplet formation, corresponds to the minimal admissible $\gamma \cong 0.00718$, which for $\gamma_1^\alpha = 2.46$, $\gamma_2^\alpha = 1.82$ brings $\max \varepsilon \cong 0.35$ and $P_e^n = m \, P_e^m \cong 1/3m$.

Potential $P_e^n$ differs from $P_e$ which generally does not support the evolution process's hierarchy.

The model's *acceptable* potential $P_e^n$ that can adapt the variations not compromising the IN hierarchy, we call *adaptive potential* $P_a = P_e^n$. Relation $P_e^n \leq P_e$ limits the variations, acceptable by the model that restrict the maximal increment of dimension and sustain the model's cooperative functions. At the variations within the $P_a$ capabilities, the generated mutations enable creating a new dimension with a trend to minimize both $\gamma$ and the system uncertainty. Extending the $P_a$ capability can bring instability, accompanied by growing $\gamma$ with a possibility to jeopardize the system cooperation. The required spectrum's *stability* at current derivations brings a direct $P_e$ dependency on the maximum *admissible deviations*, which preserve triplet's *invariants*, or a fixed $\gamma$ and the triplet code. If the fluctuations generate the admissible deviations, the model enables generating the *feedback-controls* that support a *structural robustness* for each IN's node. In this case, $P_e(\gamma) = P_r$ represents the model's *potential of robustness,* which expresses its ability to counterbalance the admissible eigenvalue's deviations (acting within the $P_r$ capacity). The *triplet's robustness*, which preserves the triplet's invariants under admissible maximal error (at current and fixed $\gamma = \gamma^*$), is satisfied if the related adaptive potential holds the relation analogous to (4.15):

$$P_a = \max \varepsilon(\gamma^*) . \qquad (4.17)$$

In a chain of connected information events, the appearance of an event, carrying $\gamma \to 1$, leads to $\mathbf{a}_o(\gamma = 1) = 0$ when both information contributions from the regular control $\mathbf{a}(\gamma = 1) = 0$ and the needle control $\mathbf{a}_o^2(\gamma = 1) = 0$ turn to zero. At a locality of $\gamma = 1$, both the event's information $\mathbf{a}_o$ and the related time undergo a jump, which could be the indicators of approaching $\gamma = 1$. This means that the appearance of an event, carrying $\gamma \to 1$, leads to a *chaos and decoupling* of the chain, whereas the moment of this event's occurrence could be *predicted* (A.3).

These results represent the mathematical formulas and the regularities, related to Darwinian evolution, where each local variance is fixed by the cooperative binding at the localities of the punched points and it is encoded in the DSS.

The cooperative information complexity, defined by a ratio of the average information speed (4.12), in the cooperative evolution dynamics, *to* a cooperated volume, *increases* with growing the dimension of a cooperating system (Lerner 2006). The evolution process accompanies an *increase* of complexity for each new formation and *decrease* of its speed of evolutionary advancement, related to a previous formation.

5. *The law of evolutionary hierarchy, stability, potential adaptation, adaptive self-controls and a self-organization; coping, genetic code, and the error correction*.

The evolutionary dynamics, created by the multidimensional eigenvalues' spectrum, forms a chain of interacting extremal segments, which are assembled in an *ordered organization* structure of the information hierarchical network (IN).

The requirements of preserving the evolutionary hierarchy (4.14-4.16) impose the restrictions on the maximal potential of evolution $P_e$ and limit the variations, *acceptable* by the model.

The model's adaptive potential $P_a \leq P_e$, which adapts the variations, not compromising the IN's hierarchy, *restricts* the maximal increment of dimension, contributed to the adaptation.

The punched evolution's nonequilibrium accumulates the variations by the hierarchy of selections and adaptations, with following a local equilibrium at each hierarchical level.

The adaptive model's *function* is implemented by the adaptive *feedback-control*, acting within the $P_a$ capabilities.

The self-control function is determined by the conditions (3.5a-b) and (3.4) of a proper *coordination* for a chain of superimposing processes, where each preceding process adopts the acceptable variations of each following process.

The above optimal controls are synthesized, as an *inner feedback*, by the *duplication* of and *copying* the current macrostates at the beginning of each segment, which are *memorized* and *applied* to the segment (A.3).

Let us show that at the equal deviations of the model parameter $\gamma_i^\alpha$: $\pm\Delta\gamma_1^\alpha, \pm\Delta\gamma_2^\alpha$, the model threshold $|\varepsilon(\Delta\gamma)|$ is asymmetric. Indeed. The macromodel with $\gamma = 0.5$ has $\gamma_1^\alpha = 2.21$, $\gamma_2^\alpha = 1.76$ and get $\varepsilon_o = 0.255$.

Admissible deviations $\Delta\gamma_1^\alpha = 0.25, \Delta\gamma_2^\alpha = 0.08$ correspond to the macromodel with $\gamma \cong 0.01$, $\gamma_1^\alpha = 0.246$, $\gamma_2^\alpha = 1.82$, which determines $\varepsilon_1 \cong 0.35$ and $\Delta\varepsilon(\Delta\gamma_1^\alpha, \Delta\gamma_2^\alpha) = \varepsilon_1 - \varepsilon_o = 0.095$.

At $\Delta\gamma_1^\alpha = -0.25$, $\Delta\gamma_2^\alpha = -0.08$ we have the macromodel with $\gamma = 0.8$, $\gamma_1^\alpha = 1.96$, $\gamma_2^\alpha = 1.68$, which determines $\varepsilon_2 \cong 0.167$ and $\Delta\varepsilon(-\Delta\gamma_1^\alpha, -\Delta\gamma_2^\alpha) = \varepsilon_2 - \varepsilon_o = -0.088$.

It is seen that at the equal deviations, the model potential, defined by

$$\max \varepsilon = \max \varepsilon_1 \cong 0.35, \tag{4.18}$$

tends to *increase at a decreasing* of $\gamma$, and vice versa. An essential *asymmetry* of $|\varepsilon(\Delta\gamma)|$ and therefore the dissimilar $P_m$ are the result of the macromodel fundamental quality of *irreversibility*.

The adaptive potential's *asymmetry* contributes the model's evolutionary improvement.

The space distributed IN's *structural robustness* is preserved by the feedback actions of the inner controls, which provide *a local stability at the admissible* variations.

This control supports a limited $P_e(\gamma) = P_r$ that determines the *potential of robustness*.

An individual macromodel *lifetime* (4.11a) is *limited* by both the number *n* of the ranged eigenvalues (the local entropy's speeds) and a maximal value of this entropies. At growing these information speeds and preserving the system's invariants, the macromodel's *life-time is shortening*. The IN's life-time depends only on the entropy speed of the IN's finally cooperating segment. A sequence of the sequentially enclosed IN's *nodes*, represented by a *discrete control logic*, creates the IN's *code* as a *virtual communication language and* an algorithm of minimal program to *design* the IN.

The optimal IN's code has a *double spiral* triplet structure (DSS), shaped at the localities of the sequential connected cones' spirals, which form the time-space path-line of transferring the IN's information through the triplet's hierarchy.

The applied control adds the forth letter to the initial minimal three triplet's code letters, which provides the model's *error correction mechanism* to the IN and its DSS code; it also provides discrete filtering of the randomness, acting at the DP-window (Lerner 2001). The IN information geometry holds the node's binding functions and an asymmetry of triplet's structures. In the DSS *information geometry*, these binding functions are encoded, in addition to the encoded nodes' dynamic information. The DSS specialization depends on the structure of the EF functions drift and diffusion. The IN's geometrical border forms an external surface Fig.5, (sec.A5), where the macromodel is open for the outside interactions and the interacting states *compete* for delivering a maximum of information. Selected states are copying and memorized by the model's control, contributed to the code. The IN's evolutionary dynamics accumulate the cooperative contributions, leading to a renovation of the information invariants and DSS code.

The control provides a *directional* evolution with the extraction of a maximum information from the environment of competing systems, while the acquired DSS code can be passed to a successor (sec.1.5).

Therefore, evolution dynamics, which follow from VP, is accompanied by the following main *features*:

-A tendency of decreasing $\gamma$ that *diminishes* the influence of randomness on the macrolevel dynamics;

- A decrease of the contribution from diffusion with the increase of the dynamic gradient that intensifies a *growing impact of the each following dynamics* on the evolutionary development;

-An average evolutionary information speed (for a population of subsystems) that declines during the time of evolution with a limited individual life-time;

-A nonsymmetrical adaptive potential that leads to both rising the system's adaptive capability (at decreasing $\gamma$) with expanding the potential ability to correct a current error *and* increasing of the impact of a dynamic prehistory on current changes;

-Forming the organization structures, ordered in the evolution hierarchy of information dynamic networks with growing information values and complexity along the hierarchy. Such organization structures should be capable of adaptive self-functioning, which limits their minimal IN's nodes number and the model's dimension by eight.

More complex organization are composed from the set of these minimal ones, collected from a single and/or different discrete ensembles, which satisfy the conditions of their cooperation.

Therefore, *emergence of such organizations is possible by discrete evolutionary steps, having a nonreducable hierarchical cooperative complexity* (Lerner 2006).

Thus, the VP is *a single form of mathematical law that defines the above regularities, which prognosis the evolutionary dynamics and its specific components: potential, diversity, speed, complexity*.

*1.5. About the information macroprocess' cyclicity*

The macrosystem, which satisfies the VP information law, is adaptive and self-controlled.

The controlled macroprocess, after a complete cooperation, is transformed into an one dimensional consolidated process

$$x_n(t) = x_n(t_{n-1})(2 - \exp(\alpha^t_{n-1} t)), \quad (5.1)$$

which at the moment $t_{n-1} = \dfrac{\ln 2}{\alpha^t_{n-1}}$ is approaching the final state $x_n(t_n) = x_n(T) = 0$ with an infinite phase speed

$$\frac{\dot{x}_n}{x_n}(t_n) = \alpha^t_n = -\alpha^t_{n-1} \exp(\alpha^t_{n-1} t_n)(2 - \exp(\alpha^t_{n-1} t_n))^{-1} \to \infty \quad (5.2)$$

The model cannot reach the zero final state $x_n(t_n) = 0$ with $\dot{x}_n(t_n) = 0$, and therefore, to finish a stable movement.

A periodical process arises as a result of alternative movements with the opposite values of each *two* relative phase speeds $\dfrac{\dot{x}(t_{n+k})}{x_n(t_{n+k})}$, $k = 1, 2$ and under the control switches. The process is limited by the admissible error $\varepsilon^* = \dfrac{\Delta x_n}{x_n}$ and/or the related time deviations of switching control $\varepsilon^*_m \leq \Delta t_i / t_i = \alpha^*_{io}$.

Such a nonlinear fluctuation can be represented (Kolmogorov 1978) by a superposition of linear fluctuations with the frequency spectrum ($\omega^*_1, ..., \omega^*_m$) proportional to the imaginary components of the eigenvalues ($\beta^*_1, ..., \beta^*_m$), where $\omega^*_1$ and $\omega^*_m$ are the minimal and maximal frequencies of the spectrum accordingly.

In A.4. we show that the nonlinear fluctuations are able to generate a new model with the new real and imaginary eigenvalues, which can give a start to a new born macroprocess and the following IN cooperative macrodynamics.

This leads to the model's *cyclic functioning*, initiated by the two mutual controllable processes, which should not consolidate by the moment of the cycle renovation.

In such a cyclic macromodel functioning, after the model disintegration, the process can repeat itself with the state integration and the transformation of the imaginary into the real information during the dissipative fluctuations.

The initial interactive process may belong to different macromodels (as "parents"), generating a new macrosystem (as a "daughter"), at the end of the "parents" process and the beginning of the "daughter's generation.

The macrosystem, which is able to continue its life process by renewing the cycle, has to transfer its coding life program into the new generated macrosystems and provide their secured mutual functioning.

The genetic code can reproduce the encoded macrosystem by both decoding the *final* IN's node with a reproduction of a related IN, and decoding the specific position of each node within the IN structure. An invariant information unit of this cycle $h_{oo} = \mathbf{a}_o(\gamma) = C_p$ defines channel capacity $C_p$, which accumulates and encodes the "parents" model's parameters, transferring the inherited code's genetics to a "daughter".

The structural stability, imposed by the VP, affects a restoration of the system structure in the cyclic process through a reproduction. A new born system sequentially evolves into anew IN that embraces its dynamics, geometry, and the DSS code. The optimal control, which *preempts* and determines the process dynamics at each extremal segment, is memorized at the segment's starting time moment. This means that such a memorized control *sequence* (a code) can remember both the previous and the prognosis segment's dynamics. Therefore, the DSS code, represented by the digits of these control sequence, is able to remember a preceding behavior of a complex system and pass it to a successor.

As a result, the successor might do both repeats some behavior of its predecessor and anticipates its own future behavior.

Among the behavioral set, remembered by the code (and initiated by the VP), is also the system's *reproduction* with transferring its code to the successor. That is why a successor, possesing this code (which memorizes the previous behavior and can encode the future dynamics), is able to *anticipate and preempt* its behavior including the reproductive dynamics with *a more valuable information*. From that also follows that the path functional's punctuated information dynamics is also responsible for a prediction of the reproductive dynamics.

According to the Second Law, each following cycles has a tendency of fading and chaotization of the periodical processes. And the generated *random* initial eigenvalues at the beginning of the cycle *do not exactly replicate* the ones at cycle end. The model of the cyclic functioning includes the generator of random parameters that renovates the macromodel characteristics and peculiarities, constrained by the VP (Lerner 2004). The adaptive potential in this cycle increases under the actions of the adaptive control even if the randomness carries a positive entropy. Such an adaptive, repeating, self-organizing process is the *evolutionary cycle* with potentially growing valuable information.

The above conditions for the existence of the adaptation potential and the model's cyclic renovation impose a *limitation* on the minimal admissible macromodel's diversity. With similar reproductive dynamics for all population, the reproduction, applied to a set of the macromodels, having a maximum diversity, brings a maximum of adaptive potential, and therefore, would be more adaptive and beneficial for all set, potentially spreading through the population.

Such an evolution trend is confirmed in a latest publication (Greig 2008).

An *association* of INs with different ($\mathbf{n}, \gamma$) can mutually interact as a *system*, creating the negentropy-entropy exchanges between them analogous to a crossover. Their ranged frequencies could be wrapped up in each other changing each IN's invariants and inner codes. The systems' communication involves both an individual IN's communication code $C_{pi}$ and a common one $C_p$ that encodes the association's joint information.

Such a code might transfer the DSS genetics to others using the model's wave frequency communication.

During the evolution dynamics, the macromodel spectrum of eigenfunctions, bounded by the VP invariants relations (A 3.9a,b), evolves toward a minimal $\gamma \to 0$ in the process of adaptation, while the evolution increases the number $\mathbf{n}$ of consolidated subsystems. With increasing $\gamma \to 1$, $\mathbf{n}$ is decreasing, while the model's cooperation vanishes at $\gamma = 1$.

At $\gamma \to 0$, both equations (A3a,b) have the nearest solutions: $\mathbf{a}_o(\gamma \to 0) \approx 0.768$, $\mathbf{b}_o(\gamma \to 0) \approx 0.7$.

Other nearest solutions correspond to $\mathbf{a}_o(\gamma \to 1) \approx 0.3$, $\mathbf{b}_o(\gamma \to 1) \approx 0.3$. The max $\gamma \cong 0.8$ still supports *cooperation*.

The consolidation process, preserving invariant $\mathbf{a}_o$, develops along the line of switching controls, corresponding $\mathbf{a}_o(\gamma) = \alpha_o^i t_i = inv$ in a region of the real entropy's (RS+) geometrical locality. The macrosystem can evolve along this line at a constant total entropy $S = S_e + S_i$ if the internal entropy ($S_i > 0$) is compensated by an external negentropy ($S_e < 0$), delivered from the environment. Until the entropy production is positive, the time direction is also positive, and the entropy is a real $S^a$. The irreversible states create the systemic order.

Another character of the macrodynamics takes place along a line $j\beta_{n+k}^i(t_{n+k})t_{ik} = \beta_{n+k}^i(t_{ik})t_i^-$, where $t_i^- = jt_{ik}$ is an imaginary time in a region of the geometrical locality (RS−), and $S^b = |\beta_{n+k}^i(t_{n+k})t_{ik}|$ evaluates an imaginary entropy in this region. The second law along this line, as well as any real physical law, is not fulfilled.

Considering the start of the imaginary movement at the moment $t_{n+k-1}$ (A.4), we have at this moment the parameter $\gamma_n = |\beta_{on}^i / \alpha_{on}^i| \cong 1$, where $\beta_{on}^i = \beta_{n+k-1}^i$, $|\alpha_{on}^i| \cong |\beta_{n+k-1}^i|$. This brings the ratio $S^b / S^a = |\beta_{n+k}^i(t_{n+k})t_{ik}| / |\alpha_o^i t_i|$ to the form $S^b / S^a = |\alpha_{on}^i| / |\alpha_o^i| = h$, assuming $t_{ik} \cong t_i$. This ratio we can evaluate by the ratio of invariants $\mathbf{a}_o(\gamma \to 1)/\mathbf{a}_o(\gamma \to 0) \cong 0.4$, where $\mathbf{a}_o(\gamma \to 1) \cong \alpha_{on}^i t_i$.

The invariant's ratio $h$ differs from a minimal achievable ratio for such invariants: $\min h = h_o$, by the value $h = h^o + \Delta h$, which brings the entropy ratio to the form

$$S^b / S^a = h^o(1 + \Delta h^* / h^o), \Delta h^* = \Delta h / h^o. \qquad (5.5)$$

This allows representing the evolution of the above invariants at $\gamma \to (0 \rightleftarrows 1)$ by Figs. 4a,b.

The minimal $h_o = (\mathbf{a}_o(\gamma = 0) - \mathbf{a}_o(\gamma^*)) / \mathbf{a}_o(\gamma = 0)$ \qquad (5.6)

is defined by such $\mathbf{a}_o(\gamma^*)$, which serves a threshold of changing the macromodel dimension $m_n$.

This means that at such a feasible $\gamma = \gamma^* > 0$, a *minimal* increment of the model dimension would be

$\Delta m(\gamma^*) = m(\gamma = 0) - m(\gamma^*) = 1$, \qquad (5.7)

and any $\gamma < \gamma^*$ will not affect the model dimension, being an admissible within a given dimension.

Because the macromodel's dimension depends on the number of cooperations, the condition (5.7) *limits* both the actual increment of dimension and the model's ability to cooperate.

We evaluate $\Delta m(\gamma^*)$ using the model invariants that determine (5.6), and find the minimal $h_o$, satisfying (5.6, 5.7):

$h_o$ =0.00729927=1/137, with $\mathbf{a}_o(\gamma^*)$=0.762443796, $\mathbf{a}(\gamma^*)$=0.238566887, and $\gamma^*$=0.007148.

This realtions after substitution to $m(\gamma^*) = m_1 = (\mathbf{a}_o^2(\gamma^*) + \mathbf{a}(\gamma^*))/(\mathbf{a}_o^2(\gamma^*) + \mathbf{a}(\gamma^*) - \mathbf{a}_o(\gamma^*))$ (5.8) bring $m_1$ =14.2729035 and allow us to estimate a minimal elementary uncertainty $\mathbf{a}(\gamma^*)$ separating the model's dimensions. The minimal $\Delta\gamma^* = \gamma^* - \gamma_o = \gamma^*, \gamma_o = 0$ that restricts reaching $\gamma$=0 is able to provide both invariant quantities of information $\mathbf{a}_o(\gamma^*)$ and $\mathbf{a}(\gamma^*)$, which can generate an elementary negentropy

$s_h = \mathbf{a}_o^2(\gamma^*) + \mathbf{a}(\gamma^*)$ =0.819887424 that conceals a *minimal cooperative uncertainty*.

This limits *minimal uncertainty* achievable *in evolution development and imposes the restriction of forming cooperative structures in evolution dynamics and their complexity*. In a three dimensional space, this $\gamma^*$ provides information necessary to form a corresponding triplet $m_k^3 = 3$, which carries 4 bits of genetic DSS code. The minimal three dimensional model $m_1^3 \cong 42.8187105 \cong 43$ (corresponding threshold $h_o$) brings the IN with a potential genetic code carrying 172 bits of a non redundant information.

A *non removable uncertainty* is inherent part of any interaction (cooperation). Its minimal relative invariant $h_o$ evaluates an elementary *increment* of the model's dimensions in (5.7), while the absolute quantity of the hidden invariant information $s_h$ is able to produce an elementary triple code, enclosed into the hyperbolic structure Fig.5, with its cellular geometry in sec.A(5). This hidden *a non removable uncertainly also enfolds a potential DSS information code.*

The above results impose important *restrictions* on both a *general evolution progress* and *each step of cooperative formation*.

*I. 6.Schema of the evolutionary information functional mechanisms.*

The considered laws and regularities determine the unified functional informational mechanisms of evolution presented on Fig. 6, which include:

-the system macrodynamics MS, defined by the model operator $A(t, x(\tau), v)$ that is governed by the inherited double spiral structure $DSS^o$;

-the control replication mechanism RE1 that transforms the $DSS^o$ code into the initial controls $v_o$ which delivers $v_o$ as the MS input programmable controls;

-the IN, formed in the process of the macrostates' cooperation and the macromodel renovation, generating a renovated $DSS1$;

-the mechanism of mutations MT, delivering external perturbations, which act on the total system;

-the adaptative and self-organizing mechanisms AS, stimulated by the MT, which generate (G) the fluctuations $\xi$;

-the replication control mechanism RE2, which selects the macrostates $x(\tau)$ at DP $t' = \tau$ and forms the current control $v(\tau) = -2x(\tau)$ by the duplication of $x(\tau)$;

-the coupling of the two macrostates CP that carry both parents' $DSS1$ and $DSS2$ invariants;

-the generation of stochastic dissipative fluctuations SDF after coupling, forming new macrosystemic invariants $(\gamma^o, n^o)$ that define a new $DSS_o^o$, initiating the new MS and IN, which are renovating under the MT and the AS, in the process of functioning (a previous inherited $DSS2$ minimizes a possible SDF set, generating a new $DSS_o^o$);

-repeating the whole cycle after coupling and transferring the inherited invariants to a new generated macrosystem.

The IMD software packet (Lerner 2010) simulates the main mechanisms of the evolutionary cycle.

## II. Information observers and their evolutionary dynamics

### 2.1. Notion of an information observer and its features

There are two kinds of information observers, objective and subjective. Generally, an objective information observer is a formal *performer* of the VP through emerging interactions. Each interaction sequentially transforms the EF random information in the IPF partitioned (including quantum) information. It could be a result of a natural quantum information entanglement "at locality, inherited by a decay of correlation functions" (Eisert *et all* 2010), or from a sharp artificial impact, analogous to the action of an impulse control.

A subjective information observer, *in addition* to the objective information functions, is capable of measuring, selecting, accepting, evaluating, and self-organizing to understand the received information and make a prognosis. Each observer separates itself from the environment during acquisition of information, as a result of the segment's internal information dynamics. In both observers, we presume that the information process, as *the object of observation,* is conveyed via energy and/or material carriers, which are not considered here. For example, ideas have no specific physical properties, but convey information if it satisfies the definitions (A0).

The *observer obtains, delivers and selects* external information through its quantum *measurement and acquisition,* which *transforms* the observed random information into the observer dynamic information, thereby creating the information macrodynamic (IMD) *regularities*, which include an internal control. The cognitive system converts the objective information to subjective information, compares and classifies it. For example, by applying Bayesian inference to the probability of the observed process (A.0.3e) with a deductive logic, by analogy with Jaynes 1998.

It integrates observed information (letters, words, images, etc.) in a bound *perception*.

As it follows from sec.1.2, the observer's internal dynamics, described by IPF, are characterized by the spectrum of the IMD operator's eigenfunctions (Prop.1.1), being interactively acquired. Thus, an information observer is *defined* by EF-IPF operations, creating specific information structures for each particular observer.

Assuming that each observer satisfies the EF-IPF regularities, the observation of the *same* information, collected by different observers (having a *similar* EF that is transformed to the *related* IPF), will determine the *similar* invariant information measures of the observers. (The specific parameter $\gamma \in (0.0 - 0.8)$ of the invariants depends on each observer's *interval of observation* (A3), ending at the moment of applying control, thus generating the observer's inner dynamics). The similarities of the observers' invariant information measures produce the *similar information images* for all observers, which collect the same EF information; that guarantees its *identical* reflection.

Thus, observer, as *information interactions* (an *objective* observer), gets information from environment (other *multiple observers)* and uses it for *building its inner information structures (Networks) by creating objects-nodes,* which *communicate (within the IN border) and outer surface,* and *originate a cognition in a subjective observer.*
Specialization of observer's functions provides the EF structure depending on its functions drift and diffusion.

### 2.2. Information acquisition

The IPF minimax provides acquisition of *optimal information*, through an observer's *interactive dynamic process*, initiated by an the observer's *controls* (which depict its interactive functions).
Obtaining delivered information requires using its part on the acquisition of received information.

We assume that the received discrete information is delivered by a step function (an external, or internal action), applied during the time of the information acquisition.

The question is how shall we evaluate the quantity of the acquired information if both the quantity of the received information and the quantity, delivered with a step function, are given?

We will solve this problem by first evaluating a quantitative impact of the step function's information invariant $\mathbf{a}(\gamma)$ on the receiving information.

*Proposition 2.1.* To formalize the above evaluation, we use Riemann-Stieltjes integral (Hildebrandt 1938, Shilov, Gurevich 1978, Beckenbach 1961): $I_s = \int_{-\infty}^{\infty} f(t-s) dg(s)$ applying it to an information function $f(t-s) \to \mathbf{a}_\tau(t-s)$, delivered information $\mathbf{a}_\tau$ during interval $t-s$, and to an information function $g(s) \to \mathbf{a}(\gamma) u(s)$, providing a fixed quantity of information $\mathbf{a}(\gamma)$ by the *step* function $u(s)$. •

*Proposition 2.2.* The above impact is evaluated by the following information form of the Riemann-Stieltjes integral:
$$I_s = \int_{-\infty}^{\infty} \mathbf{a}_\tau(t-s) \mathbf{a}(\gamma) \delta(s) ds = \mathbf{a}_\tau(t) \mathbf{a}(\gamma),$$
whose solution is found for function $dg(s) \to \mathbf{a}(\gamma) du(s)$, while the derivation of the step function forms the impulse function: $du(s) = \delta(s) ds$. •

*Proposition 2.3.* In the considered model, where information $\mathbf{a}_\tau(t)$ is delivered during each quantum time interval $t = \tau_k$, $k = 0, 1, 2, ..., m$ and a step control function provides invariant information $\mathbf{a}(\gamma)$, the above integral in the form
$I_s = \mathbf{a}_\tau(\tau_k) \mathbf{a}(\gamma)$, $k = 0, 1, 2, ..., m$
evaluates the impact of the step control function on the delivered information $\mathbf{a}_\tau(\tau_k)$. •

*Proposition 2.4.* At each moment $t = \tau_{k+1}^o$ of the step control action, the impact of the step control function is evaluated by $I_{sa} = \mathbf{a}^2(\gamma, \tau_{k+1}^o)$, which binds the invariant quantity of the information provided by this control.

*Indeed.* Since the step control applies at the moment $t = \tau_{k+1}^o$, this impact should also start at the same moment. This brings the integral at this moment $I_s = \mathbf{a}_\tau(\tau_{k+1}^o) \mathbf{a}(\gamma)$. However, at this very moment, the only delivered information is information $\mathbf{a}(\gamma, \tau_{k+1}^o)$ provided by the step control function. Fixing the above function at the moment $t = \tau_{k+1}^o$, $k = 0, 1, 2, ..., m$ of applying the step control, we have $\mathbf{a}_\tau(t = \tau_{k+1}^o) = \mathbf{a}(\gamma, \tau_{k+1}^o)$, and $I_{sa} = \mathbf{a}^2(\gamma, \tau_{k+1}^o)$, which is also fixed, evaluating the information bound at this moment. •

*Proposition 2.5.* The step control, which initiates the model's dynamic process of acquiring the delivered information $\mathbf{a}_\tau(t)$ and are fixed during the time interval of the process acquisition $t_{k+1}$, $k = 0, 1, 2, ..., m$, *binds* information $I_{sk} = \mathbf{a}(\gamma) \mathbf{a}_\tau(t_{k+1})$; or this amount of the delivered information is conserved by the step control.

*Indeed.* Since, the step control is fixed on the time interval $t_{k+1}$ *and* the acquisition of the delivered information continues during the same time, the information integral holds the above form. •

*Specifically*, the model's dynamic process of acquiring the delivered information $\mathbf{a}_\tau(t)$, initiated by the step control action, starts at the moment $\tau_{k+1}^o + o_{ko}$ and ends at the moment $\tau_{k+1}^1$ after the control is turned off at the moment $\tau_{k+1}^1$ (for each $k$-extremal segment). Thus, the interval of fixing the step-wise control is $t_{k+1} = \tau_{k+1}^1 - (\tau_{k+1}^o + o_{ko})$, and the acquisition of the delivered information continues during the same time.

In this dynamic model, with $\mathbf{a}_\tau(t) = \alpha_\tau t$, where $\alpha_\tau$ is initial speed of delivered information (defining a starting frequency), we have the information consumed during $t_{k+1}: \mathbf{a}_{\tau o}(t_{k+1}) = -\alpha_\tau t_{k+1} = -\mathbf{a}_o(\gamma) > 0$, and the information impact $I_{sk} = -\mathbf{a}_o(\gamma)\mathbf{a}(\gamma) > 0$ is measured by the model invariants.

In *Summary*: P.2.1 evaluates a quantitative impact of a step function $g(t)$ on a delivered information $f(t)$;

P.2.2 applies P.2.1 when the impact is provided by an invariant, which evaluates the step control action;

P.2.3.applies P.2.2 when information is delivered during each quantum time interval;

P.2.4 applies P.2.3 when the delivered information is evaluated by a step function and the impact of other step control (starting the macrolevel process) is evaluated at the quantum interval;

P.2.5 applied P.2.4 with the invariant evaluation of information, bound by both the step action of the delivered information and the step control, which starts the macrodynamics.

Therefore, the acquired quantity of information encloses an *additional* quantity of a conserved information, which depends multiplicatively on both quantity the delivered information and one spent on the acquisition.

Rising the acquired information increases hidden information; more information spent on the acquisition of delivered information will bring more conserved information. However, even at these assumptions, a total quantity of acquired information depends on *how* does external information would be acquired: by dividing it in the portions, or an entirely once. (In both cases, the quantum character of the encoded information is presumed).

From the above analysis it follows that acquiring information by the divided portions sequentially would bring *less* conserved information than if the same quantity where acquired once. (It is supposed that a sum of the time intervals of the portioned acquisitions is equal to a total time of acquiring the entire information.)

For *Example*, let be $m_o = x + y + z$ a total quantity of receiving information, which could be acquired once by $m$, or by the portions $x, y, z$. Assuming that a quantity of binding information is proportional to each quantity of the delivered information, then $h_o = \alpha m_o^2$ with a coefficient $\alpha$; and the total quantity of the hidden information, in first case, would be $h_o = \alpha(x + y + z)^2$, in the second case, at $h_1 = \alpha(x^2 + y^2 + z^2)$, we have $h_o > h_1$ with their difference $\Delta h = (h_o - h_1) = 2\alpha(xy + xz + yz) > 0, x > 0, y > 0, z > 0$.

It seen that at a portioned acquisition, the information contributions from the mutual influences of the portions are lost. As a result, portioning of receiving information brings less information than acquiring this information undivided, which connects the selected portions.

We can evaluate the needle control's required information by applying the Stieltjes *convolution* integral in the form $I_{s1} = \int_{-\infty}^{\infty} f(t-s)g(s)ds$ to both function $f(t-s) \to \mathbf{a}_\tau(t-s)$ and function $g(s) \to \mathbf{a}_o(\gamma)\delta(s)$; where the last one provides a fixed information $\mathbf{a}_o(\gamma) = 2\mathbf{a}(\gamma)$ by a control impulse $\delta(s)$ (evaluated by sum of a two step control's information). We get the integral value at the moment $t = \tau_k^1$:

$$I_{sa_o} = \int_{-\infty}^{\infty} \mathbf{a}_\tau(t-s)\mathbf{a}_o\delta(s)ds \Big|_{t=\tau_k^1} = \mathbf{a}_\tau(\tau_k^1)\mathbf{a}_o(\gamma) = \mathbf{a}_o^2(\gamma),$$

when information at a previous segment's end is $\mathbf{a}_\tau(\tau_k^1) = \mathbf{a}_o(\gamma)$. This amount of $I_{sa_o}$ evaluates the information effect, produced by the needle (impulse) control's delivered information, while transferring the segment information $\mathbf{a}_\tau(\tau_k^1) = \mathbf{a}_o(\gamma)$ to another segment. (The needle control form has both limited high and width).

Since the control consumes information from an observed random information process, this process is an external information source, which provides information $\mathbf{a}_o^2(\gamma)$ to this control to be able to connect the segments at transferring information $\mathbf{a}_o(\gamma) < 0$, consumed between the extremal segments, to each following segment.

Thus, the needle controls do both connect an external information process (by imposing a maximum of its probability at the DP, sec.1.2, A1-3) with an internal (consumer) macroprocess, and connect its segments (at the DP) to a chain.

*We outline creation, evolution, life and death of an information observer by the following summary.*

The evolution process, starting from randomness, produces *space-time evolutionary macromodels*, following the path functional's VP as an *information law*. The macromodel, performing regularities of this law, generates discrete *control, code and cooperative processes*, forming the cooperative *structures minimizing* path functional, which enables further consolidation, enlargement, and complexity. Starting from particular dimension, such complex structures allow adaptation, renovation, and self-organization, which sustain their existence at the interaction with environment and evolutionary development through the selection and enhancement of the acceptable code mutations; and ability of both inherit the genetic code and model's reproduction (secs.1.4,1.5).

The regularities specifics impose the followings limits on the macromodels: uncertainty, invariants, and dimensions. These restrict the complexity, stability and the potentials of robustness and adaptation, limit lifetime, and brings the restrictions on renewal the cyclic functioning (sec.1.5).

Ability of model's self-assembling creates a *system* of interacting complex information *cooperatives* and *subsystems*, whose connections provide the system's mutual *communications* (sec.1.4) and produces the system's *social and economic* environment (Lerner 2005,2006), supporting the system existence and survival.

The interactive processes of the information observers requires both a *reflection* of the environment and adequate observer's *reaction*, according to the VP law of EF-PF transformation. This leads to a *necessity* of creation of an observer's inner *cognition*, which performs the above transformation and provides the observer's controls, acting as information digits for both encoding information and information communications.

Thus, the macromodels, created *according the law*, continue their *evolution, life and death* following the law.

Moreover, multiple interactions of the information observers, acting according to the VP law, are able to *create* an environment following this *law*.

To analyze the *environment of interacting observers* we consider an observer's external surface (Fig.5), created by the informational geometry-dynamics of the IN objects-nodes (Figs.1). Such an outer surface originates from the IN space-time cellular information geometry (Lerner 2006, 2007, 2010) (details in A5).

Each surface is composed by the cells, while each of them holds information unit of the DSS code with the invariant measure $\mathbf{a}_o \cong 1$ bit (at $\gamma \to 0$). Each IN node on the surface encloses four cells $m_c = 4$ (each with $f_o^c = 1 bit$ of the DSS code), and a total surface area $F$ contains information $F_{im} = m_c f_o^c S_m$, where $S_m$ is information delivered by $m$ numbers of the IN nodes on this surface. Assuming that each observer's interacts with environment via the node's surface area, we get all observers' information available for the interaction $S_m = 1/4 F_{im}$, with total cells' number on the surface $N = F_{im} / f_o^c$. $S_m$ depends only on the cell's surface area and is not dependable on the surface curvature (A5) (because each cell holds 1 bit information independently of the cell's area curvature).

We also suppose that other observers also interact with the considered observer through *each* of these observers' single IN node. Then, we can evaluate the number of the interacting observers by $N_{em} = 1/4 N$ with a total maximal information collected during the interactions $S_{em} = N_{em} = 1/4 N f_o^c$ bits. Because each node encloses a control information, in addition to the triplet's information in three bits, which is enfolded in the cooperative IN binding, only this additional information can be transferred to the interacting observers at each interaction. Here we assume that each interacting observer acts according to the VP minimax principle, keeping a balance of the consumed and internal information.

The observers' interactions "inherit and originate evolution of developmental innovation, being a path for evolution of novel adaptations in complex multi-cellular organisms" (Bodyaev 2011). The macromodels inherit these and other peculiarities above through acquisition of a maximum needed information along the IN hierarchy, which is associated with increasing of valuable information and emergence of growing complexity. In the environment with limited information sources, the observers compete for maximum available information, which leads to selecting the ones with more valuable information, having a maximal dimension and complexity.

These change the observer's numbers, but restricting both their maximal dimension and the numbers.

It has been shown (A3.17-3.19) that the *quantity of information*, accumulated by the inner connections of a process, is *encoded* by the code word's length of an *optimal algorithm program*, determined by the IPF invariants which evaluate these connections. Building of the IN hierarchy with the information evaluation of its sequential causal relationships (at the end of A.2) leads to *logical operations* and the optimal algorithm, implemented by the IN's double spiral (DSS) code (Fig.2) with the optimal codeword's length (A.3.19). This allows encoding the EF-IPF *mathematical and logical operations* of the *information process* (following from the VP) in a form of the IMD software(Lerner 2010), which generate the IN *structures, related* to these operations.

Schema Fig. A shows the basic structure of the EF-IPF Approach, Fig.B details the main operations in this formalism, and Fig.C illustrates their implementations by the information observer.

*2.3. An information observer's cognitive and neurodynamics*

We outline the common features of the *subjective observer's* both cognitive dynamics and neurodynamics, based on the following assumptions.

- Cognitive dynamics *include* the selection of the most informative events, ordering, memorizing, cooperating, encoding, and integrating them into an IN hierarchy. Following these steps will establish a base of knowledge and allow its renovation, control, synthesis, creativity, learning, and decision-making.

- *The process* of cognition, formalized by a minimization of its observed uncertainty, is described by a piece sequence of the IPF dynamic macromodel's extremals, identified during observation and built it sequentially to maximize a reception of information (according to VP). An observer's cognition enables transforming unobserved imaginary information into observed real information, after an interactive "collapse" of the wave functions.

- A cognitive system *feeds itself* with valuable information and provides an automatic improvement via the *evolutionary cooperative dynamics* (part I), as the IMD parts. In the considered *hybrid* nonlinear dynamics (sec.1.3), the segment's reversible processes are joined at the DP's irreversible intervals, forming an *information analogy of cooperative dissipative structures*. Such *transient* information dynamics, arising between each following IN's levels of hierarchy, self-maintains this hierarchy in the evolving cooperative structures.

- A notion of *intelligence* belongs to an observer as its ability to accept, use, and create information.

This ability we evaluate by an increasing *quantity and quality* of accepted information through its use and creation. Such an increase becomes possible if an observer's inner (accumulated) information is utilized through forming cooperative dynamic structures (with an IN) while processing the accepted information.

In these cognitive processes, different forms of encoding information are instrumental for its reception, recognition, distinction, and creation. The observer's knowledge is limited by its experience of observations (interactions), including a prediction made by this knowledge (delivering a *source* of information).

That is why any knowledge beyond this experience and valuable projections are insufficient.

*An intelligence level is connected to an observer's ability of creating a highest level of its IN hierarchy, measured by the highest quality* of accepted information being transformed during the cooperative dynamics. It also measures a *structural* information of an accepted message, associated with its *meaning*, as a particular observer is able to recognize while building such a structure. Consequently, an observed semantics (enclosed in the message) can be measured by a definite quality of information of this observer's *highest* IN's level. Thus, even though such a structure does not carry a meaning and/or semantics, but the structure's building requires spending the observer's meaningful information, which evaluates its personal intellect needed for that in the process of recognition of this information. This information, being sequentially enfolded in the IN levels, evaluates the IN level of complexity, and therefore also the considered meaning (which could be reflected differently by each observer's IN while getting the identical information). Preserving the observer invariants' $\mathbf{a}_o(\gamma)$, $\mathbf{a}(\gamma)$ information structure guarantees objectivity (identity) of each observer's personal reflection.

A *maximum cooperative complexity* (Lerner 2006) of *the most valuable information* could become a *scale* for the comparison of intelligence with a possibility of the *intelligence's increasing* at growing information values in the evolution. Such an observer builds a *harmonic* structure of the IN during the interaction with environment.

In this IN (created according to the VP), the hierarchy of its node replicates external *musical rhythms in an observer's cognitive reflection*. Because the information observer performs a cognitive dynamic transformation of an observed information through its neurodynamics (which provides also the above control), we analyze both the *observer's cognitive dynamics and neurodynamics in their connection to the IMD*.

A comprehensive *review* of the physical, information, and dynamic principles in neuroscience (Arbib1987, John 2002, Freeman 2001, Penrose 1994, Scott 1995, Glynn 1999, Rabinovich et al 2006, Izhikevich 2006), its specific neurophysical and informational features (Panzeriet et al 1999, Nirenberg and Latham 2003, Lerner 2004, Salinas and Sejnowski 2001, Liu 2004, Buehlmann and Deco 2010, Sacktor 2008, Tsien 2007, others) allows us to provide the following *comparison* of both the principles of information macrodynamics and neuroscience and the information functions of their specific mechanisms.

- The physical dynamics of propagation through a neuron's axon is modeled by the IMD information dynamics at a particular extremal segment, while the neuron can be modeled by a three-dimensional dynamic system with an applied control (in both step function and impulse function forms). Spike generation is modeled by the impulse control during a window between the segments, while a spike is generated upon overcoming a threshold. The impulse control rises at the end of the extremal segment, after the information, carried up by the propagating dynamics, compensates the segment's inner information. The control joins the corresponding segments, bringing new information $\mathbf{a}_o^2 = \mathbf{a}_o - \mathbf{a}$ during an irreversible time interval *between* the segments.

It is determined by both a quantity of information $\mathbf{a}_o$ collected during the related reversible time interval within a segment, and a step control action $\mathbf{a}$. This establishes, first, the direct connection between the information analogies of both spike and the threshold, secondary, brings the spike information measure for each its generation, evaluated by $\mathbf{a}_o^2$ in bit (sec 2.2, A3). The impulse control conveys digital units of the impulse with its amplitude, measured by the value of double states $2x(\tau)$ (A3), generating the impulse and its invariant information measure $\mathbf{a}_o^2(\gamma, l_o)$. Both *quantity and quality* of information in the time-space dynamics are determined by the invariant $\mathbf{a}_o(\gamma, l)$ ($\gamma$ is the parameter of the dynamics in (A3.9), $l$ -is a space parameter, $l = l_o$ is a space location of the impulse, specifically within the IN hierarchy).

- A spike, reaching the axon's terminal, triggers the emission of transmitter, which binds the receptor of the following neuron. At the consideration of information transmission between the segment-transmitted and a segment-receiver (receptor), the needle control connects them and initiates the signal propagation between them. The propagation is associated with adding a macrostate (carried by a control) to transmitter and removing it after transmission occurs. At the receptor side, the propagation's and the control's actions initiate the addition of a macrostate to the current receptor's macrostate, indicating the occurrence of the transmission, and release the macrostate after the propagation occurs.

The needle control serves only as a messenger, which does not carry the macrostate, but rather induces the macrostate's formation by the receptor (Fig.3), the details are in Lerner 2003.

- The theoretical and experimental studies by Liu 2004 show that the excitatory and inhibitory (E/I) synaptic inputs are regulated across related dendrite trees to maintain its *constant ratio* by a "'push-pull" feedback regulatory mechanism. In the IMD neuronal model, communication of the E/I synapses proceeds by transferring information from the IN upper node to a lower node (Fig.1) applying the impulse control. The IN node location satisfies holding a *constant* model's eigenvalues ratio, or the ratio of related spectral frequencies, as the IN basic parameter $\gamma_i^\alpha \sim \gamma$, fixed at the current model's identification (sec.1.4, A3).

- The interspike intervals carry the encoded information, the same way that the intervals of discretization between the applied needle controls do. Conductivity of an axon depends on the between neuron's electrical conductance, which, in the IMD model, is determined by the diffusion conductivity of the cooperative connections, computed via a derivation of the correlation functions. A signal, passing through this conductivity, might modify a topology of a single, as well as a multiple connection, changing its macrodynamic functions (and a possibly leading to distinct networks) under different input.

- The axon's branching geometrical structure (Scott 1995) is an example of the connections of the triplet cones (Figs.1,3). At each triple point, the two cone's vertexes are connected with the base of a third cone like two inputs and one output. A neuron communicates by the *triplet* code of cooperative dynamics simulated in Fig.2, whose existence has been experimentally confirmed in the recent publications (Tsien 2007, other).

Cooperation of interacting neurons leads to their dependent ordering in neuron groups, multiple encoding of an input signal, as a result of coordinated activity and the neuron's synchronization in a harmony. Such cooperative dynamics are characterized by its specific cooperative complexity with the hierarchical dependency (Lerner 2006).

- Both neuronal dynamics and macrodynamics are strongly dissipative, based on stochastic dynamics of the controlled diffusion processes, whose macrodynamics includes chaotic dynamics with a possibility of different types of local and global bifurcations, associated with changing the structure, instabilities, and singularities (Freeman and Vitiello 2008, Rabinovich *et al* 2006, Lerner 2010). A three dimensional triplet's dynamics generally generate such bifurcations as the limited orbits, saddles and attractors. A *structural stable* information macrosystem encloses a set of the stable attractors, forming the triplet's nodes of the information network**.**

Neural oscillatory networks, measured by the brain oscillation's frequency, power and phase, dynamically reduce the high-dimensional information into a low dimensional code, which encodes the cognitive processes and dynamic routing of information (Schyns *et al* 2011).

- Secs.1.1, 1.2, 2.2 demonstrate the significance of the observer's external correlations, accumulated by EF-IPF and transferred into hidden information at its acquisition by an observer.

Panzeri *et al* 1999, Salinas, Sejnowski 2001, Nirenberg and Latham 2003 show the importance of correlations at the *interspike* communications for both generation of a neural code and decoding of the neural responses.

Segev *et al* 2004 revealed the existence of *hidden synchronized correlations* for the distinguishable neuron's subgroups of an interneuron spatiotemporal network, possibly united by the observed information *process*. These findings serve as a template for coding, storage, and retrieval of an observed information, support the observer's cognitive functions above.

- A neural system, organized as the IN hierarchical dynamic structure (in both the phase time-state and three-dimensional spatial state), utilizes such neural phenomena as the states' coordination, synchronization, and cooperation, which are probabilistically dependent on the input signal, transformed into the specific output.

Each external event forms a starting condition for the next widow of time, which connects the events also in a space while the sequent of information states is generated by quanta at the widows between the information model's segments.

In the neurodynamic structures (Rabinovich *et al* 2006), each of the saddle point represents an event in a state's sequence to be remembered. Coexistence of the multiple nodes-attractors at a given initial conditions characterizes a multistability of both neurodynamic and macrodynamic systems. In the triplet's network, each previous triplet node enables sequential attraction of the following triplet's node, with a possibility of their synchronization, generation of adaptive rhythms, and forming a transient behavior. In such *evolving* IN, the attracted node with the increased information values is a *fittest* one.

- Vicente *et al* 2008 experimentally confirm that when two neuronal populations are coupled to third population they become self-synchronized with zero phase interval across large distances.

The "polychronization" proposed by Izhikevich 2006 are consistent with the IN cooperative dynamics.

In particular, the macromodel eigenvalues are capable of assembling into a cooperative if its information frequencies operate with the sequential delay of incoming information, determined by the time intervals required for the following cooperations (Lerner 2010). The IN triplet's connections can be changed depending on the current segment's sequence, its information quantity, and their number, which leads to an alteration of network structure, reacting on incoming inputs.

The IN encoding precedes the decupling of the correlations (sec.1.2,A1,A3).

Wiechert *et al* 2010 have proven that a decorrelation emerges from the generic neuronal nonlinearities and the neuronal adaptive mechanism provides an efficient coding and storage of information.

- The IMD mechanism of the IN building includes an *automatic ordering* of the model segments evaluated by the information quantity of the segment's eigenvalues (and the eigenvectors) in the process of this information acceptance for a specific sequence (secs1.2,2.2 and A3). These information quantities involve the automatic generation of the triplet's structures with the corresponding space movement of their local space coordinate systems and forming a global IN's coordinate system at the triplet's cooperation.

- The global coordinate system is determined by the input of the IN starting node's eigenvector and the output of the IN ending node's eigenvector. This automatic procedure not only transforms the spatial-temporal input to its spatial form and finally to spatial-temporal output, but also establishes an ordered geometrical mapping relationship between them, allowing the exact localization of the transformed inputs. We assume that this transformation is carried by the spiral wave modes, represented by the model's spirals on the cones.

- The IMD model is characterized by the sequential growth of the information *effectiveness* of the impulse and step controls along the IN spatial-temporal hierarchy. This is connected with changing the *quality* of the IN node's information depending on the node's *location* within the IN geometry. These changes increase the intensity of the cooperative coupling and its competitive abilities, which make the segment's synchronization more stable and robust against noise, adding an error correction capability for decoding. It also affects the length of the discretization intervals, and an ability to adjoin more cooperating elements. The growing intensity of the coupling increases the model's hierarchical cooperative complexity (Lerner 2004, 2010), considered in a curved phase space, whose curvature, created by the intensive cooperation, is measured by the information invariants.

In the IMD neuron model, this leads to spike strengthening along the formed networks, which affects the interspike intervals, and increases the neuron's chain ability for connectivity and self-organization.

*Long Term Potentiation* (Cohen-Matsliah *et al* 2011, Bliss, Collingridge 1999) allows a synapse to increase *strength* with *increasing numbers of transmitted signals between the two neurons*. These cells also organize themselves into groups, specializing in different kinds of information processing.

- Growing the network effectiveness, quality, and intensity allows conserving and concentrating *more valuable* invariant information with increasing information volume (Yomdin 1987, Lerner 2010).

Neuronal communication includes the transmitted information with the spike strength's dependency on a prehistory, a flexible spike threshold, and the accumulation of information at the attractors, or at the saddles (depending on the specifics of the bifurcation).

As spike information intensity grows, its generation requires overcoming more information thresholds.

Increasing the node number in the IN model's hierarchy develops growing the node's enclosed information, which raises each following spike threshold. Nodes can be added from nearby or multiple distant locations.

- By collecting, ordering and concentrating the observed information in the observer's individual IN, the observer is able to produce more control information, conserved in its final IN node.

Therefore, the sequential growing of the information effectiveness of the impulse (and step-wise) controls' actions along the IN spatial-temporal hierarchy (associated with an individual ability to build an extended IN hierarchy) increases the intensity of the cooperative coupling and generates more valuable internal information, which is able to attract the equivalent external information.

The final IN node's information, which was not spent on the current IN assembling, evaluated by the invariants $\mathbf{a}_o^2 - \mathbf{a}_o = \mathbf{a}$ can be generated outside, or can be used for assembling other individual INs, or parts, producing, as a result, new IN connections that could have not existed before. This final node's information (negentropy), which generates a "request" for its compensation by an information input, is a source of the anticipating actions of observer cognitive dynamics' discussed in Rosen 1985.

Such actions work as a control (a stimulus), initiating the propagation dynamics, which generate a spike. Moreover, the informational intensity strengthens the spike's ability to bind information, depending on the specific node location, which can provide a needed quality of information. A stimulus might also be created by an observed external event, which identifies a state $x(\tau)$ and produces a related internal control $-2x(\tau)$ (associated with a sensory registration). Thus, the final IN node's propagation dynamics, with invariant information $\mathbf{a}_o$, generates the control's information $\mathbf{a}$, and the spike produces information $\mathbf{a}_o^2$, which binds the interacting node's information.

Aldworth *et al* 2011 have confirmed that the stimulus quantities can be sequentially encoded by a single neuron.

The information binding is accompanied by fixing the information, which is associated with a memory formation. Gaining and assembling information, accompanied by its fixing, are the components of the IMD cooperative dynamics, memorized by the IN. The neuron's ability to fix an information value is determined through the formation of proteins, RNA synthesis, and encoding in an observer's DNA (Miller, Sweatt 2007). This DNA- encoded information is translated to RNA that synthesizes *four* proteins (Powell 2008), which both bind and memorize the information by fixing it during the binding. These proteins encode information during the process of memory formation by determining a sequence of distinct transient conductivities. When a bio-current is transmitted in the brain, such a sequence indicates the jumps of the voltage dependent current, which encode the transferred information. The voltage activated K+ channels are the membrane integral proteins that control neuronal excitability through re-polarization of the action potential and modulation of repetitive firing (Ferroni *et al* 1992).

Neurons form a heterogeneous population of K+ channels with diverse voltage-dependence characteristics and pharmacological properties. This protein sequence's code corresponds to the IN triplet's four-symbol code.

- Protein memory code is also a brain *communication* code (Tsien 2007).

Kunes and Ashraf 2006 found that a messenger RNA (mRNA, a genetic photocopy that conveys information from DNA to a cell's translation machinery) is transported to synapses when a memory begins to form. This mRNA transport and the following protein synthesis are facilitated by components of the biochemical pathway, called RISC, that determines *if and where* this protein synthesis happens. This pathway interacts with RNA at the synapses using very short RNA molecules called microRNAs to guide their activity at the protein synthesis, forming a stable memory. Therefore, a neuronal memory code might work like the DNA's four-letter triplet code, but with essentially more options. In addition to the four different proteins digital forms, these digits are also distinguished by their height values, exposing an intensity of the encoding signal ($-2x(\tau)$) generated by the spike that binds and fixes the communicated neurons during memory formation. This signal also connects both the neuron's groups and their codes, while the RISC pathway traces a location where the memorized information is physically stored.

As multiple neurons consolidate, each following cooperation requires strengthening the connections between neurons, associated with an intensification of both spike generation and the consolidation of encoded signals.

A triplet that encodes this ensemble conceals more information than each of its components do (because the concealed information also encloses information that binds the elements). Sequential encoding of a neuron's population automatically creates a successively-enclosed hierarchy of such triplets (Fig. 1).

An ordered sequence of the memorized and encoded signals serves to organize the ensembles in a hierarchy of collective units with a corresponding group (population) code (Fries 2005), and the IN code.
Each collective unit in the group responds with a signal that encodes the group. Preserving a triple consolidation and triplet code for all organization also produces the invariant information measure for both the quantity information for each triplet and its quality, evaluated by the triplet's location within IN hierarchy.

● According to Rabinovich *et al* 2006, "experimental evidence shows the existence of a population code that collectively expresses a complex stimulus better than individual neurons" by a spatiotemporal code. And … "the presence of network coding", i.e. "spatiotemporal dynamic representation of incoming message, has been confirmed in several experiments." Thus, the IN network's cooperative code (Fig.2), portrayed through the triplet's genetics, is a valid element of informational neurodynamics.

Each instant of the IN code has a dual representation: by the time moment (or the space location) of firing the impulse control, *or* by the time (space) intervals between these impulses.

The code's universal time (or space location) is set up by the system of model equations following from the VP that defines the invariant relations for the time-space intervals and the controls (A3).

● As a neuronal subsystem, the information cooperative subsystem is ensembled from the interconnected units at each IN's hierarchical level, all of which are similarly organized. This preserves a local autonomy and a self-regulation at a diversity of the key-lock cooperative connections (Fig. 3,4) (Lerner 2001). The multiple triplets and the IN nodes created by various sensory inputs do not assume establishing between them the *specific* encoder-decoder relationships. There could be many simultaneous senders and different receivers with distinct messages and transmission processes, but with a universal coding language, determined by the triplet's genetics.

The universal time scale (and/or corresponding space locations) is established for a given system by its local clock time course (Lerner 2004a).

● The cognitive extraction, selection, and ranging of the events depend on its informativeness (both the quantity and quality of the event's information content), which build the event's *hierarchical connections* as a key for understanding their causal-consequence relationships (A3).

These cooperative connections cohere and organize this information, targeting its meaning and remembrance. Motivated by this organization, the large groups of neurons in the brain synchronize oscillatory activity in order to achieve coherence (Buehlmann, Deco 2010, Sacktor 2008). Another way of understanding may aid memory by reducing the amount that has to be remembered via "chunking". Chunking is the process that organizes objects into meaningful groups, memorized as a *cooperative unit* rather than separate objects. For example, a text's letters are chunking by perceiving and remembering their meaningful units: words. The same way, multiple numbers of items are remembered by creating meaningful "packets" in which many related items are stored as one, for example.

H. Simon 1964 has shown that the ideal size for chunking letters and numbers, meaningful or not, *is three*, which coincides with the IN optimal triplet's size, whose final node encodes a chunk cooperative.

- The above analysis shows that many features of neural systems can be modeled by the IN. In particular, Short-Term Memory (STM, or Working Memory) is modeled by fixing the double states at the segments' cooperation, provided via the action of the current control (both regular and impulse). The Long-Term Memory (LTM) fixes a sequence of the cooperating triplet's nodes-attractors, which guarantee memory robustness. Following the IMD model, such cooperative structures are activated *only* at the locations of the "windows", which restricts the LTM. LTM proceeds after the STM formation, which has a limited capacity. Placing the LTM into a spatial IN occurs after removing the current multiple STMs which are not composing the IN nodes. STM is formed in a dynamic process, while LTM is forming the *structural* connections, produced by dynamics. STM has a limited capacity of four to seven items. It follows from the triplet's cooperative dynamics, joining the extremal segments by three, or coupling sequentially seven segments in a joint triplet. These numbers correspond to Miller's numbers (Miller 1954).

STM works until structural connections have not been made, and until the STM limited information capacity has not been exceeded. After that, LTM starts, and it relies on the particular triplet's element, which keeps the latest STM connections.

- The cognitive *applications* of the IMD formal model bring *observer cognitive dynamics* (OBCD) (both reversible and irreversible) with the above results. The OBCD starts with eight *stimuli* initiated, as a minimum, by eight internal IN's ends including its final control. Such eight stimuli represent the OBCD potential self-motivation. The stimuli provide eight starting step-wise controls, which launch the neuron's propagation processes.

The activated neuron's ending state generates a spike (by the state doubling control's action on the neuron's ends). These minimal ends, as the IN subsystems, are required to guarantee the system's ability to restore an adaptive cyclic process (sec1.5). New (sensory scanned) information, which appears within the interspike windows of the inter communication nodes, is collected during information transmission between the segment-transmitted and a segment-receiver (receptor). Fixing this collection is associated with a sensory registration, which corresponds to encoding of the scanned information. The observer's bound information might include some symbols (quanta) of sensory information, which had been collected a *priori* by the specific communication neurons.

The observer's external informational quant delivers information necessary for forming the controls, which is spent on binding with existing information by connecting some neurons.

- Encoding begins with any *novel* situation (A0). The selected information group is remembered in STM, being enhanced by a minimum of *four neurons'* cells axons with a specific metabotropic glutamate receptor (Rosen 1985), which is able to fix this group. Such receptor's ability is marked by a DNA gen (as a knot) at the cell protein's formation (Miller, Sweatt 2007). This binding ability builds up the DNA code, which (in its DSS equivalent) arises by carrying the discrete controls.(The code discreteness follows from the dynamic constraint imposed on the VP). From new information in STM, the most valuable are chosen to encode in LTM. LTM is modeled by the OBCD set of various INs, having a range of different dimensions with a choice of ending nodes.

These nodes had enclosed a memorized defect of negentropy (evaluated by invariant $\mathbf{a}_o$ and associated with effectiveness of the node's attracting ability). The STM encoding forms a group code, which fits to the specific IN's node code to be attached to this node. By adding the new node to this IN, the IN automatically both evaluates the information value of the attaching node and increases its total self-information value.

This means that a more valuable IN attracts information that is more valuable. That provides an automatic selection of the STM groups simultaneously with a long-term memorization of them in the IN.

It is assumed that the LTM cells inherit from DNA this ability during a specialized embryonic development.

An individual neuron's ability for structural connections depends on specific proteins which form the internal medium of Long-Term Memory, also limiting the memory volume. Therefore, the OBCD stimuli's motivation initiates not only scanning information and its extraction, but also the process of connecting the extracted information in STM, its automatic selection by the information values, and storing in LTM.

● Changing the neural network's connectivity is associated with learning (Rabinovich *et al* 2006, Lerner *et al* 1993). Learning needs a precise spike synchronization and the sequential memory encoding that requires a sequential generation of the temporal asymmetry, following from the properties of the network's connections.

Non-stationary activities between the attractors initiate the transient responses and their competing connections through the information communications, which produce the cooperative behavior. Thus, through overcoming some threshold, each IN's node-attractor, with memory storage, enables the transmission of information and attracting a following doublet or a triplet. Each triplet, attracting others, is capable of producing ordered cooperative connections, counteracting instabilities and singularities of the connections. An asymmetrical coupling goes through a stochastic synchronization and geometry, which after cooperation keeps robustness of the renovated structure.

● The cooperative connections also arise between different networks (INs), as well as within each ne*t and* the element-nodes, which are located at the end one net and the beginning of the next (Lerner 2005).

According to Pang and Maslov 2011: "the metabolic network of an organism evolves by horizontal transfer of the branched cyclic metabolic pathways from other species". These distributed pathways are part of a larger "universal network formed by the union of all species-specific networks". And following Daisuke *et al* 2011: "a small local change in migratory pathways provides single neurons with the ability to migrate to a new brain, where they play a role in *information recursion*, which makes possible a higher intelligence".

The experimental study (Dubey and Ben-Yehuda, 2011) reveal a "bacterial communication, mediated by nanotubes that bridge neighboring cells, serving as a route for exchange of intracellular molecules".

Therefore, *information drives evolution of intelligence*.

- The learning process is known to engage a variety of modulator transmitters in order to create and consolidate memories. These transmitters cause the nucleus to initiate the processes required for neuronal growth and long-term memory, mark specific synapses for the capture of long-term processes, regulate local protein synthesis, and even appear to mediate attentional processes required for the formation and recall of memories (Gobe*t et al* 2001).

While short-term memory encodes information acoustically, long-term memory encodes it semantically (Kandel 2004, Demb *et al* 1999). Studies indicate that lexical, semantic and phonological factors interact in verbal working memory. The phonological similarity effect (PSE) is modified by word concreteness.

PSE emphasizes that a verbal working memory performance cannot exclusively be attributed to phonological or acoustic representation but also includes an interaction of linguistic representation (Conrad 1964). STM encoding relies primarily on acoustic rather than semantic encoding. Important role plays a perceptible encoding via the process of a personal inner filling (Wincler *et al* 2009) . Neurons in the primary somatosensory cortex react to vibrotactile stimuli by activating in synchronization with each series of vibrations (Baddeley 1966, Crawley *et al* 1998).

- The IMD model admits the creation of inner information without a specific external input. It is possible by two ways: when an impulse control connects the extremal segments by closing access to external information. In this case, previously memorized information produces the control that maintains new information structures by cooperation between the segments, which were not directly initiated by external inputs. An instable chaotic activity between the segments is also a potential generator of new information at any of the considered dynamic moves. Another way uses the information triplet's surpluses, generated by the triplet's interactions.

Physical processes, associated with these mechanisms, are connected to the phenomena of *superimposing* processes (sec.1.4, Lerner 2010) and their connection to *control theory*.

More powerful three-dimensional visual information, compared with scalar auditory information, might trigger the production of new information. Actually, the IN is activated by a string (of the eigenvalues) with initial information, which is sequentially amplified through the contributions of cooperating triplets. Such a network is described by a system of the ordinary differential equations (A3), which are joint by the double and/or triple cooperative connections. The connection's activities are synchronized, generating the network's periodic rhythms, which possess the inner harmony and anticipative functions.

The reviewed references, related to cognition, neurodynamics, and macrodynamics, have shown that their information regularities and specific information mechanisms are governed by a general information principle, or a law-the same as the VP, imposed on the delivered information for an observer. The simplified schematics Fig.C illustrates the *main* functional relationships between Neurodynamics and Macrodynamics based on the cited references.

*2.4. Conclusion*

The EF information measure integrates and encodes the *inner connections* between the information states during an observed random process, revealing its concealed information. This measure encloses hidden information with unknown regularities. The IPF variation principle allows transforming the EF information to the process' *dynamic* regularities via the informational macrodynamics (IMD) (with its recursive controls) on the basis of proven mathematical foundations. The IMD-disclosed regularities include: the optimal time-space intervals of extracting maximum information from the observed process and its acquisition; forming the cooperative information structures with the defined geometrical borders; their causal-consecutive relationships, hierarchical information networks (IN) and a discrete control logic with the IN code, as a virtual communication language and an algorithm of minimal program to design the IN and provide the evolution cyclic processes with increasing information values.

The applications of these formal information regularities are demonstrated for an information observer's actual cognitive and neurodynamic processes.

The practical question is: What can IMD bring to neuroscience and understanding intelligence?

The answer is summarized by what the IMD does. It:

- Integrates the neurosystem information's features, based on the path functional variation principle and its specific applications. This brings a complex information systemic approach to the cognitive dynamic system, defining not only the main functions of its components, but also their locations in the IN modeled information hierarchy and geometry. Connecting statistical dynamics and information macrodynamics provides us with the optimal window for both measuring and the IMD modeling of many biological interactions directly from real data.
- Explains the information specifics of some neuronal functions, allowing the prediction of their behavior, including the origination of neuron stochastic dynamics and its local and global bifurcations; the memory formation; details of the information transmission during neuronal communication, the dependency of spike strength on a prehistory; the flexibility of the spike's threshold in multiple spatial locations, and the connection between time delay and the neuron's internal and cooperative dynamics;
- Explains the origin of neuronal code from both dynamic and information points of view, and identifies the universal double spiral code with its information dynamic and geometrical structure;
- Presents the formal information mechanisms which create the triplet information structure and the hierarchical information network (IN), based on the ordered cooperative triplet's information dynamics and the formation of the IN node attractors;
- Explains the IN node's sequential attraction, synchronization, competitions, self-organization, anticipation of the future local operations, creativity, and forming cognitive structures, which provide an evolution improvement based on the subjective observer's ability to attract and encode the progressively increased information values.
- Describes the *information analogies* of important physical features and specific neuronal mechanisms; and unifies them with a common concept and model, establishing the *similarities* of these mechanisms.

These features and mechanisms are governed by the law, formulated by the variation principle for a path functional, with the information macrodynamics, which models information regularities and the evolution dynamics of a system, applied to an observer. Moreover, multiple interactions of information observers, acting according to the VP law, are able to create an environment where the law operates.

Finally, both objective and subjective observers are able to implement physically the IMD information software operations as its hardware. Thus, the VP abstract knowledge is able to produce the observer's information regularities, including specific functions and mechanisms of the evolution dynamics.

The observer's formal information regularities are applicable to various specific observers, which are essential parts of the biological, economic and social processes (Lerner 2010, others).

## Appendix
### A0. Diffusion process and information measure of a random process.

A diffusion process is defined as a continuous Markov process. satisfying a finite diffusion and drift .

The process' simple example is a Brownian interactive movement, Levy walks, others, which models many physical, chemical and biological phenomena, and has important applications in economics and finances.

Notion of *information process* is based on the *definition of information*, applied to a random process and modeled by the diffusion process. Considering information as a substance being different from energy and matter, we need to find its origin and a measure, satisfying some definition and/or a law.

On intuitive level, information is associated with diverse forms of *changes* (transformations) in material and/or non-material objects of observation, expressed *universally and independently* of the changes' cause and origin. Such universal changes are brought potentially by *random* events and processes as a set, identified via the related probability in a *probability space*, which are studied in the theory of probability, founded as a *logical* science (Kolmogorov [35]). This randomness with their probabilities we consider as a *source* of information, which means that some of them, but not all of such randomness produces information. The source specifics depend on its particular random events in the *probability space.*

*The questions* are: How to describe a change formally and evaluate information generated by the change?

We consider generally a random process (as a continuous or discrete function $x(\omega,s)$ of random variable $\omega$ and time $s$), described by the *changes* of its elementary probabilities from one distribution (a *priory*) $P_{s,x}^a(d\omega)$ to another distribution (*a posteriori*) $P_{s,x}^p(d\omega)$ in the form of their transformation

$$p(\omega) = \frac{P_{s,x}^a(d\omega)}{P_{s,x}^p(d\omega)}. \tag{A0.1}$$

Such a probabilistic description *generalizes* different forms of specific functional relations, represented by a sequence of different transformations, which might be extracted from the source, using the probabilities ratios (A0.1). It is convenient to measure the probability ratio in the form of natural logarithms: $\ln[p(\omega)]$, where the logarithm of each probability $\leq 1$ would be negative, or zero. That is why the considered function holds the form a logarithmical distance

$$-\ln p(\omega) = -\ln P_{s,x}^a(d\omega) - (-\ln P_{s,x}^p(d\omega)) = s_a - s_p = \Delta s_{ap}, \tag{A0.1a}$$

represented by a difference of a *priory* $s_a > 0$ and a *posteriori* $s_p > 0$ *entropies, which* measures *uncertainty,* resulting from the transformation of probabilities for the source events, that satisfies the entropy's additivity.

A *change* brings a certainty or *information* if its uncertainty $\Delta s_{ap}$ is removed by some equivalent entity call information $\Delta i_{ap}: \Delta s_{ap} - \Delta i_{ap} = 0$. Thus, information is delivered if $\Delta s_{ap} = \Delta i_{ap} > 0$, which requires $s_p < s_a$ and a positive logarithmic measure with $0 < p(\omega) < 1$. Condition of zero information: $\Delta i_{ap} = 0$ corresponds to a redundant change, transforming a priory probability to the equal a posteriori probability, or this transformation is identical–informationally undistinguished. In the same manner, each of the above entropies can be turned into related information. The removal of uncertainty $s_a$ by $i_a$: $s_a - i_a = 0$ brings an equivalent certainty or *information* $i_a$ about entropy $s_a$.

In addition, a *posteriori uncertainty* decreases (or might removes) a priory uncertainty. If $s_p = s_a$, then $s_p$ brings information $i_p = i_a$ regarding $s_a$ by removing this uncertainty, but such a change does not deliver information.

At $s_p < s_a$, the $s_p$ still brings some information about $s_a$ but holds a non zero uncertainty $s_a - s_p = \Delta s_{ap} \geq 0$, which can be removed by the considered equivalent certainty-information $\Delta i_{ap}$. Uncertainty $s_a$ could also be a result of some a priory transformation, as well as uncertainty $s_p$ is resulted from a posteriori transformation.

The logarithmic measure (A0.1a) also fits to the ratio (A0.1) of Markov diffusion process' probabilities having exponential form (Stratonovich 1966,1975), which approximates the probability ratios for many other random processes. Because each above probability and entropy is random, to find an average tendency of functional relations (A0.1,A1a) it is logically to apply a mathematical expectation:

$$E_{s,x}\{-\ln[p(\omega)]\} = E_{s,x}[\Delta s_{ap}] = \Delta S_{ap} = I_{ap} \neq 0, \quad (A0.2)$$

which we call the mean *information of a random source*, being averaged by the source *events* (a probability state), or by the source *processes*, depending on what it is considered: a process, or an event, since both (A0.2) and (1.1) include also Shannon's formula for information of a states (events).

Therefore, *information* constitutes a *universal nature and a general measure of a transformation*, which conveys the changes decreasing uncertainty. Having a *source* of information (data, symbols, relations, links, etc.) does not mean we get *information* and might evaluate its quantity and or a quality. A source only provides changes.

To obtain information, a subset of probability space, selected by formula (A0.2) from the source set of probability space, should *not be empty*, which corresponds $I_{ap} \neq 0$.

*Definition 1.* Information, selected by formula (A0.1a) from the source set of probability space, is defined by *not an empty subset of probability space*, which chooses only *a non-repeating (novel) subset* form the source.

*Definition 2.* Numerical value (in Nat, or Bit), measured by formula (A0.2), determines the *quantity* of information selected from the source.

While, the notion of information formally separates the *distinguished from the undistinguished* (repeating ) subsets (events, processes), formula (A0.2) *evaluates numerically this separation*.

From these definitions it follows that both information, as a *subset* (a string) in probability space, and its measure are *mathematical* entities, *resulting* generally from logical operations, while both *delivering* a source and the *selection* from the source *do not define* notion of information.

In our approach, these actions are performed by an individual's *information observer* (Secs.3-4).

Some of operations, as well as transmission and acquisition of information, require spending energy, that leads to binding information with energy, and to the consideration of related physical, thermodynamic substances, associated with this conversion. A classical physical measurement of this information value (or a "barrier" of the above separation) requires an energy cost to decrease the related physical entropy at least by $k_B$ (at temperature $T$, $k_B$ is Boltzmann constant), which is used for transforming the information measure to the physical one, or for the erasure of this information.

A quantum measurement involves the collapse of a wave function needed to transform the information *barrier*, separating the distinguished from the undistinguished subsets, up to a quantum physical level, which corresponds to an interaction, or an observation. In both cases, such operations *bind* the required equivalent energy with information, or the information logics with physics. The information subset, following from the definitions, is standardized by an encoding this information in different form of information language, which assigns a specific code-word from some chosen language's alphabet to each sequence of the subset symbols. An *optimal* encoding allows transforming the encoded information with preservation of its information quantity. Perhaps the code is a natural product of an observer's cognitive system, which allows it to select specific external information for building each individual observer's body system (Sec.4), with a variety of material implementations for both observer's code and a body. For example, a DNA is a set of information symbols (a code), exposing information, collected by an observer's interaction with environment, which is resulted from various evolutionary informative changes, described by analogy with transformations (A0.1-2). The formula for quantity of information for information process (1.1), modeled by the diffusion process, follows directly from the definitions and basic formulas (A0.1-2).

Indeed. Let have the $n$-dimensional controlled stochastic Ito differential equation:

$$d\tilde{x}_t = a(t,\tilde{x}_t,u_t)dt + \sigma(t,\tilde{x}_t)d\xi_t, \quad \tilde{x}_s = \eta, \; t \in [s,T] = \Delta, \; s \in [0,T] \subset R_+^1 \qquad (A.03)$$

with the standard limitations (Dynkin 1960, others) on the functions of a controlled shift $a(t,\tilde{x}_t,u_t)$, diffusion $\sigma(t,\tilde{x}_t)$, and Wiener process $\xi_t = \xi(t,\omega)$, which are defined on a probability space of the elementary random events $\omega \in \Omega$ with the variables located in $R^n$; $\tilde{x}_t = \tilde{x}(t)$ are solutions of (A.03)) under applied control.

Let us have a diffusion process $\tilde{x}_t^a = \int_s^t \sigma(v,\zeta_v)d\zeta_v$ as the (A.03) solution at $a(t,\tilde{x}_t,u_t)=0$ (that models an uncontrollable noise with $E[\tilde{x}_t^a] = O$), which is defined on a priory probability distribution $P_{s,x}^a(d\omega)$.

Suppose that the control function $u_t$ provides the transformation of $P_{s,x}^a(d\omega)$ to a posteriori probability $P_{s,x}^p(d\omega)$, where a posteriori process $\tilde{x}_t^p$, as a solution of (A.03) at $a(t,\tilde{x}_t,u_t) \neq 0$, is defined.

The above transformation has a density measure (A.01) in the form

$$p(\omega) = \frac{P_{s,x}^a(d\omega)}{P_{s,x}^p(d\omega)} = \exp\{-\varphi_s^t(\omega)\}, \qquad (A.03a)$$

which for the above solutions of (A.03) is determined through the additive functional of the diffusion process (Dynkin1960):

$$\varphi_s^T = 1/2 \int_s^T a^u(t,\tilde{x}_t)^T (2b(t,\tilde{x}_t))^{-1} a^u(t,\tilde{x}_t) dt + \int_s^T (\sigma(t,\tilde{x}_t))^{-1} a^u(t,\tilde{x}_t) d\xi(t). \quad (A.03b)$$

Using the definition (A.02) for (A.0.2a), we get the entropy for these processes

$$\Delta S_{ap} = E_{s,x}\{\varphi_s^t(\omega)\}, \quad (A.03c)$$

and after substituting the math expectation of (A.03a) (at $E[\tilde{x}_t^a] = O$) in (A.03b) we obtain the *Entropy Functional* (EF):

$$\Delta S[\tilde{x}_t]|_s^T = 1/2 E_{s,x}\{\int_s^T a^u(t,\tilde{x}_t)^T (2b(t,\tilde{x}_t))^{-1} a^u(t,\tilde{x}_t) dt\} = \int_{\tilde{x}(t)\in B} -\ln[p(\omega)] P_{s,x}(d\omega) = -E_{s,x}[\ln p(\omega)] \quad (A.03d)$$

EF is an *integral measure* of the process trajectories, expressed through the drift function $a^u(t,\tilde{x}_t) = a(t,\tilde{x}_t,u_t)$, depending on a control $u_t$, and covariation function $b(t,\tilde{x}_t)$, describing its diffusion component; $E_{s,x}$ is a conditional to the initial states $(s,x)$ math expectation, taken along the $\tilde{x}_t$ trajectories. On the right side of (A.03d) is the EF equivalent's formula expressed via probability density $p(\omega)$ of random events $\omega$ integrated by the probability measure $P_{s,x}(d\omega)$ along the process trajectories $\tilde{x}(t) \in B$, defined at the set $B$.

The parameters of Ito equations (A.03) are a *partially observable* via measuring only a current covariation function of the process' trajectories, which allows identify the corresponding shift function (A3) and find an optimal control minimizing the EF (A2, A3). Changing any of drift's and covariation's functions will modify the process' EF.

Entropy functional (A.03c,d) is an *information indicator* of a *distinction* between the processes $\tilde{x}_t^a$ and $\tilde{x}_t^p$ by these processes' measures; it measures a *quantity of information* of process $\tilde{x}_t^p$ regarding process $\tilde{x}_t^a$. For the process' equivalent measures, this quantity is zero, and it takes a positive value for the process' nonequivalent measures.

The definition (A.01, A.3a) specifies *Radon-Nikodym's density* measure for a probability density measure, applied to entropy of a random process (Stratonovich). The *quantity of information* (A.02) is an equivalent of Kullback–Leibler's divergence (KL) for a continiuos random variables (Kullback 1959):

$$D_{KL}(P_{x,s}^a \parallel P_{x,s}^p) = \int_X \ln\frac{dP^a}{dP^p} dP^p = E_x[\ln\frac{P_{x,s}^a(d\omega)}{P_{x,s}^p(d\omega)}] = E_x[\ln p(\omega)] = -\Delta S_{ap}, \quad (A.03e)$$

where $\frac{dP^a}{dP^p}$ is Radon–Nikodym derivative of probability $P^a$ with respect to probability $P^p$:

$$P^a(X) = \int_X P_{x,s}^a(d\omega), \quad P^p(X) = \int_X P_{x,s}^p(dx),$$ which are expressed via elementary probabilities (A.01) and also has a nonsymmetrical distance's measure between the above entropies $s_a$ and $s_p$.

The KL measure is connected to both Shannon's conditional information and Bayesian inference (Jaynes 1998) of testing a priory hypothesis (probability distribution) by a posteriori observation's probability distribution.
The observer (Sec.2.3-2.4) implements the *Bayesian inference* by *collecting* a priory evidences.

The considered information path functional (IPF), defined on its each $i-$ component of the VP extremal's segment between two points $x_i(\tau_o), x_i(\tau_1)$, has the simplified form (Lerner 2010):

$$I_{x_i}^p = 1/8 \int_{\tau_o}^{\tau_1} r_i^{-1} \dot{r}_i dt = 1/8[\ln r_i(\tau_1) - \ln r_i(\tau_o)], \quad (A0.4)$$

where $r_i(t) = E(\tilde{x}_i^2)$ is a covariation (correlation) function, determined for each random process' component $\tilde{x}_i$, which is connected with the corresponding component of the diffusion matrix function by relation $\dot{r}_i = 2b_i$.

The IPF for a total process is

$$I_{x_t}^p = 1/8 \int_s^T \sum_{i=1}^n \dot{r}_i r_i^{-1} dt = 1/8 \int_s^T Tr[\dot{r}r^{-1}] = 1/8 Tr[\ln r \big|_s^T], (s = \tau_o, \tau_1,...,\tau_n = T). \qquad (A0.4a)$$

For the comparison, we provide below Shannon's information measure for the *Gauss random states*.

Let's have $n$ dimensional vector of random states $x = (x_i), i = 1,...,n$, described by a normal distribution

$$p(x) = (2\pi)^{-n/2} \det^{-1/2} r \exp[-1/2(x - E(x))r^{-1}(x - E(x))^T] \qquad (A0.4b)$$

with a vector of mean $E(x)$ and correlation matrix $r = E[(x - E(x))(x - E(x))^T]$.

Then, as it is shown in (Stratonovich 1975), the quantity of information, following from applying (A0.2) to this distribution (at a subsequent scale of the form (A0.4b)), is $I_x = 1/2 \ln \det r$, and after using the matrix eigenvalues $r = (\lambda_k), k = 1,...n$, it acquires the form

$$I_x = 1/2 \ln \sum_{k=1}^n \lambda_k; \text{ or } I_x = 1/2 Tr[\ln r], \qquad (A0.4c)$$

while each $k$-matrix component gets information
$I_x^k = 1/2 \ln \lambda_k$.

Information exists ($I_x \neq 0$) if $\sum_{k=1}^n \lambda_k \neq 1$, or for each $I_x^k$ if $\lambda_k \neq 1$.

The EF-IPF approach converts the *uncertainty* of a random process into the *certainty* of a dynamic information process, with its code and the IN structure (Secs. 1.2,1.3,1.5). The considered cut-off pieces of information (Sec. 1.1) can be encoded through measuring the related correlations (A. 04a,c) at the DP locations of diffusion process.

**A1.** *The step-wise control's action on the entropy functional* **(A.03d)** *of diffusion process* $\tilde{x}_t$.

The considered control $u_t$ is defined as a piece-wise continuous function of $t \in \Delta$:

$$u_+ \overset{def}{=} \lim_{t \to \tau_k + o} u(t, \tilde{x}_{\tau_k}), u_- \overset{def}{=} \lim_{t \to \tau_k - o} u(t, \tilde{x}_{\tau_k}), \qquad (A1.1)$$

which is differentiable, excluding a set
$$\Delta^o = \Delta \setminus \{\tau_k\}_{k=1}^m, k = 0,...,m. \qquad (A1.2)$$

The jump of the control function $u_-$ in (A1.1) from a moment $\tau_{k-o}$ to $\tau_k$, acting on a *diffusion* process $\tilde{x}_t = \tilde{x}(t)$, might "cut off" this process after moment $\tau_{k-o}$.

The "cut off" diffusion process has the same drift vector and the diffusion matrix as the initial diffusion process.

Functional (A1.1) can be expressed via the process additive functional $\varphi_s^T$ in the form $\Delta S[\tilde{x}_t]\big|_s^T = E_{s,x}[\varphi_s^T]$, and the additive functional related to this "cut off" has a form (Prochorov, Rozanov 1973):

$$\varphi_s^{t-} = \begin{cases} 0, t \leq \tau_{k-o}; \\ \infty, t > \tau_k. \end{cases} \qquad (A1.3)$$

The jump of the control function $u_+$ (A1.2) from $\tau_k$ to $\tau_{k+o}$ might cut off the diffusion process *after* moment $\tau_k$ with the related additive functional

$$\varphi_s^{t+} = \begin{cases} \infty, t > \tau_k; \\ 0, t \leq \tau_{k+o}. \end{cases} \qquad (A.1.4)$$

At the moment $\tau_k$, between the jump of control $u_-$ and the jump of control $u_+$, we consider a control *impulse*

$$\delta u_{\tau_k}^{\mp} = u_-(\tau_{k-o}) + u_+(\tau_{k+o}). \tag{A1.5}$$

The related additive functional at a vicinity of $t = \tau_k$ acquires the form of an impulse function

$$\varphi_s^{t-} + \varphi_s^{t+} = \delta\varphi_s^{\mp}. \tag{A1.6}$$

The entropy functional at the localities of the control's switching moments (A1.2) takes the values

$$S_- = E[\varphi_s^{t-}] = \begin{cases} 0, t \leq \tau_{k-o}; \\ \infty, t > \tau_k. \end{cases} \text{ and } S_+ = E[\varphi_s^{t+}] = \begin{cases} \infty, t > \tau_k; \\ 0, t \leq \tau_{k+o}, \end{cases} \tag{A1.7}$$

changing from 0 to $\infty$ and back from $\infty$ to 0 and acquiring an *absolute maximum* at $t > \tau_k$, between $\tau_{k-o}$ and $\tau_{k+o}$.
The related multiplicative functionals are

$$p_s^{t-} = \begin{cases} 0, t \leq \tau_{k-o} \\ 1, t > \tau_k \end{cases} \text{ and } p_s^{t+} = \begin{cases} 1, t > \tau_k \\ 0, t \leq \tau_{k+o} \end{cases}, \tag{A1.7a}$$

which determine probabilities $\tilde{P}_{s,x}(d\omega) = 0$ at $t \leq \tau_{k-o}, t \leq \tau_{k+o}$ and $\tilde{P}_{s,x}(d\omega) = P_{s,x}(d\omega)$ at $t > \tau_k$.

For the "cut-off" diffusion process, transitional probability (at $t \leq \tau_{k-o}, t \leq \tau_{k+o}$) turns to zero.

Then, the states $\tilde{x}(\tau - o), \tilde{x}(\tau + o)$ become independent, and the mutual time *correlations are dissolved*:

$$r_{\tau-o,\tau+o} = E[\tilde{x}(\tau-o)\tilde{x}(\tau+o)] \to 0. \tag{A1.7b}$$

The entropy $\delta S_-^+(\tau_k)$ of the additive functional $\delta\varphi_s^{\mp}$, produced within, or at a border of the control impulse (A1.5), is define by the equality

$$E[\varphi_s^{t-} + \varphi_s^{t+}] = E[\delta\varphi_s^{\mp}] = \int_{\tau_{k-o}}^{\tau_{k+o}} \delta\varphi_s^{\mp} P_\delta(d\omega), \tag{A1.8}$$

where $P_\delta(d\omega)$ is a probability evaluation of impulse $\delta\varphi_s^{\mp}$.

Taking integral of the $\delta$-function $\delta\varphi_s^{\mp}$ between the above time intervals, we get at the border: $E[\delta\varphi_s^{\mp}] = 1/2 P_\delta(\tau_k)$ at $\tau_k = \tau_{k-o}$, or $\tau_k = \tau_{k+o}$. The impulse, produced by the controls, is a non random with $P_\delta(\tau_k) = 1$, which brings the EF estimation at $t = \tau_k$:

$$S_{\tau_k}^{\delta u} = E[\varphi_s^{\mp}] = 1/2. \tag{A1.9}$$

This entropy increment evaluates an information contribution from the impulse controls (A1.5) at a vicinity of the above discrete moments. Since that, each information contribution

$$E[\varphi_s^{t-}]_{\tau_k} = S_{\tau_k}^{u_-} \text{ and } E[\varphi_s^{t+}]_{\tau_k} = S_{\tau_k}^{u_+}, \tag{A1.9a}$$

at a vicinity of $t = \tau_k$, is produced by the corresponding controls' step functions $u_-(\tau_k), u_+(\tau_k)$ in (A1.1), (A1.5) accordingly, which can be estimated by

$$S_{\tau_k}^{u_-} = 1/4, u_- = u_-(\tau_k), \tau_{k-o} \to \tau_k; S_{\tau_k}^{u_+} = 1/4, u_+ = u_+(\tau_k), \tau_k \to \tau_{k+o}, \tag{A1.10}$$

where the entropy, according to its definition (1.1), is measured in the units of Nat (1 Nat $\cong$ 1.44bits).
Estimations (A1.9), (A1.10) determine the entropy functional's cut-off values at the above time's borders under actions of these controls, which decreases the quantity of the functional's information by the amount that had been concealed before the cutting the process correlations (A7b). According to the evaluation of an upper bound entropy per an English character (token) by Brown et al 1992, its minimum is estimated by 1.75 bits, with the average amount between 4.66-7 bits per character.

The evaluation includes the inner information bound by a character. At minimal entropy per symbol in 1 bit, a minimal symbol's bound information is 0.75, which is close to our evaluation at the cut-off.

The EF definition by the probability density measure (A1.1),(A1.7a), which specifies *Radon-Nikodym's density* measure (Stratonovich 1975), *holds* also the considered cut-off peculiarities of the controllable process' EF.

The impulse control of Markov processes with a discrete process' intervention were studied by Yushkevich 1983, Dempster and Ye 1995, and others.

**A2. The considered *variation problem*** (VP):

$$\min_{\tilde{x}_t(u)} \tilde{S}[\tilde{x}_t] = S[x_t], \ \tilde{S}[\tilde{x}_t] = S(\tilde{x}_t / \varsigma_t), \tag{A2.1}$$

connects the entropy functional (A1.1) (defined on a random trajectories) with the IPF integral functional

$$S = \int_s^T L(t, x, \dot{x}) dt = S[x_t], \tag{A2.1a}$$

*Proposition* A2.1. The *solution* of variation problem (A2.1) for the entropy functional brings the following equations of extremals for a vector $x$ and a conjugate vector $X$ accordingly:

$$\dot{x} = a^u, \ (t, x) \in Q, \tag{A2.2}$$

$$X = (2b)^{-1} a^u; \tag{A2.3}$$

and the *constraint*

$$a^u(\tau) X(\tau) + b(\tau) \frac{\partial X}{\partial x}(\tau) = 0, \tag{A2.4}$$

imposed on the solutions (A2.2), (A2.3) at some set

$$Q^o \subset Q, \ Q^o = R^n \times \Delta^o, \Delta^o = [0, \tau], \tau = \{\tau_k\}, k = 1, ..., m; \tag{A2.5}$$

where controls (A1.8), (A1.9), applied at the discrete moments $(\tau_k)$ (A2.5), implement these VP solutions. •

Here function $a^u = a^u(t, x_t)$, defined on extremals $x_t$, is a dynamic analogy of the drift function $a^u = a(t, \tilde{x}_t, u_t)$ in (A1.1), and $b = b(t, x_t)$ is a dynamic analogy of the dispersion matrix $b(t, \tilde{x}_t)$ in (A1.1).

On the extremals $x_t = x(t)$, both $a^u$ and $b$ are nonrandom, after their substitution to the conditional math expectation (A1.1) we get the integral functional $\tilde{S}$ on the extremals:

$$\tilde{S}_e[x(t)] = 1/2 \int_s^T (a^u)^T (2b)^{-1} a^u dt = S[x_t]. \tag{A2.6}$$

Solution of this variation problem's (VP) (with the detailed proofs in Lerner 2007) *automatically brings the dynamic constraint* (DC), applied *discretely* at the states' set DP (A2.5) by the controls (A1.2) (synthesized in A3). The DC equation follows from the VP specifics (which join the conjugate vector in (A2.3) with the operator of the extremal (A2.1)). This brings the DC solution to an extremal solution of the functional (IPF), which provides a minimum for the entropy functional (EF); while both operator and DC are the attributes of the diffusion process' equation, identified on the process.

The DC imposition, implemented by the control action, is limited by the punched points $x(\tau_k), x(\tau_{k+1})$ when the control, starting at the moment $\tau_k^o = \tau_k + o$, should be *turned off* at a moment $\tau_k^1 = \tau_{k+1} - o$ preceding $\tau_{k+1}$, when the extremal's solution of (A2.2) approaches the punched point $\tau_{k+1}$. To continue the extremal movement, the control should be turned *on* again at the moment $\tau_{k+1}^o = \tau_{k+1} + o$, following $\tau_{k+1}$, to start a next extremal segment. This determines a discrete action of a step-wise control, imposing the constraint during the extremal movement within each interval $t_k = \tau_k^1 - \tau_k^o, t_{k+1} = \tau_{k+1}^1 - \tau_{k+1}^o, k = 1, ..., m$ ($m$ is the number of applied controls), which limits a time length of the extremal segment between the punched localities. The process continuation requires connecting the extremal segments between the punched localities by the joint step-wise control action: with a $k$-control *turning off* at the moment $\tau_k^1$, while transferring to $\tau_{k+1}$ locality, and a next $k+1$-control, which, transferring from $\tau_{k+1}$ locality, is *turning on* at the moment $\tau_{k+1}^o$ at the following extremal segment. Both step-wise controls form an impulse control function $\delta(x(\tau_k^1), x(\tau_{k+1}), x(\tau_{k+1}^o))$ (by analogy with A1.5) acting between moments $\tau_k^1, \tau_{k+1}, \tau_{k+1}^o$ (or $\tau_{k-1}^1, \tau_k, \tau_k^o$) and implementing (A1.7, A1.7a), which brings the EF peculiarities (A1.14), (A1.15) at these moments. Through this control action, applied at these points, the extremal segments interact, modeling the interactive dynamics in Lerner 2010. Following (A2.4), (A2.6), the constraint integral's increments along an extremal segment between the punched points, is preserved:

$$E[\Delta S_{ic}(x(t_k))] = inv, \ k = 1,...,m. \quad (A2.7)$$

The invariant condition (A2.7) allows selecting and evaluating an extremal *segment*, which approximates a related segment of the random (microlevel) process (between these punched localities), as a *macroprocess*. This fact also leads to a *prediction* of each following process movement based on information, collected at each previous DP (sec.A3), and as a result, the macroprocess' extremals provide a dynamic prognosis of the microprocess' evolution.

*The law of information dynamics,* following from the Hamilton equation (A.2.2, A2.3):

$$\dot{x}_i = 2b_i X_i, X_i = \frac{\partial \Delta S_i}{\partial x_i} \quad (A2.8)$$

defines a dynamic speed (information flow, sec.1.4) for the macroprocess' information states $x(t) = \{x_i(t)\}, i = 1,...,n$, which is a carrier of information (novelty) within each $\Delta S_{ic}(x(t_k))$. Each local increment of an *information potential* $\Delta S_i$ is limited by the segment's total information potential $\Delta S_{ic}(x_i(t_k))$ at the points $x_i(t_k)$ of stopping the information movement on the segment's end. A field of information force $X_i$ is aimed at preservation of the information process' integrity, impeding the separation of a given process on its portions. According to (A2.8), a *flow* of information $\dot{x}_i$, carried by the states' information speed, is *proportional* to a gradient of information dynamic potential $X_i$ and a dispersion $b_i$ of each random state $\tilde{x}_i(t)$, which model their *causal-consequence relationship*. At a fixed $X_i$ and an increased dispersion, this novelty is distributed with growing speed, which corresponds to a practical dissemination of novelties when, at growing $b_i$, the sets of their carriers, with the process' probabilities, are spreading. The information law is extended on the space moving and cooperating systems (Lerner 2008). We may also specify the information force $X_\delta = \frac{\partial \Delta S_\delta}{\partial x}$, which we will get using $\partial x = 1/2cr^{-1/2}\partial r$ from (A3.13)(below) and $\frac{\partial \Delta S_\delta}{\partial r} = -1/8r^{-1}$ from (A.03). Then we obtain

$$X_\delta = \frac{\partial \Delta S_\delta}{1/2cr^{-1/2}\partial r} = -1/4c_o r^{-1/2}. \quad (A.29)$$

This is a finite information force, which keeps the states of process bounded (sec.1), counterbalancing to the destructive force that arises at dissolving correlations, at $r \to 0$ (A1).

**A3. *The identification, optimal controls, information invariants, and an optimal prediction***

Writing equation of extremals $\dot{x} = a^u$ in a dynamic model's traditional form (Alekseev *et al* 1979):
$$\dot{x} = Ax + u, u = Av, \dot{x} = A(x+v), \quad (A3.1)$$
where $v$ is a control reduced to the state vector $x$, we find the control $v$ that solves the initial variation problem (VP) and identify matrix $A$ under this control's action.

<u>Proposition</u> <u>A3.1.</u> The reduced control is formed by a feedback function of macrostates $x(\tau) = \{x(\tau_k)\}, k = 1,...,m$:

$$v(\tau) = -2x(\tau) \qquad (A3.2)$$

measured at the moments $\tau = (\tau_k)$ (A2.5), and the matrix $A$ is identified by the equation

$$A(\tau) = -b(\tau)r_v^{-1}(\tau), r_v = E[(x+v)(x+v)^T], b = 1/2\dot{r}, r = E[\tilde{x}\tilde{x}^T] = r_v(\tau) \qquad (A3.3)$$

through the above correlation functions, or directly, via the dispersion matrix $b$ from (A1.1):

$$|A(\tau)| = b(\tau)(2\int_{\tau-o}^{\tau} b(t)dt)^{-1} > 0, \ \tau - o = (\tau_k - o), k = 1...,m. \bullet \qquad (A3.3a)$$

*Proposition* A3.2. Relation $A^v = A(x(t) + v(\tau))$, under the control $v(\tau_k^o) = -2x(\tau_k^o)$, applied during interval $t_k = \tau_k^1 - \tau_k^o$, by the moment $\tau_k^1$ acquires the form

$$A^v(\tau_k^1) = -A(\tau_k^o)\exp(A(\tau_k^o)\tau_k^1)[2 - \exp(A(\tau_k^o)\tau_k^1)]^{-1}. \bullet \qquad (A3.4)$$

*Proposition* A3.3. (1)-Each moment $\tau_k^1$ of the control turning off is found from the fulfillment of the constraint Eq.(A2.4) (at this moment) under applying control $v(\tau_k^o) = -2x(\tau_k^o)$ starting at $\tau_k^o$;

(2)-Between the moments $\tau_k^1 - \tau_k^o = t_k$, both the random process and its dynamic model, running under the same control $v(\tau_k^o)$ (that imposes the constraint during each $t \leq t_k$), have the equal dynamic trends of functions $a^u(\tau_k^o, t), b(\tau_k^o, t), t \leq t_k$ in (1.1)- *for the extremal* in (A2.1),(A2.2), *and* in (A3.1), which provide a *dynamic equivalence* for the macroprocess and the diffusion process. $\bullet$

*Proposition* A3.4. The constraint's information invariant $E[\Delta S_{ic}]$ (A2.7) during interval $t_k$ of applying controls $v(\tau_k^o) = -2x(\tau_k^o)$ leads to the following three invariants:

$$\lambda_i(\tau_k^o)\tau_k^o = inv = i_1 ,(a): \ \lambda_i(\tau_k^1)\tau_k^1 = inv = i_2, \ (b); \ \lambda_i(\tau_k^o)\tau_k^1 = inv = i_3 ,(c) \qquad (A3.5)$$

and their connections in the forms

$$i_1 = 2i_2 ,(a); i_2 = i_3 \exp i_3 (2 - \exp i_3)^{-1} ,(b) \qquad (A3.6)$$

where $\lambda_i(\tau_k^o)$ and $\lambda_i(\tau_k^1)$ are the model matrix $A = (\lambda_i)_{i=1}^n$ eigenvalues taken at the moment $\tau_k^o$ and $\tau_k^1$ of a segment's time interval $t_k$ accordingly. $\bullet$ Detailed proof is in Lerner 2007.

Considering the real eigenvalues for the complex eigenvalues at each of the above moments: $\text{Re}\lambda_i(\tau_k^1) = \alpha_{ik}$, $\text{Re}\lambda_i(\tau_k^o) = \alpha_{iko}$, we come to the real forms of invariant relations (A3.5),(A3.5a-b):

$$\alpha_{ik}\tau_k^1 = \mathbf{a}_i, \alpha_{iko}\tau_k^1 = \mathbf{a}_{io}, \alpha_{iko}\tau_k^o = a_i^o, \qquad (A3.7)$$

where $a_i^o = -a_i^\tau, a_i^\tau = \text{Re}\lambda_i(\tau_k + o)\tau_k^o; \lambda_i(\tau_k + o) = -\lambda_i(\tau_k^o)$. $\qquad (A3.7a)$

Applying relations (A3.5a-c), (A3.5a-b) for the complex eigenvalue' imaginary parts: $\text{Im}\lambda_i(\tau_k^1) = \beta_{ik}, \text{Im}\lambda_i(\tau_k^o) = \beta_{iko}$ brings the related invariants:

$$\beta_{ik}\tau_k^1 = \mathbf{b}_i, \beta_{iko}\tau_k^1 = \mathbf{b}_{io}, \beta_{iko}\tau_k^o = b_i^o. \qquad (A3.8)$$

Here, $\alpha_{iko}$ measures a fixed entropy derivation at each segment's DP punched locality, invariants $\mathbf{a}_i, a_i^o$ evaluate the control's information contribution by step-wise controls (A3.2), (A1.8), which turns the constraint on *or* off.

Invariants $\mathbf{b}_{io}$ and $\mathbf{b}_i, b_i^o$ evaluate the model segment's information, generated by the eigenvalues imaginary components, while a total real information contribution is evaluated by $\mathbf{a}_{io}$.

Eq. (A3.3a), determined by an interval of observation, imposes the restriction on the solution of (A3.4), which we express through the equations for both invariant $\mathbf{a}_{io}$ and $\mathbf{b}_{io}$ and the starting eigenvalue's ratio $\gamma_i = \beta_{iko}/\alpha_{iko}$:

$$2\sin(\gamma_i\mathbf{a}_{io}) + \gamma_i \cos(\gamma_i\mathbf{a}_{io}) - \gamma_i \exp(\mathbf{a}_{io}) = 0(a); \ \mathbf{b}_{io}2\cos(\gamma\mathbf{b}_{io}) - \gamma_i \sin\gamma(\mathbf{b}_{io}) - \exp(\mathbf{b}_{io}) = 0(b). \qquad (A3.9)$$

Eq (A3.9a) allows us to find *function* $\mathbf{a}_{io} = \mathbf{a}_{io}(\gamma_i)$ for each starting eigenvalue $\alpha_{iko}$ satisfying (A.3.4), and get also $\gamma_i(\alpha_{iko})$, indicating the dependence of $\gamma_i$ on the identified $\alpha_{iko}$. A stable process within a time interval holds $\mathbf{a}_{io} < 0$ at $\gamma_i \cong 0.5$, and $\gamma_i \to 0$ corresponds to a real $\alpha_{iko}$. The computation, using the Eqs (A3.9a) and (A3.6a), brings $\mathbf{a}_{io}(\gamma_i \to 0) \cong 0.75$ and $\mathbf{a}_i(\gamma_i \to 0) \cong 0.25$, which is consistent with (A1.15). The information values of the above invariants $\mathbf{a}_{io}$ and $\mathbf{a}_i$ correspond to 1 bit and 0.35 bits accordingly. (These results, following from the IPF, do not apply classical information theory). *If* the model segment's eigenvalues $\lambda_i(\tau_k^o) = \alpha_i(\tau_k^o) \pm \beta_i(\tau_k^o)$ is currently identified, *then* $\gamma_i$ is known, and using (A3.9) can be found $\mathbf{a}_{io}(\gamma_i)$ which determines $\tau_{k-o}$ for $\alpha_i(\tau_k^o) = \alpha_{iko}$.

At the known invariants, the eigenvalues (identified at the punched localities) determine the segments time interval and the model solutions. The model dynamics are initiated by applying a starting step-wise control (to a random object) in the form

$$v(\tau_o^o) = -2E_{\tau_o^o}[\tilde{x}_t(s)], \qquad (A3.10)$$

at $\tau_o^o = s + o$, where $\tau_o^o$ is the moment of the control's starts, $\tilde{x}_t(s)$ are the object's initial conditions, which also include given correlations

$$r(s) = E[\tilde{x}_t(s)\tilde{x}_t(s)^T] \text{ and/or } b(s) = 1/2\dot{r}(s). \qquad (A3.11)$$

These initial conditions also determine a starting *external* control (in (A3.1)):

$$u(\tau_o^o) = b(\tau_o^o)r(\tau_o^o)^{-1}v(\tau_o^o), \qquad (A3.12)$$

where $v(\tau_o^o) = -2x(\tau_o^o)$, and a nonrandom state can be defined via

$$x(\tau_o^o) \cong |r^{1/2}(\tau_o^o)|. \qquad (A3.13)$$

Control (A3.12) imposes the constraint (in the form A(2.13)) that allows starting the dynamic process.
The above initial conditions identify

$$A(\tau_o^o) = b(\tau_o^o)r(\tau_o^o)^{-1}, A(\tau_o^o) = (\lambda_i(\tau_o^o)), i = 1,...,n \qquad (A3.14)$$

which is used to find a first time interval $t_1 = \tau_1^1 - \tau_o^o$ between the punched localities, where the next matrix's elements are identified, and so on.

The impulse control (IC) (Fig.1b), (A.15) (with two step-controls SP1, SP2, forming the IC's left and right sides accordingly), performs the following functions:

- The IC, applied to diffusion process, provides a sharp maximum of the process probability, which maximizes an influx of external information to the starting at the same moment of the model's dynamic process; implementing at this moment the principle of maximal entropy;
- The start of the model's dynamic process, which is associated with imposing the dynamic constraint (DC) on the equation for the diffusion process with the SP2 aid, acting along the extremal;
- At the moment when the DC solution finishes the imposing constraint, the SP2 stops by applying the control reverse step action SP1, and the dynamic process' extremal solution meets the punched ("bounded") locality of diffusion process;
- At this locality, the IC is applied (with both SP1, stopping the DC, and the SP2, starting the DC at the next extremal segment), and all processes sequentially repeat themselves, concurrently with the time course of the diffusion process;
- The starting impulse control binds the process' maximum probability state (holding its maximal entropy), with the following start of the extremal movement. The IC also joins the extremal segments in their sequential chain with building the IN.

A primary step control is determined by initial conditions $x_o$ of the diffusion process, in particular, by its dispersion $b_o$ (A3.13), which allows the identification the initial model's dynamic operator for the starting at this moment DC. Knowing the operator's eigenvalues leads to finding the moment of turning the control off, at which time the following impulse control is applied. While the control's left side is turning off, its right side starts the DC and the next extremal movement, beginning with the new punched location of the diffusion process $x_1(\tau_1)$ and corresponding $b_1$. The model's renovated operator can be found during the dynamic process movement on this extremal, because of its dynamic approximation with a maximal probability of the related trajectory of the diffusion process. In particular, using the identification eq. $\lambda_1 = 1/2\dot{r}_1(t)r_1^{-1}(t), r_1 = E[x_1^2(t)]$ on the model's extremal $x_1(t)$, enables finding the interval of applying the above control. After that, the next impulse control is applied, whose IC left side turns the DC off, and its right side turns the DC on, starting the next extremal segment. Under these conditions, the direct identification of each following $b_k, x_k(\tau_k)$ and the models operator is not needed. The quantity of information delivered at each punched locality is determined by the model's invariants, depending on both the eigenvalue and the interval of the DC applying.

*The specific* of the considered optimal process *consists of the computation of each following time interval* (where the identification of the object's recent operator will take place and the next optimal control is applied) *during the optimal movement under the current optimal control, formed by a simple function of dynamic states.*

In this *optimal dual* strategy, the IPF optimum predicts each extremal's segments movement not only in terms of a total functional path goal, but also by setting at each following segment the renovated values of this functional's controllable drift and diffusion, identified during the optimal movement, which currently correct this goal.

Fixing and memorizing the cooperative dynamics (by the optimal controls) produce the IN *structural* information.

An influx of a local maximum of external information is reduced by growing the structural information that minimizes the system information according to the VP.

*Connection to Shannon's information theory.*

Considering a set of discrete states $\tilde{x}(\tau^o) = \{(\tilde{x}_i(\tau_k^o))\}, i = 1,...,n; k = 1,...,m$ at each fixed moments $\tau_k^o$ along $n$-dimensional random process, and using definition of the entropy functional (1.1), we get the conditional entropy *function* for the conditional probabilities (corresponding to (1.1)) at *all* moments $\tau_k^o$ at each process dimension $S_{\tau_k^o}^i$ and for the whole process $S_{\tau_k^o}$ accordingly:

$$S_{\tau_k^o}^i = -\sum_{k=1}^{m} p_k[\tilde{x}_i(\tau_k^o)]\ln p_k[\tilde{x}_i(\tau_k^o)], S_{\tau_k^o} = \sum_{i=1}^{n} S_{\tau_k^o}^i, \quad (A3.15)$$

which *coincides with the Shannon entropy* for each probability distribution $p_k[\tilde{x}_i(\tau_k^o)]$, measured at each fixed $\tilde{x}_i(\tau_k^o)$. Function (A3.15) holds all characteristics of Shannon's entropy, following from the initial Markov process and its additive functional for these *states*. For the comparison, the controllable IPF entropy, measured at the related DP's punched localities $\tilde{x}(\tau) = \{\tilde{x}(\tau_k)\}, k = 1,...,m$

$$\tilde{S}_{\tau m}^i = \sum_{k=1}^{m} \Delta S_k[\tilde{x}(\tau_k)], \quad (A3.16)$$

(where $\Delta S_k[\tilde{x}(\tau_k)]$ is the entropy at each $\tau_k, k = 1,...m$) is *distinctive* from the entropy $S_{\tau_k^o}^i$ (A3.15), because: (1)- The IPF entropy $\Delta S_k[\tilde{x}(\tau_k)]$ holds the macrostates' ($x_i(\tau_k - o), x_i(\tau_k), x_i(\tau_k + o)$) *connection* through the punched locality (performed by applying the two step-wise controls), while the EF binds all random states, including the punched localities;

(2)-The distribution $p_k = p_k[\tilde{x}(\tau_k)]$ is selected by variation conditions (A2.1, 2.1a) (applied to (A1.7a)), as an extremal probability distribution, where a macrostate $x(\tau_k)$ estimates a random state $\tilde{x}(\tau_k)$ with a maximum probability $p_k = p_k[\tilde{x}(\tau_k)]$. At this "sharp" probability maximum, the information entropy (A1.7) reaches its local maximum at $\tau_k$-locality, which, for the variation problem (A2), is associated with turning the constraint (A2.4) off (by the controls) when function A(2.5) takes its maximum (at switching to the random process).

Selecting these states (with an aid of the dynamic model's control (A3.2)) allows an optimal discrete filtration of random process at all $\tau_k, k = 1,...m$, where the macromodel is identified and external information is delivered. Thus, $x(\tau_k)$ emerges as the *most informative* state's evaluation of the random process at the $\tau_k$-locality. Even though the model identifies a sequence of the most probable states, representing the most probable trajectory of the diffusion process, the IPF minimizes the EF entropy functional, defined on a *whole* diffusion process.

Therefore, the IPF implements the *optimal process' functional information measure in the evolution dynamics.*

Each of the entropy $\Delta S_k[x_i(\tau_k - o)]$ that measures the process segment's undivided information, and the entropy $\Delta S_k[x_i(\tau_k)]$ delivers *additional* information compared to the traditional Shannon entropies, which measure the process' sates at the related discrete moments. This information, connecting the IPF segments, compensates one for a punched locality. Evaluating each segment's *total* information contribution from both entire segment's inner entropy $\Delta S_k[x_i(\tau_k - o)]$ and segments' $\tau_k$-locality $\Delta S_k[x_i(\tau_k)]$ by the invariant's information measure $\mathbf{a}_o(\gamma_k)$, we get the total process $\tilde{S}_{\tau m}^i$ *estimation* by the sum of the invariants, which count both the inner segment's and control inter-segment's information:

$$\tilde{S}_{\tau m}^i = \sum_{k=1}^{m}(\mathbf{a}_o(\gamma_k) + \mathbf{a}_o^2(\gamma_k)), \tilde{S}_\tau = \sum_{i=1}^{n}\tilde{S}_{\tau m}^i,, \tag{A3.17}$$

(delivered by $\delta$ − control), where $m$ is the number of the segments, $n$ is the model dimension (assuming each segment has a single $\tau_k$-locality). However, to *predict* each $\tau_k$- locality, where $\Delta S_k[\tilde{x}(\tau_k)]$ should be measured, we need only invariant $\mathbf{a}_o(\gamma_k)$ (evaluating (A2.7)) and a sum of process's invariants

$$\tilde{S}_{\tau m}^{io} = \sum_{k=1}^{m}\mathbf{a}_o(\gamma_k), \tilde{S}_\tau^o = \sum_{i=1}^{n}\tilde{S}_{\tau m}^{io} \tag{A3.17a}$$

estimates the IPF entropy with a maximal process' probability. Knowing *this* entropy allows encoding the *random process* using the Shannon formula for an average optimal code-word length:

$$l_c \geq \tilde{S}_\tau^o / \ln D, \tag{A3.18}$$

where $D$ is the number of letters of the code's alphabet, which encodes $\tilde{S}_\tau^o$ (A3.17a).

An elementary code-word to encode the process' segment is

$$l_{cs} \geq \mathbf{a}_o(\gamma_k) \text{ [bit]}/ \log_2 D_o, \tag{A3.19}$$

where $D_o$ is a segment's code alphabet, which implements the macrostate connections.

At $\mathbf{a}_o(\gamma_k \to 0.5) \cong 0.7$, $D_o = 2$, we get $l_{cs} \geq 1$, or a bit per the encoding letter.

Therefore, the invariant $\mathbf{a}_o = \mathbf{a}_o(\gamma_i)$ for each *i*-segment allows us to encode the *process* using (A3.18),(A3.19) *without* counting each related entropy (A3.16), and to compress each segment's random information to $\mathbf{a}_o$ bits.

Whereas, to estimate the entropy $\tilde{S}_\tau$ (A3.17) we need also counting $\mathbf{a}_o^2(\gamma_k) \cong 2\mathbf{a}(\gamma_k)$.

*In a system of information transmission, each sender, encoding the process' information, and a receiver, decoding the transmitted information, include also the related observers.*

**A4**. *Mathematical specifics of cyclic evolution*

<u>Proposition</u> 4.1. A nonlinear fluctuation is able to generate a new model with the parameter

$$\gamma_{lo} = \frac{\beta_{lo}(t_o)}{\alpha_{lo}(t_o)} = \frac{\beta_n^i(t_{n+k})}{\alpha_n^i(t_{n+k})} = \frac{2\cos(\beta_{n+k-1}^i t) - 1}{2\sin(\beta_{n+k-1}^i t)}, \tag{A4.1}$$

where $\alpha_{lo}(t_o) = \alpha_i^*(t_{n+k})$, $\beta_{lo}(t_o) = \beta_i^*(t_{n+k})$ are the new model's starting real and imaginary eigenvalues.

*Proof.* During the oscillations, initiated by the control action, a component $\beta_i^*$ is selected from the model's imaginary eigenvalues $\text{Im}\lambda_n^i(t) = \text{Im}[-\lambda_{n-1}^i(2 - \exp\lambda_n^i t)^{-1}]$, at each $t = (t_{n+k-1}, t_{n+k})$.

We come to relation

$$\text{Im}\lambda_n^i(t_{n+k}) = j\beta_n^i(t_{n+k}) = -j\beta_{n+k-1}^i \frac{\cos(\beta_{n+k-1}^i t) - j\sin(\beta_{n+k-1}^i t)}{2 - \cos(\beta_{n+k-1}^i t) + j\sin(\beta_{n+k-1}^i t)}, \tag{A4.2}$$

at $\beta_i^* = \beta_{n+k}^i, \beta_i^* \neq 0 \pm \pi k$. It seen that $\beta_n^i(t_{n+k})$ includes a real component

$$\alpha_n^i(t_{n+k}) = -\beta_{n+k-1}^i \frac{2\sin(\beta_{n+k-1}^i t)}{(2 - \cos(\beta_{n+k-1}^i t))^2 + \sin^2(\beta_{n+k-1}^i t)}, \tag{A4.3}$$

at $\alpha_i^* = \alpha_i^*(t_{n+k}) \neq 0$, with the corresponding parameter

$$\gamma_i^* = \frac{\beta_n^i(t_{n+k})}{\alpha_n^i(t_{n+k})} = \frac{2\cos(\beta_{n+k-1}^i t) - 1}{2\sin(\beta_{n+k-1}^i t)}. \tag{A4.4}$$

These eigenvalues $\lambda_i^*(t_{n+k}) = \alpha_i^*(t_{n+k}) \pm j\beta_i^*(t_{n+k})$, at some moment $t_0 > t_n$, could give a start to a new forming macromodel with $\lambda_{lo} = \alpha_{lo}(t_o) \pm j\beta_{lo}(t_o)$, with the initial real $\alpha_{lo}(t_o) = \alpha_i^*(t_{n+k})$, the imaginary $\beta_{lo}(t_o) = \beta_i^*(t_{n+k})$, and the parameter

$\gamma_{lo} = \frac{\beta_{lo}(t_o)}{\alpha_{lo}(t_o)}$ equals to (A4.1). •

*Comments* A4.1. This new born macromodel might continue the consolidation process of its eigenvalues. Therefore, returning to some primary model's eigenvalues and repeating the cooperative process is a quite possible after ending the preceding consolidations and arising the periodical movements. This leads to the cyclic macromodel functioning when the state integration alternates with the state disintegration and the system decay. (The time of the macrosystem decay increases with an increase in the accuracy $\varepsilon^* = \frac{\Delta x_n}{x_n}$ of reaching the given final state). Because the system instability corresponds to $\gamma \geq 1$, while $\gamma \to 0$ corresponds to the start of the cooperative process, we might associate the cycle *start* (at the end the invariant loop (Fig.5.1b)) with the model's oscillations at $\gamma \geq 1$ (generating the spectrum $\lambda_i^*(t_{n+k})$), and the cycle *end* with the start of the new born macromodel at $\gamma \to 0$. In this case, $\gamma \to 0$ can be achieved at $2\cos(\beta_{n+k-1}^i t) \to 1$, or at

$(\beta_{n+k-1}^i t) \to (\pi/3 \pm \pi k), k = 1, 2, \ldots$, with $\alpha_n^i(t_{n+k}) = \alpha_{lo}^m(t_o) = \lambda_{lo}^m \cong -0.577 \beta_{n+k-1}^i, \beta_{lo}(t_o) \cong 0$,

whereas $\gamma = 1$ corresponds to $(\beta_{n+k-1}^i t) \approx 0.423 rad(24.267^o)$, with $\beta_i^*(t_{n+k}) \cong -0.6\beta_{n+k-1}^i$.

Here $\beta_i^*(t_{n+k}) \cong \alpha_{lo}^m(t_o)$ determines the maximal frequency $\omega_m^*$ of the fluctuation by the end of the optimal movement. The new macromovement starts with this initial frequency.

*Proposition* 4.2. The maximal *frequency's ratio*, generated by the initial (*n*-1) dimensional spectrum with an imaginary eigenvalue $\beta_{n-1,o}(t_{n-1,o})$ (at the end of the cooperative movement at ($\gamma = 1$)):

$$\frac{\beta_i^*(t_{n+k})}{\beta_{n-1,o}(t_{n-1,o})} = l_{n-1}^m \tag{A4.5}$$

is *estimated by the invariant*

$$\frac{0.577\pi/3}{a_o(\gamma)/a(\gamma)\ln 2} = l_{n-1}^m(\gamma = 1). \tag{A4.6}$$

*Proof.* Let us estimate (A4.5) using the following relations.

Because $\beta_{n-1,o}(t_{n-1,o}) = \alpha_{n-1,o}(t_{n-1,o})$ at $\gamma = 1$ and $\beta_i^*(t_{n+k}) \cong \alpha_{lo}^m(t_o)$, we come to $l_{n-1}^m = \frac{\alpha_{lo}^m(t_o)}{\alpha_{n-1,o}(t_{n-1,o})}$.

Applying relations $\alpha_{n-1,o}(t_{n-1,o}) = \alpha_{n-1,t}(t_{n-1}) a_o(\gamma)/a(\gamma)$ and using $\alpha_{lo}^m(t_o) = -0.577\pi/3$, we have

$$\frac{\alpha_{lo}^m(t_o)t_o}{\alpha_{n-1,o}(t_{n-1})t_o} = \frac{0.577\pi/3}{a_o(\gamma)/a(\gamma)\alpha_{n-1,t}(t_{n-1})t_o} = l_{n-1}^m(\gamma = 1). \tag{A4.7}$$

Assuming that a minimal $t_o$ starts at $t_o = t_{n-1} + o(t) \cong t_{n-1}$, and using (8.42): $\alpha_{n-1,t}(t_{n-1})t_o = \ln 2$, at $\alpha_{n-1,t}(t_{n-1}) = \alpha_{n-1}$, we get the invariant relation (A4.6). •

*Comments* A.4.2. The ratio (A4.6) at $\mathbf{a}_o(\gamma = 1) = 0.58767$, $\mathbf{a}(\gamma = 1) = 0.29$ brings $l_{n-1}^m(\gamma = 1) \approx 0.42976$.

Because $\alpha_{n-1,o}(t_{n-1,o})$ is the initial eigenvalue, generating a starting eigenvalue $\alpha_{lo}^m(t_o)$ of a new model, their ratio $\frac{\alpha_{n-1,o}(t_{n-1,o})}{\alpha_{lo}^m(t_o)} = \gamma_l^{\alpha m}$ determines the parameter of the eigenvalue multiplication for the new formed model, equal to $\gamma_l^{\alpha m} = (l_{n-1}^m)^{-1}$, or at $\gamma = 1$, we get $\gamma_l^{\alpha m} \approx 2.327$. For the new model, the corresponding parameter $\gamma_l \to 0$, at which the cooperative process and the following evolutionary development might start, holds true. For a new formed potential triplet, satisfying relations (8.16), we get the eigenvalues and their parameters of multiplication in the forms:

$\alpha_{l-1,o}^m(t_o) = \alpha_{n-1,o}(t_{n-1,o}) \approx -1.406$, $\alpha_{lo}^m(t_o) \approx -0.60423$, $\alpha_{l+1,o}^m(t_o) \approx -0.26$, with

$$\gamma_{l-1}^{\alpha m}(\gamma) = \frac{\alpha_{l,o}^m(t_o)}{\alpha_{l+1,o}(t_o)} \approx 2.3296 \text{ and } \gamma_{l-1,l+1}^{\alpha m}(\gamma) = \frac{\alpha_{l-1,o}^m(t_o)}{\alpha_{l+1,o}(t_o)} \approx 5.423. \tag{A4.8}$$

If the optimal cooperation in the new formed model is continued, both parameters $\gamma_l$ and $\gamma_l^{\alpha m}$ will be preserved, which determines both model's invariants and the information genetic code.

The transferred invariants are the carriers of the evolutionary hierarchical organization, self-control, and adaptation (while the adaptation process could change both $\gamma_l$ and $\gamma_l^{\alpha m}$). •

*A5. Analysis of the surface structure formed by the IN cellular geometry: the cell-space area, its curvature and rotation dynamics*

The surface area (Fig.5), defined by the function of its current radius $y$ in the form $F = \pm \pi y^2$ (Lerner 2010), is shaped by rotation of a hyberbolic function $y = a/x$ around axis $0 - x$.

Here $a \sim \alpha_o^n$, where $\alpha_o^n$ is a parameter of the hyperbola for a fixed model's dimension $n$.

For the considered information model $y \sim \alpha_t^i, x \sim t$, where $\alpha_t^i$ is a current entropy derivation (current model's eigenvalue) at the moment $t$. A sequence of the model's eigenvalues, located along the hierarchy of the IN's triplet's nodes, holds function

$$\alpha_t^i(t_m) = \alpha_o^n / t_m, \tag{A5.1}$$

where $\alpha_t^i(t_m)$ is the node's eigenvalue at the discrete moment $t_m$ of creating a $m$-th triplet, $\alpha_o^n = \mathbf{a}(\gamma_m^\alpha)$ is a model invariant depending on the triplet's ratio $\gamma_m^\alpha(\gamma)$ along the hyperbola (1) and $\gamma$ is a parameter of the macrodynamics.

Location of each triplet's node forms a spot with a space area $f_o$ having an information measure $\alpha_t^i(t_m)$.

Each node's spot is measured by the node's code, which consists of the number of information cells $m_c$ having a space area $f_o^c$ for each cell. Thus, $m_c f_o^c = f_o$ and the surface area $F$ which is covered by the number of $N_e$ space cell's areas at $F = N_e f_o^c$.

Let us find the spot area for each triplet's node

$$f_o^m = \Delta F^m = \pm 2\pi \alpha_m \Delta \alpha_m \text{ at } F^m = \pm \pi \alpha_m^2, \alpha_t^m(t_m) = \alpha_m. \tag{A5.2}$$

Curvature of the information phase space at the moment of forming a $m$-th triplet is defined (Lerner,2006) as

$K_\alpha^m = -3\alpha_m \dot{\alpha}_m$.

An increment of this curvature, related to the spot area, during the time interval $\Delta t_m$ of forming the spot area is

$$K_\alpha^{\Delta t_m} = K_\alpha^m \Delta t_m = -3\alpha_m \Delta \alpha_m. \tag{A5.3}$$

A space size of this spot area is changed along the curved surface. To find a cellular spot area being independent of its curvature $f_o^{Km}$, we apply the ratio of the spot area to the increment of the curvature during this area's formation:

$$f_o^m / K_\alpha^{\Delta t_m} = f_o^{Km}. \qquad (A5.3a)$$

Using relations (A5.2), (A5.3) and (A.5.3a), we get the cellular spot area $f_o^{Km} = \mp 2\pi / 3$. (A5.3b)

According to the IN cellular geometry, each node's cellular spot contains of the number of the DSS cells $m_c = 4$.

Thus the spot area belonging to each cell is $f_o^{CKm} = \mp \pi / 6$. (A5.4)

Following the DSS code structure, this spot area enfolds 1 bit of information of that code. Therefore, the number of bits $N_e^b$ that a total space area $F$ enfolds is $N_e^b = F / f_o^{CKm}$, or for each current $m$ we have the number of bits

$$N_e^{bm} = F^m / f_o^{CKm} = -6\alpha_m^2. \qquad (A5.5)$$

(Actually, because $\alpha_m$ is measured in bits per sec, both space area (A5.2) and $N_e^b$ are evaluated in $bit^2 / \sec^2$).

Now let us find how both the space area and its curvature are changed between the nearest triplets.

Because each of the $m$-th triplet is formed by joining three eigenvalues of the previous $(m-1)$-th triplet, this requirement imposes a limitation on the ratio of a nearest triplet's eigenvalues.

According to each eigenvalue's dynamics, we have for each $m$-triplet's third eigenvalue at the moment $\tau_m$ (before the triplet's formation) the following relations:

$$\alpha_{3\tau}^m = \alpha_{3o}^m (\mathbf{a}(\gamma_m^\alpha) / \mathbf{a}_o(\gamma_m^\alpha)),\ \alpha_{3o}^m = \alpha_{1o}^m / \gamma_m^\alpha \text{ and } \alpha_{1\tau}^m = \alpha_{1o}^m (\mathbf{a}(\gamma_m^\alpha) / \mathbf{a}_o(\gamma_m^\alpha)), \qquad (A5.6)$$

where $\alpha_{1o}^m, \alpha_{3o}^m$ are the triplet's first and third initial eigenvalues accordingly, and $(\mathbf{a}(\gamma_m^\alpha) / \mathbf{a}_o(\gamma_m^\alpha)) = inv$ (A5.6a) is the invariant ratio of the model's dynamic invariants. Thus, we have $\alpha_{3\tau}^m = \alpha_{1\tau}^m / \gamma_m^\alpha$. (A5.6b)

By the moment of the triplet's formation $\tau_m$ all its three eigenvalues become equal: $\alpha_{3\tau}^m = \alpha_{2\tau}^m = \alpha_{1\tau}^m$, and at the *moment* of triplet's formation $\tau_m + o$ we come to the following relations for a joint triplet's eigenvalue

$$\alpha_3^m(\tau_m + o) = 3\alpha_{3\tau}^m = \alpha_m,$$

where $\alpha_m$ holds the total $m$-th triplet's eigenvalues. Substituting (A5.6b) we get $\alpha_m = 3\alpha_{1\tau}^m / \gamma_m^\alpha$. And because the triplet's first eigenvalue $\alpha_{1\tau}^m$ holds the eigenvalues of the previous $(m-1)$-th triplet: $\alpha_{1\tau}^m = \alpha_{m-1}$, we get relation

$$\alpha_m / \alpha_{m-1} = 3 / \gamma_m^\alpha. \qquad (A5.7)$$

This allows us to write the ratio of a nearest square areas in the form

$$F_m / F_{m-1} = \alpha_m^2 / \alpha_{m-1}^2 = (3 / \gamma_m^\alpha)^2. \qquad (A5.8)$$

A curvature of a $m$-th triplet:

$K_\alpha^m = -3\alpha_m \Delta \alpha_m / \Delta t_m$ at $\alpha_m = (\alpha_m - \alpha_{m-1}), \Delta\alpha_m / \alpha_m = (1 - \alpha_{m-1} / \alpha_m) = 1 - 3/\gamma_m^\alpha$ and

$\Delta t_m / t_m = (1 - t_{m-1} / t_m) = (1 - \alpha_m / \alpha_{m-1}) = 1 - 3(\gamma_m^\alpha)^{-1}$, at $\mathbf{a}(\gamma_m^\alpha) = \alpha_m t_m$,

acquires the form

$$K_\alpha^m = -3\alpha_m(\alpha_m / t_m)(1 - 3/\gamma_m^\alpha)/(1 - (3/\gamma_m^\alpha)^{-1}) = 9\alpha_m^3 / \gamma_m^\alpha \mathbf{a}(\gamma_m^\alpha),\ K_\alpha^{m-1} = 9\alpha_{m-1}^3 / \gamma_{m-1}^\alpha \mathbf{a}(\gamma_{m-1}^\alpha). \quad (A5.8)$$

where at $\mathbf{a}(\gamma_m^\alpha) = \mathbf{a}(\gamma_{m-1}^\alpha)) = inv$ and $\gamma_{m-1}^\alpha = \gamma_m^\alpha = \gamma_{m=1}^\alpha$, we come to

$$K_{m-1} / K_m = (3 / \gamma_{m-1}^\alpha)^3. \qquad (A5.8a)$$

We get also a relative curvature for each relative area

$$K_m^F = (K_m / F_m) / (K_{m-1} / F_{m-1}) = (3/\gamma_m^\alpha).  \tag{A5.8b}$$

Because each following triplet brings the same contribution, a total area contribution, related to the contribution from the very first triplet, is $F_m / F_{m=1} = (F_{m=2} / F_{m=1})^{(m-1)} = (3/\gamma_{m=1}^\alpha)^{2(m-1)}$. (A5.9)

Here the values of $\gamma_{m-1}^\alpha = \gamma_m^\alpha = \gamma_{m=1}^\alpha < 3$ are crucial, because at $\gamma_{m-1}^\alpha = \gamma_m^\alpha = \gamma_{m=1}^\alpha = 3$ we get $\alpha_m / \alpha_{m-1} = 1$, and all model's eigenvalues will repeat the initial one; in this case, the IN structure of its $m$ triplets disappears.

Therefore, at the above conditions, limiting the eigenvalues ratio, both the space area and its curvature increase with moving from one triplet to another one along the IN space structure, which is associated with the model's time course $t \to T$. During this move the node's space is compressed by enfolding the space spot of each previous node.

However, both the node spot area $f_o^{Km}$ and the cell spot area $f_o^{CKm}$ are not changed enfolding the same numbers of bits of the DSS code. For example, with growing triplet's number, a relative curvature $K_m / K_{m-1} = (3/\gamma_{m-1}^\alpha)^3$ increases at any fixed $\gamma_{m-1}^\alpha < 3$, specifically at $\gamma_{m-1}^\alpha = 2.3$, it takes the value $K_{m+1,o} / K_{m,o} \cong 1.3^3 \cong 2.2$.

The relative areas for the nearest triplets and regarding a very first triplet are accordingly:
$F_{m+1,o} / F_{m,o} \cong 1.3^2 = 1.69$ and $F_m / F_{m=1} \cong (n/2 - 1)1.69$.

An absolute value of the space area at its final moment $T$ of creation is
$$F_m(T) = \pm \pi \alpha_\tau^2 \big|_{t_{m=1}}^T = \pm \pi (\alpha_m^2(T) - \alpha_{m=1}^2(t_{m=1})) = \pm \pi (\alpha_{m=1}^2(t_{m=1})[\alpha_m^2(T)/\alpha_{m=1}^2(t_{m=1}) - 1],$$
$$F_m(T) = \pm \pi \alpha_{m=1}^2(t_{m=1})[(3/\gamma_{m=1}^\alpha)^{2(m-1)} - 1], \tag{A5.10}$$

where $\alpha_{m=1}(t_{m=1}) = \alpha_{m=1}$ is the eigenvalues of the first triplet being formed at the moment $t_{m=1}$.

According to the triplet's structure, we have $\alpha_{m=1}(t_{m=1}) = \alpha_{om=1}(t_{om=1})/\gamma_{m=1}^\alpha$, where $\alpha_{om=1}(t_{om=1})$ is the initial eigenvalue of this structure. It's seen that the total area in determined entirely by a structure of formation of the first triplet with a known initial eigenvalue. With growing $m \to n/2$ this area is growing approaching
$$F_{m=n/2}(T) = \pm \pi \alpha_{om=n/2}^2(t_{om=1})(3/\gamma_{m=1}^\alpha)^n, \tag{A5.11}$$

where $\alpha_{n/2}(t_o)$ is also grows with increasing dimension $n$. Actually, all above relations are measured not in an metrical measure like $[M]$ (but in terms of a macrodynamic information speed $\alpha_m = [bit/\sec]$, specifically, being defined by the information speed of the initial IN's eigenvalue $\alpha_{om} = [bit/\sec]$).

According to the conditions of the cooperation of the model's eigenvectors into a triplet (Lerner 2007, 2010), each such cooperation requires the rotation of each following eigenvector regarding each previous one on the angle $\varphi_i = \pm \pi/4$, while each of such a local rotation (within a triplet formation) generates a single cell-code. (See also footnote).

Formation of a whole triplet, which generates four such cell-codes, requires the rotation on the angle $\varphi_m = \pi$ for each $m$-th triplet, whereas a pair of the triplet's cell-code areas should be turned on angle $\varphi_{2m} = 2\pi$ completing the twist.

The $n$-dimensional model, having a manifold of triplet's cell-code areas $m = n/2$, located on the rotating hyperbola, creates an evolving spatial helix dynamic coding structure Fig.2.

Let us find the time of rotation $T_R$ for the triplet's square areas, located on hyperbola, around axis 0- T to produce the space area $F_m(T)$.

The hyperbola (1), which enfolds $m = n/2$ of the triplet's areas, rotates through all rotation path $m\pi R, R = \alpha_m(T)$ with a path $\pi R$ for each of the $m$-th triplet (corresponding to the rotating angle $\varphi_m = \pi$) to produce $F_m(T)$. However, because each of such a triplet area contains 4 bits, the time of rotation for a whole cell-bit hyperbolas' area (with $4m = 2n$ bits) is

$T_R = 4m\pi\alpha_m(T)/F_m(T) = 4m\pi\alpha_m(T)/(\pm\pi\alpha_m^2(T)) = [\{bit(cell)bit/\sec\}/\{bit^2/\sec^2\}] = [\sec]$.

Substituting $\alpha_m(T) = \alpha_{on}(t_{om=1})[(3/\gamma_{m=1}^\alpha)^m - 1]$, we get $T_R = \pm 4m/(\alpha_{om=1}(t_{om=1})[(3/\gamma_{m=1}^\alpha)^m - 1])$, (A5.12)

At growing dimension $n$ and the related triplet's number $m = n/2$, this time approaches

$$T_R = \pm 2n\alpha_{on}^{-1}(t_{om=1})(3/\gamma_{m=1}^\alpha)^{-n/2}.$$  (A5.12a)

The time of a particular $m$-th triplet rotation (A5.10) depends on its explicit location within the IN hierarchical structure $\gamma_m^\alpha = (\gamma_{m=1}^\alpha)^m$, which is determined by the triplet's number $m = 1, 2, ..., n/2$ (specifying this location), while triplet $m = n/2$ finalizes the $n$ dimensional model's structure.

Because all four cells in each triplet's area are moving simultaneously, each of those cells spends the same time during this rotation. Thus, a cell area (A5.4) ( that enfolds 1 bit) also rotates during the time $T_R$ and hence it moves with a rotation speed

$$C_R = 6\pi/T_R = 3\pi n^{-1}\alpha_{on}(t_{om=1})(3/\gamma_{m=1}^\alpha)^{n/2}.$$  (A5.13)

Considering the numerical examples: for the model with $n = 22, \alpha_{1o} \cong 476.4$, we get time

$T_R = 44/1.3^{11} 476.4 \cong 0.005\sec$, which in 5 times more than a minimal time interval for this dimension

$t_{1o} \cong 0.001$ sec. For the model with three triplets and a final control, having $n = 8, \alpha_{1o} \cong 4.36, t_{1o} \cong 0.161$, we get

$T_R = 16/1.3^4 4.36.4 \cong 1.285 \sec$.

Comparing these models with a fixed initial eiegenvalue: $\alpha_{1o} = 1$, we have $T_R(n = 22) = 44/1.3^{11} \cong 2.45\sec$ and

$T_R(n = 8) = 16/1.3^4 \cong 5.6\sec$ accordingly. That requires increasing the rotation speed from

$C_R(n = 8) \cong 3.366 cell/\sec$ to $C_R(n = 22) \cong 7.69 cell/\sec$, which corresponds to an accelerating evolution spiral structure at the same initial dimension and the starting eigenvalue. Such a speed's increases is needed consequently for each following triplet' helix regarding the previous one to enclose in it the previous formed triplet's coding structure.

The acceleration allows finishing all twists by the time (A5.12a) of forming a final IN cellular structure.

Thus, each cell of the space structure undergoes two kinds of motions:

(1)-with a macrodynamic speed $C_{\alpha_m} = [bit(cell)/\sec]$ and

(2)-with a speed of rotation $C_{Rm} = [bit(cell)/\sec]$ being orthogonal to the macrodynamic motion. Following these results, evolution of the surface area, as a function of the time $F = F(T_m)$, matches to its function $F = F(\gamma_m)$ at $\gamma_m \cong 2.3$.

We illustrated this function on Fig.7.

Importance of this function consists on only in revealing the model *space-time' evolution dynamics*, but also in focusing on the observer, possessing such an external space area, where all internal-external interactions take place and the results of Prop.1-5 are employed. At a collective interactions in a limited information space area, evolution of the oberver's space area is restricted by a maximal available information, which each observer can obtain.

<u>Footnotes.</u> Turning each of the triplet's eigenvector $\alpha_{i\tau o}$ on angle $\pm\pi/4$ will bring it's a maximal increment that is measured by the ratio of the eigenvector's modules: $\alpha_{i\tau}/\alpha_{i\tau o} = k_{i\tau}, k_{i\tau} = (\cos(\pi/4))^{-1} \cong 1.154$.

Because all three triplet's eigenvectors undergo sequentially such rotations, the total increment of these three rotations is

$k_{m\tau} = (k_{i\tau})^3 \cong 1.525$. This will decrease the initial ratio of the triplet's eigenvalue $\gamma_{mo}$ (prior to the eigenvalue's cooperation) to its value considered above: $\gamma_m = \gamma_{mo}/k_{m\tau}$.

Using an example of the model's computations (Lerner 1999,2010) with $\gamma_{mo} \cong 3.495$ (which corresponds its minimal value at $\gamma \cong 0.7$) we get $\gamma_m \cong 2.292$, which is approximated by $\gamma_m \cong 2.3$ above.

The dynamic invariants (A5.6a) are not changing at these rotations, because the time intervals, related to each of the rotating eigenvectors, will take the reverse changes, which preserve the invariants in the rotating macrodynamic process.

a).

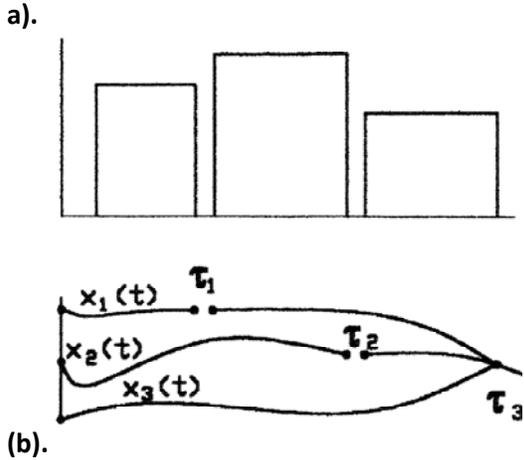

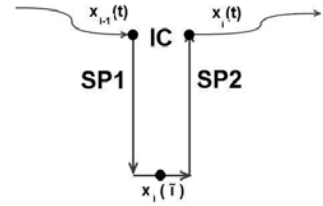

(b).

**Fig.1a. Selection of the process' information portions (segments)(a) SP2) with the windows between them, and *assembling* them in a triple, built at the windows *during* the initial flow's time-space movement(b).**

**Fig.1b. Applying controls: IC (SP1,**

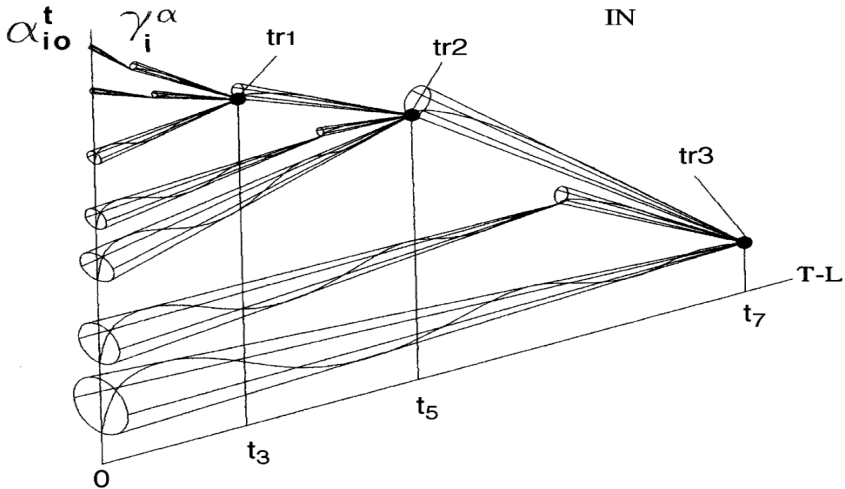

**Fig. 1. The IN time-space information structure, represented by the hierarchy of the IN cones' spiral space-time dynamics with the triplet node's (tr1, tr2, tr3, ..), formed at the localities of the triple cones vertexes' intersections, where $\{\alpha_{io}^t\}$ is a ranged string of the initial eigenvalues, cooperating around the $(t_1, t_2, t_3)$ locations; T-L is a time-space.**

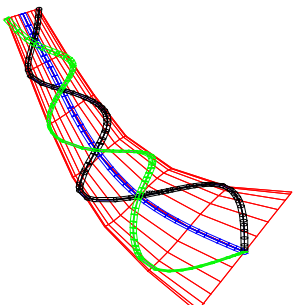

**Fig 2. Simulation of the double spiral code's structure (*DSS*), generated by the IN's nodes: the central line models the IN node's information cells; the left and right spirals encode the IN's states, chosen at the DPs by the IC control's double actions.**

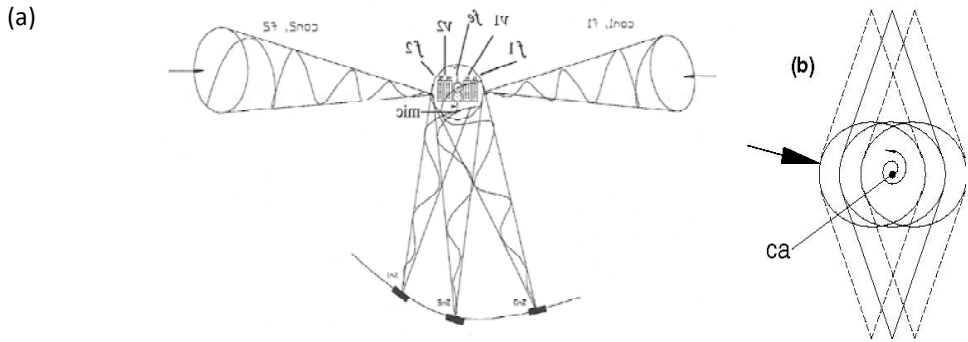

**Fig. 3.(a). An illustration of the competition and selection for the state's variations at the DP locality, brought by: the external frequencies $f1, f2$, the controls $v1, v2$, the microlevel $mic$, and the external influences $fe$, with forming the multiple node's spots $sn1, sn2, sn3$, generated by a chaotic attractors; the node, selected from the spot, is memorized. (b) Chaotic dynamics, generating an attractor (ca) at joining the cone's processes, which forms the following cone's basin.**

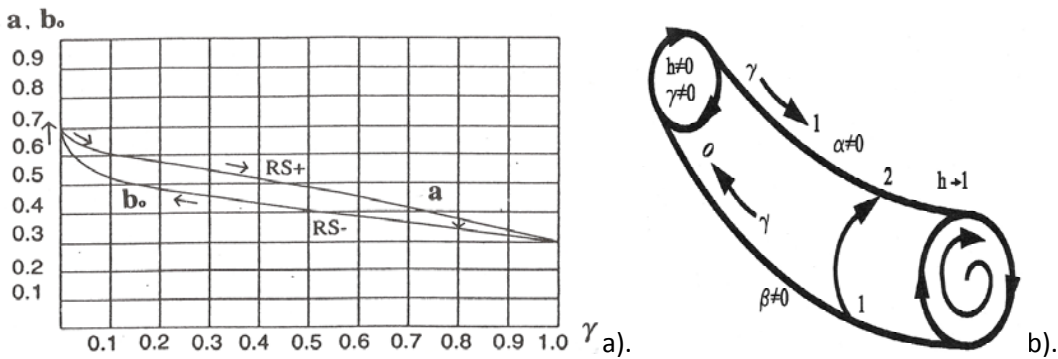

**Fig. 4. a). The evolution of the model invariants with a loop between information subspaces $RS+$ and $RS-$ b). A schematic of the cyclic process.**

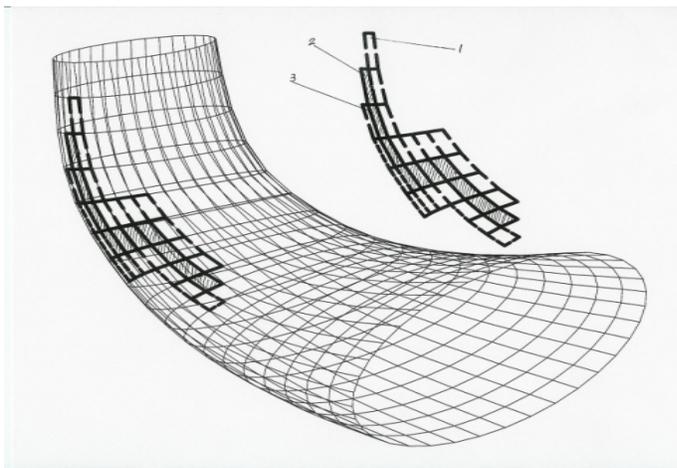

**Fig. 5. Structure of the cellular geometry, formed by the cells of the DSS triplet's code, with a portion of the surface cells (1-2-3), illustrating the space formation.**

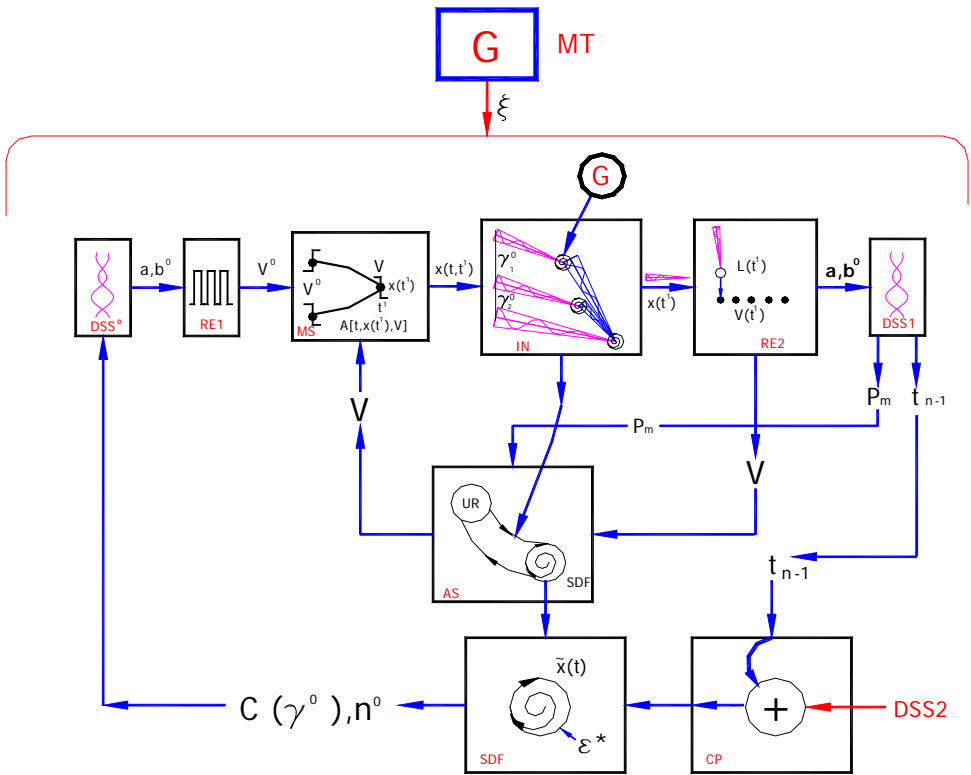

**Fig. 6. The functional schema of the evolutionary informational mechanisms**

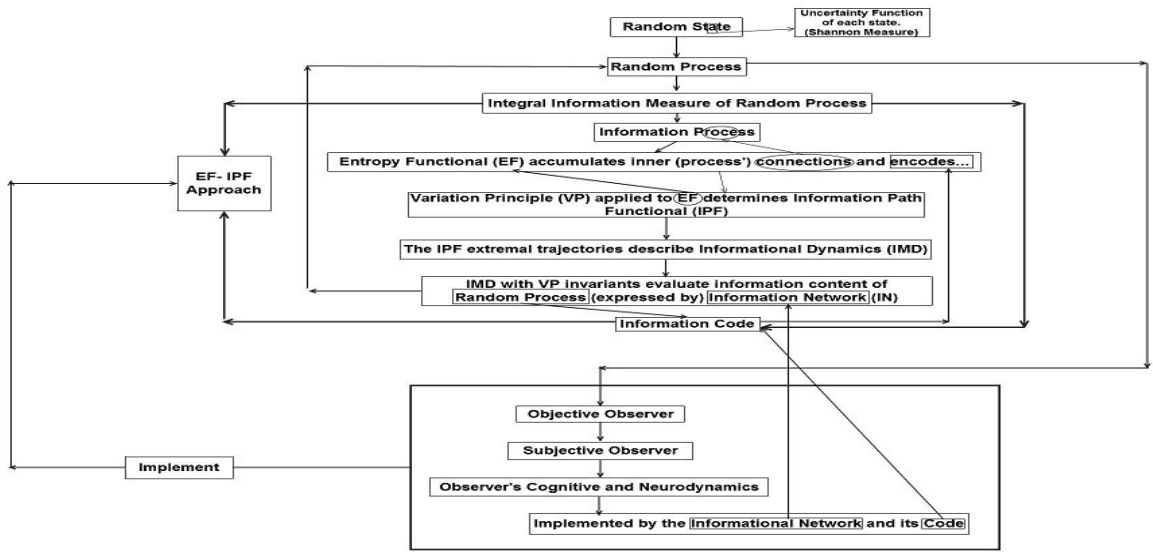

**Fig. A. Schema of the EF-IPF Approach's Structure, implemented by the information observer:** *it focuses on (a)-the difference between the Shannon's and EF information measures for the evaluation of isolated symbols (letter) and their connection in a process-by the EF ;(b)-the hierarchy of the approach's relationships by the links between the schema' objects.*

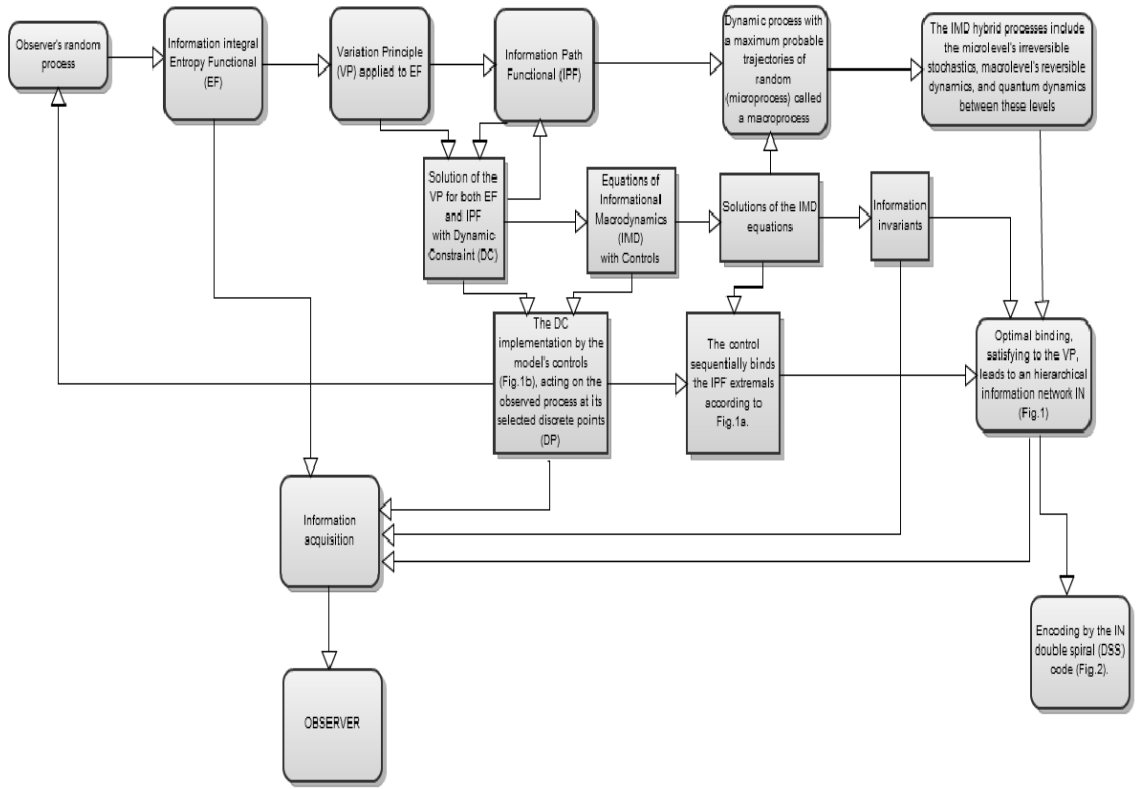

**Fig.B. The main operations in IPF-IMD formalism**

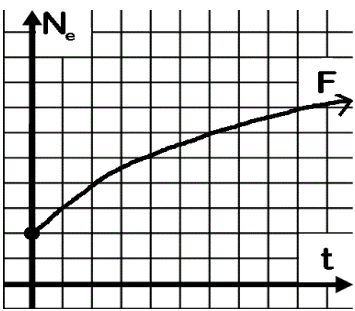

**Fig.7. The number of external elements $N_e$ as a function of obersver's external surface $F$.**

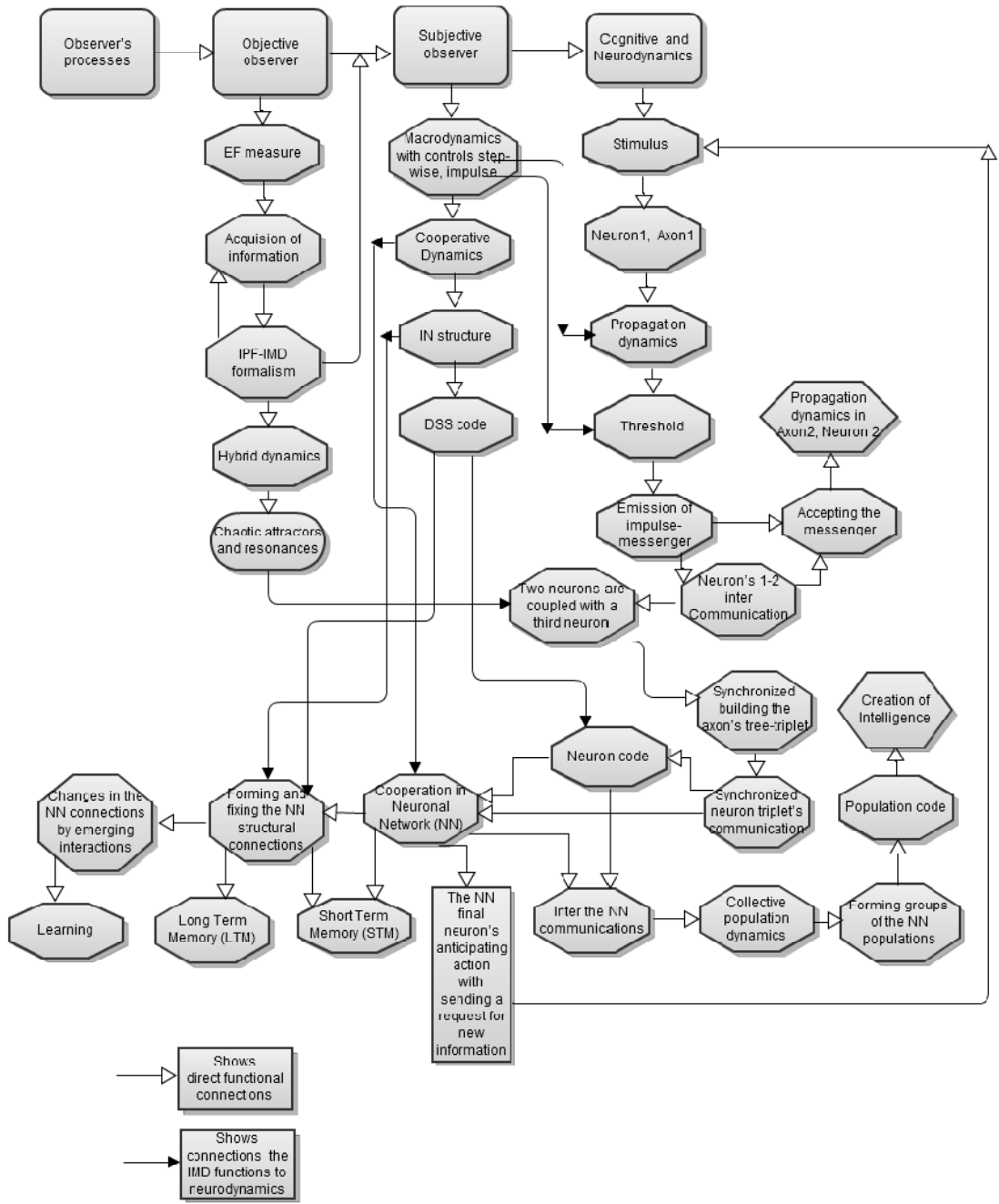

**Fig. C. The simplified schematics for the comparison of the main functional relationships between Neurodynamics and Informational Macrodynamics,** *illustrating the cited references and their connections to IMD.*